\pdfoutput=1 

\documentclass[usenatbib]{mn2e}
\usepackage{graphicx}

\usepackage[total={17.8cm,24.0cm},centering]{geometry}
\usepackage{times,enumerate,deluxetable}

\newcommand{\apj}{ApJ}           
\newcommand{\apjl}{ApJ}           
\newcommand{\mnras}{MNRAS}       
\newcommand{\nat}{Nature}
\newcommand{\aap}{A\&A}

\newcommand{\aj}{AJ}
\newcommand{\pasp}{PASP}
\newcommand{\apjs}{ApJS}           

\newcommand{\na}{New Ast.}

\newcommand{\sauron}{\texttt{SAURON}}
\newcommand{\atl}{ATLAS$^{\rm 3D}$}
\newcommand{\kms}{\hbox{km s$^{-1}$}}

\newcommand{\re}{\hbox{$R_{\rm e}$}}
\newcommand{\plotone}[1]{\includegraphics[width=\columnwidth]{#1}}
\newcommand{\refsec}[1]{Section~\ref{#1}}
\newcommand{\reffig}[1]{Fig.~\ref{#1}}
\newcommand{\refeq}[1]{equation~(\ref{#1})}

\title[The \atl\ project -- XV. Dynamical models]
{The \atl\ project -- XV. Benchmark for early-type galaxies scaling relations from 260 dynamical models: mass-to-light ratio, dark matter, Fundamental Plane and Mass Plane}

\author[M.~Cappellari et al.]
{Michele Cappellari,$^1$\thanks{E-mail: cappellari@astro.ox.ac.uk}
Nicholas Scott,$^{1,2}$
Katherine Alatalo,$^3$
Leo Blitz,$^3$
Maxime Bois,$^4$\and
Fr\'ed\'eric Bournaud,$^5$
M.~Bureau,$^1$
Alison F. Crocker,$^6$
Roger L. Davies,$^1$\and
Timothy A. Davis,$^{1,7}$
P. T. de Zeeuw,$^{7,8}$
Pierre-Alain Duc,$^{5}$
Eric Emsellem,$^{7,9}$\and
Sadegh Khochfar,$^{10}$
Davor Krajnovi\'c,$^7$
Harald Kuntschner,$^{7}$
Richard M. McDermid,$^{11}$\and
Raffaella Morganti,$^{12,13}$
Thorsten Naab,$^{14}$
Tom Oosterloo,$^{12,13}$
Marc Sarzi,$^{15}$\and
Paolo Serra,$^{12}$
Anne-Marie Weijmans$^{16}$
and Lisa M. Young$^{17}$\\
$^1$ Sub-department of Astrophysics, Department of Physics, University of Oxford, Denys Wilkinson Building, Keble Road, Oxford OX1 3RH\\
$^2$ Centre for Astrophysics \& Supercomputing, Swinburne University of Technology, PO Box 218, Hawthorn, VIC 3122, Australia\\
$^3$ Department of Astronomy, Campbell Hall, University of California, Berkeley, CA 94720, USA\\
$^4$ Observatoire de Paris, LERMA and CNRS, 61 Av. de l'Observatoire, F-75014 Paris, France\\
$^5$ Laboratoire AIM Paris-Saclay, CEA/IRFU/SAp CNRS Universit\'e Paris Diderot, 91191 Gif-sur-Yvette Cedex, France\\
$^6$ Department of Astrophysics, University of Massachusetts, 710 North Pleasant Street, Amherst, MA 01003, USA\\
$^7$ European Southern Observatory, Karl-Schwarzschild-Str. 2, 85748 Garching, Germany\\
$^8$ Sterrewacht Leiden, Leiden University, Postbus 9513, 2300 RA Leiden, the Netherlands\\
$^9$ Universit\'e Lyon 1, Observatoire de Lyon, Centre de Recherche Astrophysique de Lyon\\ and Ecole Normale Sup\'erieure de Lyon, 9 avenue Charles Andr\'e, F-69230 Saint-Genis Laval, France\\
$^{10}$ Max-Planck Institut f\"ur extraterrestrische Physik, PO Box 1312, D-85478 Garching, Germany\\
$^{11}$ Gemini Observatory, Northern Operations Centre, 670 N. A`ohoku Place, Hilo, HI 96720, USA\\
$^{12}$ Netherlands Institute for Radio Astronomy (ASTRON), Postbus 2, 7990 AA Dwingeloo, The Netherlands\\
$^{13}$ Kapteyn Astronomical Institute, University of Groningen, Postbus 800, 9700 AV Groningen, The Netherlands\\
$^{14}$ Max-Planck Institut f\"ur Astrophysik, Karl-Schwarzschild-Str. 1, 85741 Garching, Germany\\
$^{15}$ Centre for Astrophysics Research, University of Hertfordshire, Hatfield, Herts AL1 9AB, UK\\
$^{16}$ Dunlap Institute for Astronomy \& Astrophysics, University of Toronto, 50 St. George Street, Toronto, ON M5S 3H4, Canada\\
$^{17}$ Physics Department, New Mexico Institute of Mining and Technology, Socorro, NM 87801, USA}

\date{Accepted 2013 March 27. Received 2013 March 12; in original form 2012 August 16}

\pagerange{\pageref{firstpage}--\pageref{lastpage}} \pubyear{2013}

\begin{document}
\label{firstpage}
\maketitle

\clearpage
\begin{abstract}
We study the volume-limited and nearly mass selected (stellar mass $M_{\rm stars}\ga6\times10^9$ M$_\odot$) \atl\ sample of 260 early-type galaxies (ETGs, ellipticals Es and lenticulars S0s). We construct detailed axisymmetric dynamical models (JAM), which allow for orbital anisotropy, include a dark matter halo, and reproduce in detail both the galaxy images and the high-quality integral-field stellar kinematics out to about 1\re, the projected half-light radius. We derive accurate total mass-to-light ratios $(M/L)_e$ and dark matter fractions $f_{\rm DM}$, within a sphere of radius $r=\re$ centred on the galaxies. We also measure the stellar $(M/L)_{\rm stars}$ and derive a median dark matter fraction $f_{\rm DM}=13\%$ in our sample. We infer masses $M_{\rm JAM}\equiv L\times (M/L)_e \approx 2\times M_{1/2}$, where $M_{1/2}$ is the total mass within a sphere enclosing half of the galaxy light.  We find that the thin two-dimensional subset spanned by galaxies in the $(M_{\rm JAM},\sigma_e,R_e^{\rm maj})$ coordinates system, which we call the Mass Plane (MP) has an observed rms scatter of 19\%, which implies an intrinsic one of 11\%. Here $R_e^{\rm maj}$ is the major axis of an isophote enclosing half of the {\em observed} galaxy light, while $\sigma_e$ is measured within that isophote. The MP satisfies the scalar virial relation $M_{\rm JAM}\propto\sigma_e^2 R_e^{\rm maj}$ within our tight errors. 
This show that the larger scatter in the Fundamental Plane (FP) $(L,\sigma_e,R_e)$ is due to stellar population effects (including trends in the stellar Initial Mass Function [IMF]). It confirms that the FP deviation from the virial exponents is due to a genuine $(M/L)_e$ variation. 
However, the details of how both \re\ and $\sigma_e$ are determined are critical in defining the precise deviation from the virial exponents. The main uncertainty in masses or $M/L$ estimates using the scalar virial relation is in the measurement of \re. This problem is already relevant for nearby galaxies and may cause significant biases in virial mass and size determinations at high-redshift. Dynamical models can eliminate these problems. 
We revisit the $(M/L)_e-\sigma_e$ relation, which describes most of the deviations between the MP and the FP. The best-fitting relation is $(M/L)_e\propto \sigma_e^{0.72}$ ($r$-band). It provides an upper limit to any systematic increase of the IMF mass normalization with $\sigma_e$. The correlation is more shallow and has smaller scatter for slow rotating systems or for galaxies in Virgo. For the latter, when using the best distance estimates, we observe a scatter in $(M/L)_e$ of 11\%, and infer an intrinsic one of 8\%. We perform an accurate empirical study of the link between $\sigma_e$ and the galaxies circular velocity $V_{\rm circ}$ within 1\re\ (where stars dominate) and find the relation $\max(V_{\rm circ}) \approx 1.76\times \sigma_e$, which has an observed scatter of 7\%. The accurate parameters described in this paper are used in the companion Paper~XX of this series to explore the variation of global galaxy properties, including the IMF, on the projections of the MP.
\end{abstract}


\section{Introduction}

Scaling relations of early-type galaxies (ETGs, ellipticals E and lenticulars S0) have played a central role in our understanding of galaxy evolution, since the discovery that the stellar velocity dispersion $\sigma$ \citep{Minkowski1962,Faber1976} and the galaxy projected half-light radius $\re$ \citep{Kormendy1977} correlate with galaxy luminosity $L$. An important step forward was made with the discovery that these two relations are just projections of a relatively narrow plane, the Fundamental Plane (FP) \citep{Faber1987,Dressler1987,Djorgovski1987}, relating the three variables $(L,\sigma_e,\re)$. When the plane is used as a distance indicator, as was especially the case at the time of its discovery, the luminosity can be replaced by the surface brightness within \re\ as $\Sigma_e\equiv L/(2\pi R_{\rm e}^2)$ and the observed plane assumes the form
\begin{equation}
\re \propto \sigma^{1.33} \Sigma_e^{-0.82},
\end{equation}
where the adopted parameters are the median of the 11 independent determinations tabulated in \citet{Bernardi2003fp}.

It was immediately realized that the existence of the FP could be due to the galaxies being in virial equilibrium \citep[e.g.][]{Binney2008} and that the deviation (tilt) of the coefficients from the virial predictions $\re \propto \sigma^2 \Sigma_e^{-1}$, could be explained by a smooth power-law variation of mass-to-light ratio $M/L$ with mass \citep{Faber1987}. The FP showed that galaxies assemble via regular processes and that their properties are closely related to their mass. The tightness of the plane gives constraints on the variation of stellar population among galaxies of similar characteristics and on their dark matter content \citep{Renzini1993,Borriello2003fp}. The regularity also allows one to use the FP to study galaxy evolution, by tracing its variations with redshift \citep{vanDokkum1996}.

However, other reasons for the deviation of the coefficients are possible: the constant coefficients in the simple virial relation only rigorously apply if galaxies are spherical and homologous systems, with similar profiles and dark matter fraction. But both galaxies concentration \citep{Caon1993} and the amount of random motions in their stars \citep{Davies1983} were found to systematically increase with galaxy luminosity.

The uncertain origin of the tilt led to a large number of investigations about its origin, exploring the effects of (i) the systematic variation in the stellar population or IMF \citep[e.g.][]{Prugniel1996fp,Forbes1998fp}, or (ii) the non-homology in the surface brightness distribution \citep[e.g.][]{Prugniel1997fp,Graham1997fp,Bertin2002fp,Trujillo2004fp} or (iii) the kinematic \citep[e.g.][]{Prugniel1994,Busarello1997}, or (iv) the variation in the amount of dark matter \citep[e.g.][]{Renzini1993,Ciotti1996,Borriello2003fp}, on the FP tilt and scatter.
Those works were all based on approximate galaxy spherical models, trying to test general hypotheses and not reproducing real galaxies in detail, which sometimes led to contrasting results. What became clear however was that various effects could potentially influence a major part of the FP tilt. Moreover it was found that the small scatter in the FP implies a well regulated formation for ETGs.

The next step forward came with subsequent studies, which instead of testing general trends, used small samples of objects and tried to push to the limit the accuracy of measuring galaxy central masses, while reducing biases as much as possible. Those accurate total masses could be directly compared to the simple virial ones, testing for residual trends. Similar but independent studies were performed using two completely different techniques, either stellar dynamics \citep{Cappellari2006} or strong gravitational lensing \citep{Bolton2007mp,Bolton2008,Auger2010}. The results from those efforts agree with each others, and showed that the tilt of the FP is almost entirely due to a genuine $M/L$ variation.

In this paper we investigate once more the origin of the FP tilt. This new study is motivated by the dramatic increase in the size and quality of our galaxy sample, with respect to any previous similar study. We have in fact state-of-the-art \sauron\ \citep{Bacon2001} stellar kinematics for all the 260 early-type galaxies of the \atl\ sample \citep[hereafter Paper~I]{Cappellari2011a}, which constitute a volume-limited and carefully selected sample of ETGs, down to a stellar mass of about $M_{\rm stars}\ga6\times10^9$ M$_\odot$. This fact, combined with detailed dynamical models for the entire sample, allows us to test previous claims with unprecedented accuracy. The new models also include a dark matter halo and give constraints on the dark matter content in the centres of early-type galaxies. These measurements will be used in the companion \citet[hereafter Paper~XX]{Cappellari2012p20} to provide a novel view of galaxy scaling relations.

In what follows, in Section~2 we present the sample and data, in Section~3 we describe the methods used to extract our quantities, in Section~4 we present our results on the FP tilt, dark matter and the $(M/L)-\sigma$ relation, and finally we summarize our paper in Section~5.

\section{Sample and data}

\subsection{Selection}

The galaxies studied in this work are the 260 early-type galaxies which constitute the volume-limited and nearly mass-selected \atl\ sample (Paper~I). The object were morphologically selected as early-type according to the classic criterion \citep{Hubble1936,deVaucouleurs1959,Sandage1961} of not showing spiral arms or a disk-scale dust lane (when seen edge-on). The early-types are extracted from a parent sample of 871 galaxies of all morphological types brighter than $M_K=-21.5$ mag, using 2MASS photometry \citep{Skrutskie2006}, inside a local ($D<42$ Mpc) volume of $1.16\times10^5$ Mpc$^3$ (see full details in Paper~I).

\subsection{Comparison to previous samples: dynamics and lensing}

Our goal is to measure total masses, or equivalently mass-to-light ratios ($M/L$), in the central regions of galaxies. $M/L$ of significant samples of individual ETGs have been previously obtained via dynamical modelling (e.g.\ \citealt{vanDerMarel91} [37 ETGs]; \citealt{Magorrian1998} [36 ETGs]; \citealt{Gerhard2001} [21 ETGs]; \citealt{Cappellari2006} [25 ETGs]; \citealt{Thomas2007} [16 ETGs]; \citealt{Williams2009} [14 ETGs]; \citealt{Scott2009} [48 ETGs]) or strong gravitational lensing (e.g.\ \citealt{Rusin2003} [22 ETGs]; \citealt{Koopmans2006} [15 ETGs]; \citealt{Bolton2008} [53 ETGs]; \citealt{Auger2010} [73 ETGs]).
An important, and perhaps not obvious, difference between the quantities obtained with the two techniques is that the dynamical models provide masses enclosed within a {\em spherical} radius, while strong lensing measures the mass inside a {\em cylinder} with axis parallel to the line-of-sight. Care has to be taken when comparing the two methods. An illustration of this fact is given in figure~1 of \citet{Dutton2011swells}.

An advantage of the strong lensing technique is that the recovered mass inside a cylinder with the radius of the Einstein ring is nearly insensitive to the mass distribution, and completely independent on the stellar dynamics. However, the requirement of a galaxy to act as a strong lens, necessarily imposes biases in the objects selection, and in particular limits mass measurements via strong lensing to the most massive nearby ETGs ($\sigma\ga200$ km s$^{-1}$ in \citealt{Auger2010}).

The dynamical modelling technique has the significant advantage that it can in principle be applied to any bound system made of stars. However, it requires a detailed treatment of the observed surface brightness and orbital distribution, in combination with integral-field data, for robust and accurate values \citep[e.g.][]{Cappellari2006}.

In this paper we apply the stellar dynamical modelling technique to the \atl\ sample of 260 early-type galaxies. This increases the sample size for which accurate total masses have been measured by a factor of four. Moreover the sample is volume-limited and statistically representative of the nearby galaxy population with stellar mass $M_{\rm stars}\ga6\times10^9$ $M_\odot$ and in particular includes ETGs with velocity dispersion as low as $\sigma_e\approx40$ km s$^{-1}$ (see Paper~I for an illustration of the characteristics of the sample).

\subsection{Stellar kinematics and imaging}

Various multi-wavelengths datasets are available for the sample galaxies (see a summary in Paper~I). In this work we make use of the \sauron\ \citep{Bacon2001} integral-field stellar kinematics within about one half-light radius \re, which was introduced in \citet{Emsellem2004}, for the subset of 48 early-types in the \sauron\ survey \citep{deZeeuw2002}, and in Paper~I for the rest of the \atl\ sample. Maps of the stellar velocity for all the 260 galaxies were presented in \citet[hereafter Paper~II]{Krajnovic2011}.

In this paper we are not interested in the shape of the stellar line-of-sight velocity distribution (LOSVD), but we want to approximate velocity moments which are predicted by the \citet{Jeans1922} equations. In \citet{Cappellari2007} we used semi-analytic models to compute a set of realistic galaxy LOSVDs with known velocity moments, using the \citet{Hunter1993} formalism, as implemented in \citet{Emsellem1999}. The models LOSVDs were used to broaden galaxy spectral templates and noise was subsequently added. The kinematics was then extracted from the synthetic spectra using pPXF \citet{Cappellari2004} as done for the real galaxies. We found that $V_{\rm rms}\equiv\sqrt{V^2+\sigma^2}$, where $V$ and $\sigma$ are the mean and standard deviation of the best fitting Gaussian provide a better empirical approximation to the velocity second moment $\langle v_{\rm los}^2\rangle^{1/2}$ than an integral of a more general LOSVD described by the Gauss-Hermite parametrization \citep{vanDerMarel93,Gerhard1993}. This is due to the large sensitivity of the moments to the wings of the LOSVD, which are observationally ill determined. For this reason all the kinematic quantities used in the paper are extracted using a simple Gaussian LOSVD in the pPXF software (keyword MOMENTS$=$2).

The photometry used in this work comes from the Sloan Digital Sky Survey (SDSS, \citealt{York2000}) data release eight \citep[DR8][]{Aihara2011} for 225 galaxies and was supplemented by our own photometry taken at the 2.5-m Isaac Newton Telescope in the same set of filters and with comparable signal to noise for the rest of the sample galaxies \citep[hereafter Paper~XXI]{Scott2011}.

\section{Methods}

\subsection{Measuring galaxy enclosed masses}

\subsubsection{Choosing the dynamical modelling approach}

Various dynamical modelling techniques have been developed in the past. They are all characterized by their ability to reproduce in detail, in a {\em non-parametric} way, the characteristics of the galaxy surface brightness. This contrasts with a more qualitative toy-model approach \citep[e.g.][]{Tortora2009,Treu2010} that assume a spherical shape and a simpler parametrization  (e.~g. \citealt{Hernquist1990} or \citealt{Sersic1968} profile) for the surface brightness of all galaxies. An accurate description of the galaxy surface brightness is a necessary requirement for quantitative and unbiased measurements of dynamical quantities as much of the kinematic information on real galaxies is contained in the photometry alone \citep{Cappellari2008}. The state of the art in the field is currently represented by \citet{Schwarzschild1979} orbit-superposition approach, which was originally developed to reproduce galaxy stellar densities and was later generalized to produce detailed fits to the stellar kinematics \citep{Richstone1988,Rix1997,vanDerMarel98} and has been widely used for determinations of masses of supermassive black holes \cite[e.g.][]{vanderMarel1997,Gebhardt2000,Cappellari2002bh,Valluri2004,Houghton2006}, for galaxy mass determinations \citep[e.g.][]{Cappellari2006,Thomas2007} and to recover orbital distributions \citep[e.g.][]{Krajnovic2005,Cappellari2007,vanDenBosch2008,Thomas2009}. A close contender technique, but not as widely used, is the particle-based made-to-measure method of \citep{Syer1996} as implemented to reproduce kinematical observables by various groups \citep{deLorenzi07,Dehnen2009,Long2010}. When the gravitational potential is assumed to be known, and the particles are chosen to fully sample all integrals of motion, the method effectively corresponds to a particle-based analogue of Schwarzschild's method, and is expected to provide similar results. However, the method may be very useful when the potential is derived from the particles in a self-consistent way. Not much however is known about the convergence and uniqueness of the solution in this case.

The sophistication and generality of the dynamical models has reached a level that exceeds the amount of information that the observations of external galaxies can provide. As a result the observations are unable to uniquely constrain all the model parameters, which suffer from degeneracies \citep{Dejonghe1992,Gerhard1998,deLorenzi2009,Morganti2012}. A key degeneracy is in the deprojection of the observed surface brightness into a three dimensional stellar mass distribution, which has been proved to be of mathematical nature \citep{Rybicki1987,Gerhard1996} and applies even when the galaxy is assumed to be axisymmetric. However, similar degeneracies are likely to exists when higher (than zero) moments of the velocity are considered. This is expected from dimensional arguments: the current data provide at most a three-dimensional observable (an integral-field data cube), which is the minimum requirement to constrain the orbital distribution, which depends on three integrals of motion, for an assumed potential and known light distribution. It is unlikely for the data to contain enough information to constrain additional parameters, like the dark matter halo shape and the viewing angles \citep[e.g.][]{Valluri2004}. Numerical experiments confirm that even with the best available integral-field stellar kinematics, and assuming the gravitational potential is known and axisymmetric, not even the galaxy inclination can be inferred from the data using general Schwarzschild models \citep{Krajnovic2005,Cappellari2006,vandenBosch2009}. This implies that the mass distribution is also quite poorly known.

The situation becomes even more problematic when one considers the fact that the majority of early-type galaxies are likely to have bars. 30\% have obvious bars (Paper~II) in the \atl\ sample, but more must be hidden by projection effects. Bars are characterized by figure rotation which is ignored by most popular modelling approaches. The treatment of bars could be included in the models as demonstrated in the two-dimensional limit by \citep{Pfenniger1984} and as done to models the Milky Way in three dimension \citep{Zhao1996,Hafner2000,Bissantz2004}. However, no applications to external galaxies exists. This is due to the extra degeneracy that the addition of at least two extra model parameters, the bar pattern speed and position angle, will produce on an already degenerate problem. This combines with the dramatic increase in the non-uniqueness of the mass deprojection expected in a triaxial rather than axisymmetric distribution \citep{Gerhard1996triax} and in the additional unavoidable biases introduced by observational errors. All this is expected to further broaden the minima in the $\chi^2$ distributions of the fits and increase the uncertainties and covariances in the recovered parameters.

We chose a different approach. Rather than allowing for the full generality and degeneracies of the models, we adopt a modelling method that makes empirically-motivated assumptions to restrict the range of model solutions and improve the accuracy of the mass recovery. This is motivated by the finding that the kinematics of real fast-rotator early-type galaxies in the \sauron\ sample \citep{deZeeuw2002} is well approximated by models characterized by a remarkably simple and homogeneous dynamics, characterized by a cylindrically-aligned and nearly oblate velocity ellipsoid $\sigma_\phi\approx\sigma_R\ga\sigma_z$ \citep{Cappellari2008}, as previously suggested by more general Schwarzschild's models \citep{Cappellari2007,Thomas2009}. The models are called Jeans Anisotropic MGE (\textsc{JAM}), where MGE stands for the Multi-Gaussian Expansion method of \citet{Emsellem1994}, that is used to accurately describe the galaxy photometry. The \textsc{JAM} models can reproduce the full richness of the observed state-of-the-art \sauron\ integral-field kinematics of fast rotator ETGs using just two free parameters \citep{Cappellari2008,Scott2009,Cappellari2012}, providing a compact description of their dynamics. The JAM models are ideal for this work given that the nearly-axisymmetric fast rotator ETGs constitute the 86\% of the \atl\ sample (Paper~II; \citealt{Emsellem2011}, hereafter Paper~III). Moreover the JAM models only require the first two velocity moments ($V$ and $\sigma$), and not the full LOSVD, which is not available for about half of the sample (see Paper~I). The JAM models do not have the freedom to actually fit small-scale details of the kinematics, but they make a prediction based on an accurate description of the photometry and a couple of parameters. This constitutes an advantage in presence of noise and systematics in the data, as it makes spurious features easy to recognize and automatically exclude from the fit. Moreover the approach is at least three orders of magnitudes faster than Schwarzschild's approach.

Not all ETGs are well described by the JAM models however: some of the slow rotators in \atl\ are likely nearly spherical in the region where we have stellar kinematics, but about 10\% of the sample galaxies are weakly triaxial or out of equilibrium (Paper~II). For those objects the modelling results should be treated with caution. Errors of up to 20\% can arise when measuring masses of triaxial objects with axisymmetric models \citep{Thomas2007massRecovery,vandenBosch2009} and this should be kept in mind when interpreting our results. However, preliminary tests using real galaxies in the \sauron\ sample indicate excellent agreement between the $M/L$ recovery using axisymmetric models and triaxial ones with identical data \citep{vandenBosch2008PhDT}. Moreover, in what follows, unless explicitly mentioned, we verified that all conclusion are unchanged if we remove the slow rotator galaxies from the sample. Barred galaxies provide a further complication, which will be discussed in the next Section.

\subsubsection{JAM models with dark halo}
\label{sec:jam}

In practice the modelling approach we use in this paper starts by approximating the observed SDSS and INT $r$-band surface brightness distribution of the \atl\ galaxies using the Multi-Gaussian Expansion (\textsc{MGE}) parametrization \citep{Emsellem1994}, with the fitting method and \textsc{mge\_fit\_sectors} software package of \citet{Cappellari2002mge}\footnote{Available from http://purl.org/cappellari/idl}. The choice of the photometric band is a compromise between the need of using the reddest band, to reduce the contamination by dust, and the optimal signal-to-noise in the images. For barred galaxies the Gaussians of the \textsc{MGE} models are constrained to have the flattening of the outer disk, following \citet[their fig.~4]{Scott2009}. Full details of the fitting approach and illustrations of the quality of the resulting MGE fits are given in Paper~XXI. The \textsc{MGE} models are used as input for the \textsc{JAM} method$^1$ \citep{Cappellari2008} which calculates a prediction of the line-of-sight second velocity moments $\langle v^2_{\rm los}\rangle$ for given model parameters and compare this to the observed $V_{\rm rms}$.

In \citet{Cappellari2006} it was shown that, when the surface brightness distribution is accurately reproduced and good quality integral-field data are available, simple two-integral Jeans models measure masses nearly as accurate as those of Schwarzschild's models, with errors of 6\%. The agreement can be further improved by allowing for orbital anisotropy, in which case the two methods give equally accurate results \citep{Cappellari2008}. We have run an extensive set of tests using JAM to determine the $M/L$ of realistic numerical simulations \citep[hereafter Paper~XII]{Lablanche2012}. We found that for unbarred galaxies, even when the anisotropy is not accurately constant inside the region with kinematic data, the $M/L$ can be recovered with maximum biases as small as 1.5\%. The situation changes when the galaxies are barred. In this case biases of up to 15\% can be expected for the typical bar strengths we find in ETGs.

\begin{figure*}
\includegraphics[width=0.97\textwidth]{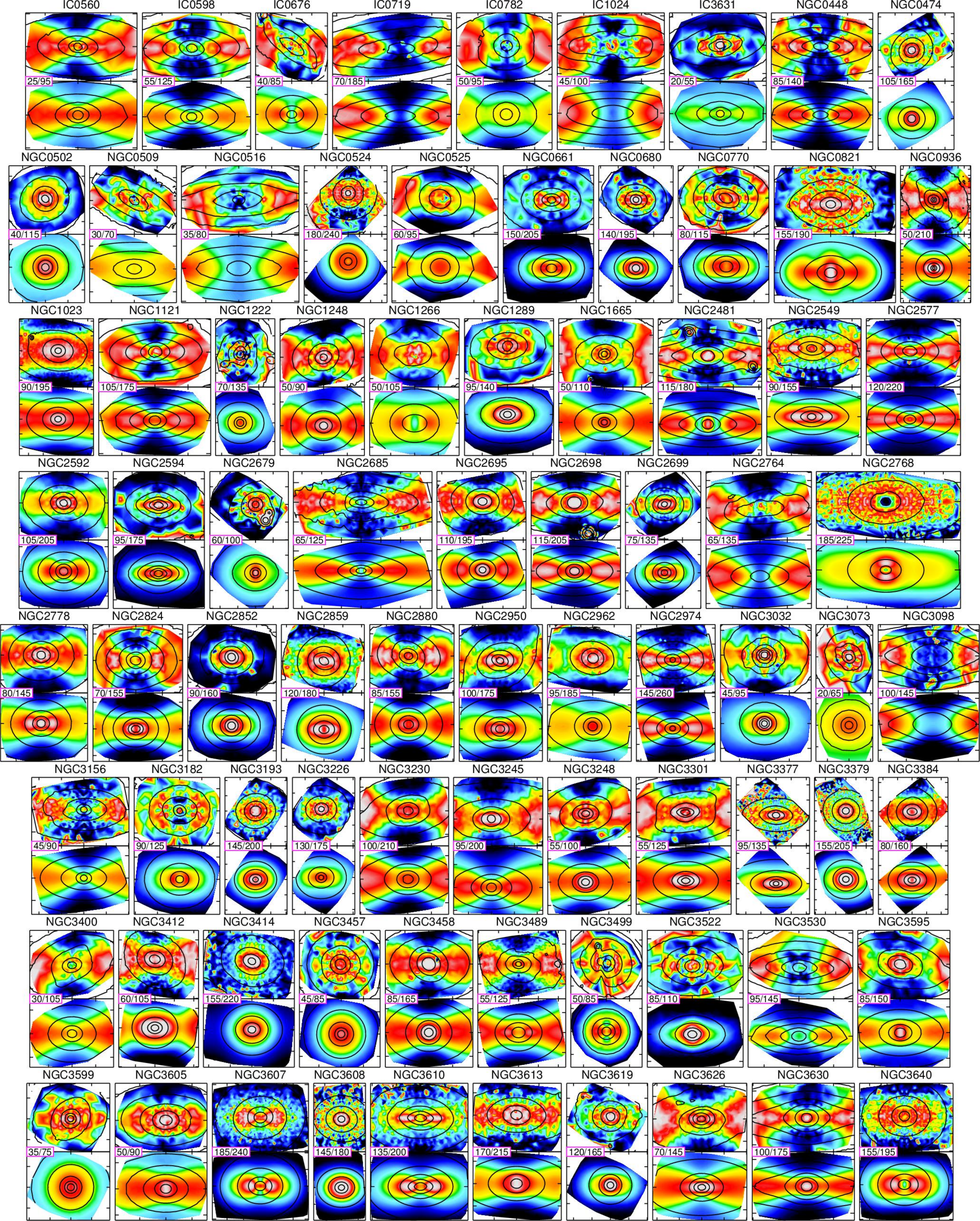}
\caption{Mass-follows-light JAM models of the \atl\ sample. In each panel, the top plot shows the by-symmetrized and linearly interpolated \sauron\ $V_{\rm rms}\equiv\sqrt{V^2+\sigma^2}$, where $V$ is the mean stellar velocity and $\sigma$ is the stellar velocity dispersion. $V_{\rm rms}$ ranges are printed. Ticks are separated by 10\arcsec. The observed galaxy surface brightness is overlaid, in steps of 1 mag. The bottom plot shows the best-fitting JAM model, and the adopted MGE surface brightness. These models (A) have just two free non-linear parameters, the inclination and the global anisotropy $(i,\beta_z)$, to reproduce the shape of the observed $V_{\rm rms}$. Yet, once the surface brightness is given, most of the variety in our maps can be reproduced. Nearly all significant deviations between data and models are due to bars, recognizable from the asymmetries in the observed surface brightness, dust, which affects both the mass model and the kinematics, or inferior data. The predictive power of these simple JAM models qualitatively suggest that the assumed total potential is not significantly in error, which implies dark matter is unimportant (or accurately follows the light). The good fits also show that ETGs have a simple dynamics within 1\re.
\label{fig:jam_vrms}}
\end{figure*}

\begin{figure*}
\includegraphics[width=0.97\textwidth]{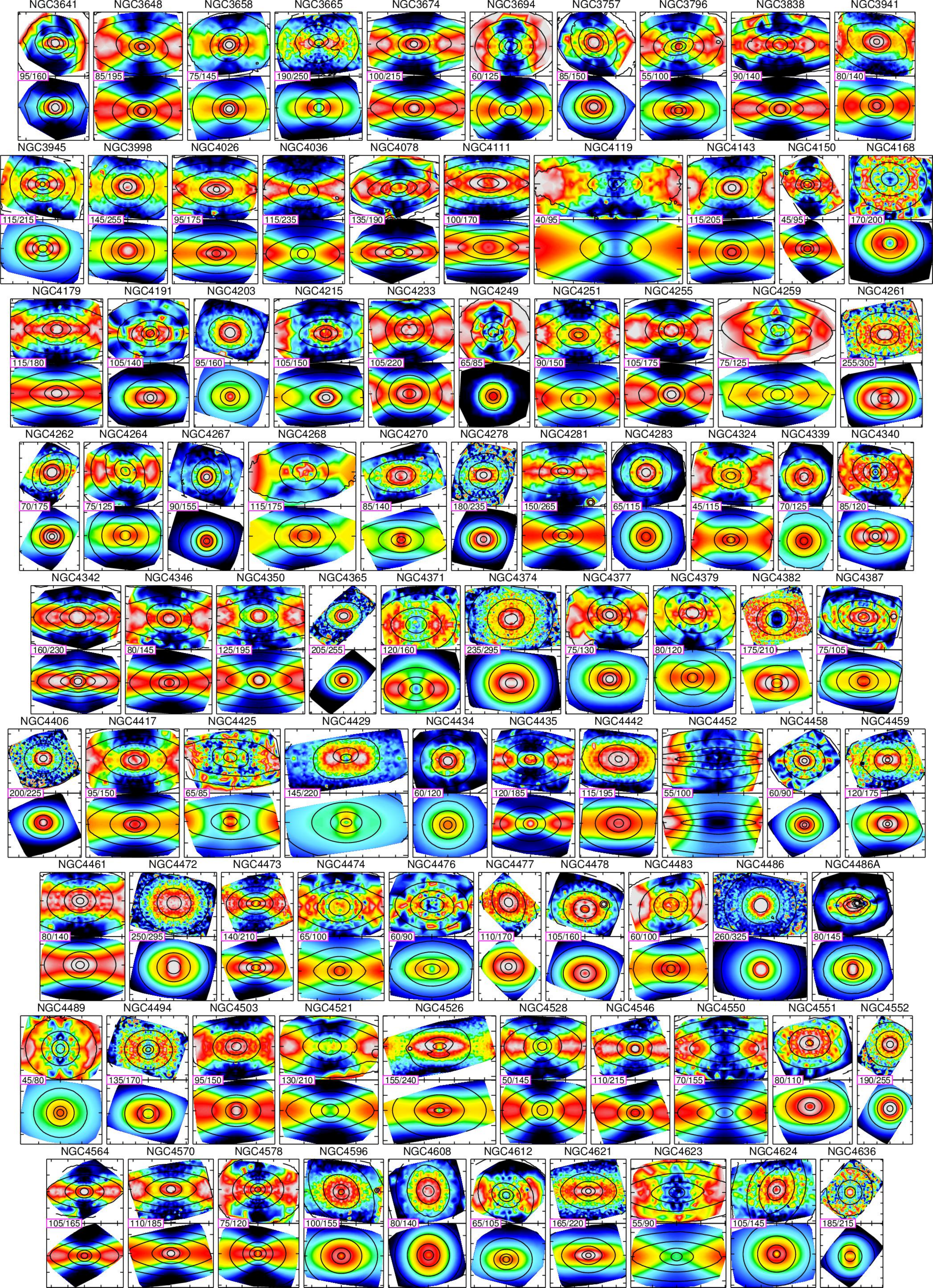}
\addtocounter{figure}{-1}
\caption{ --- continued}
\end{figure*}

\begin{figure*}
\includegraphics[width=0.97\textwidth]{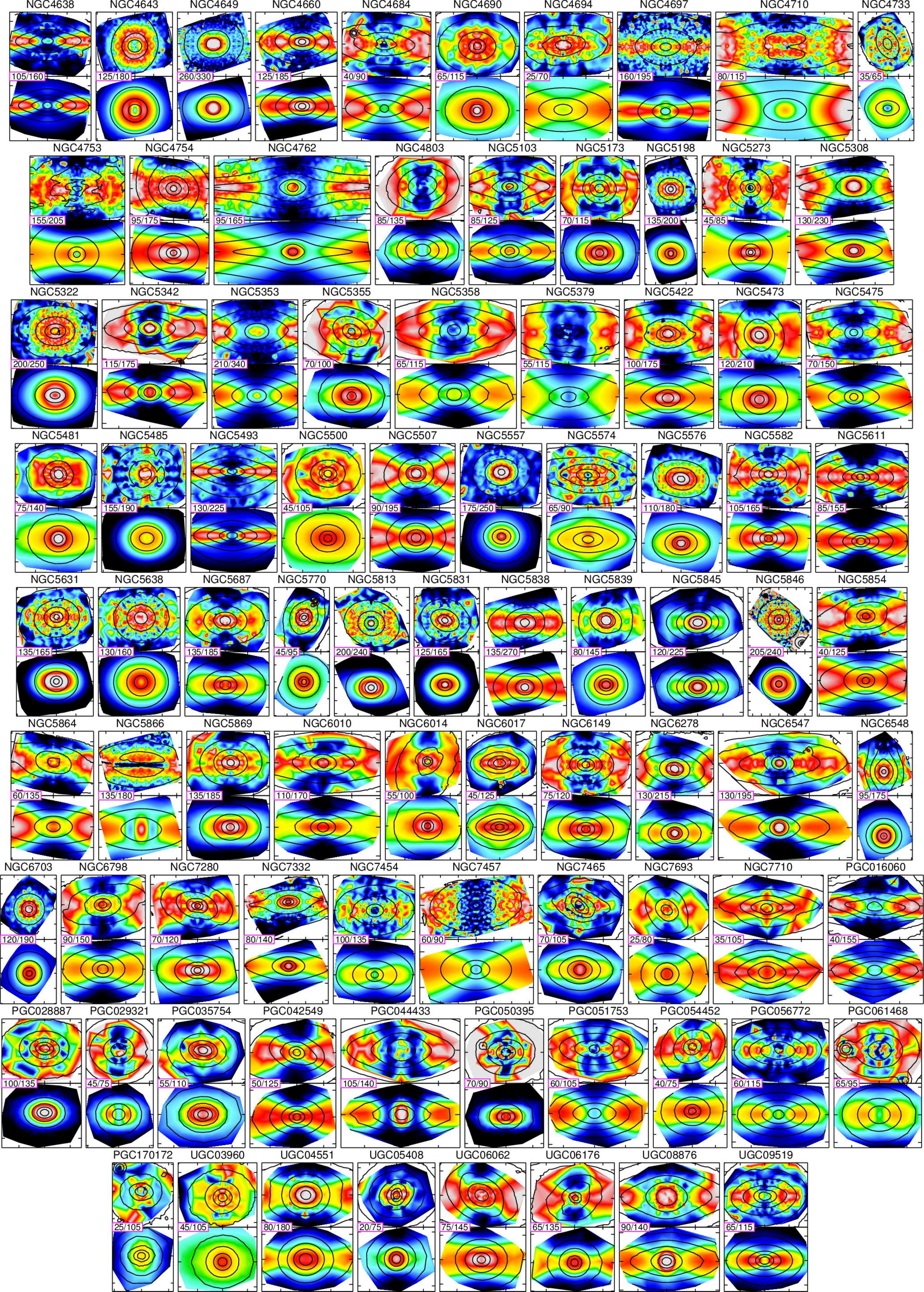}
\addtocounter{figure}{-1}
\caption{ --- continued}
\end{figure*}

\begin{figure}
\plotone{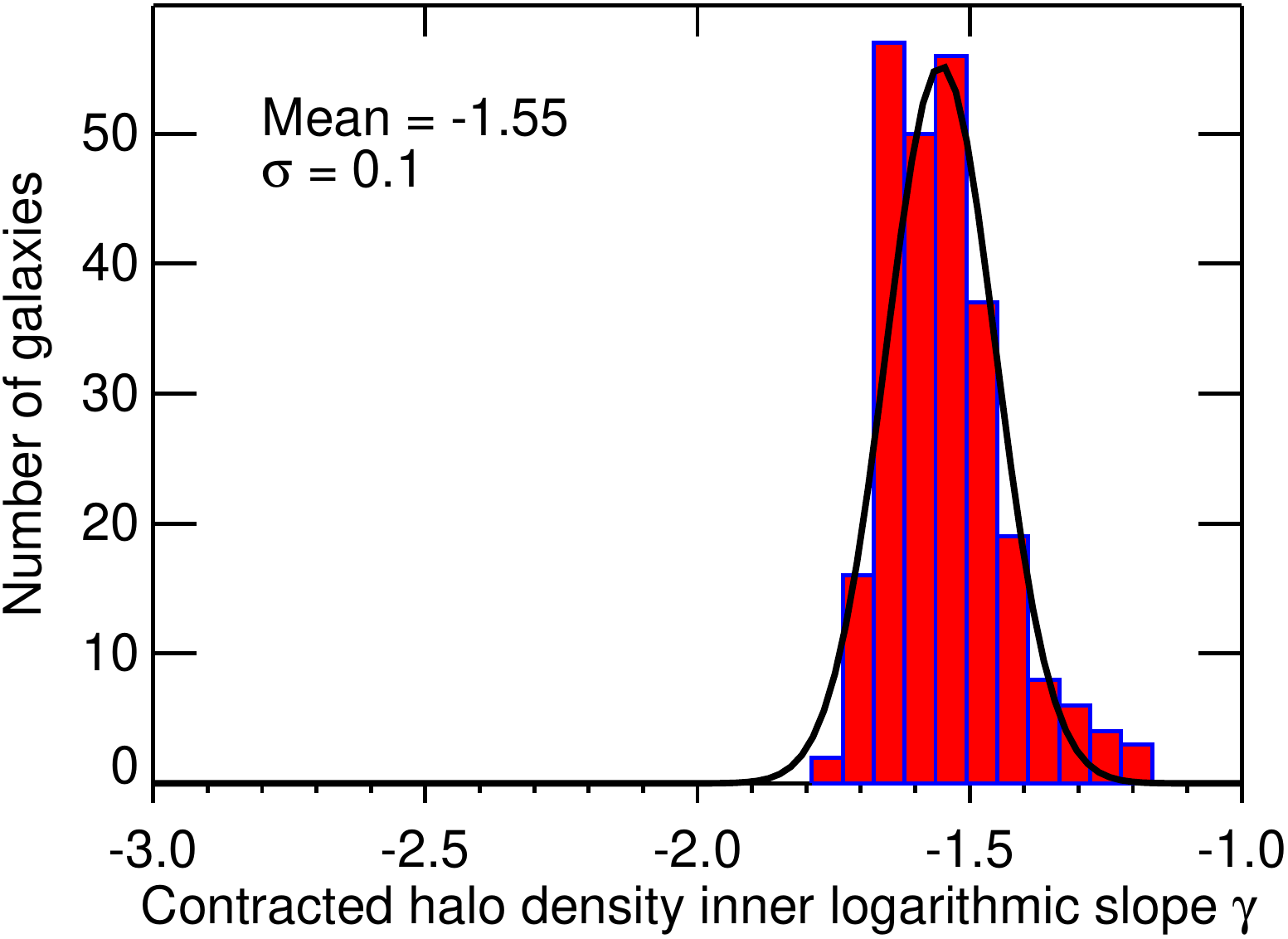}
\plotone{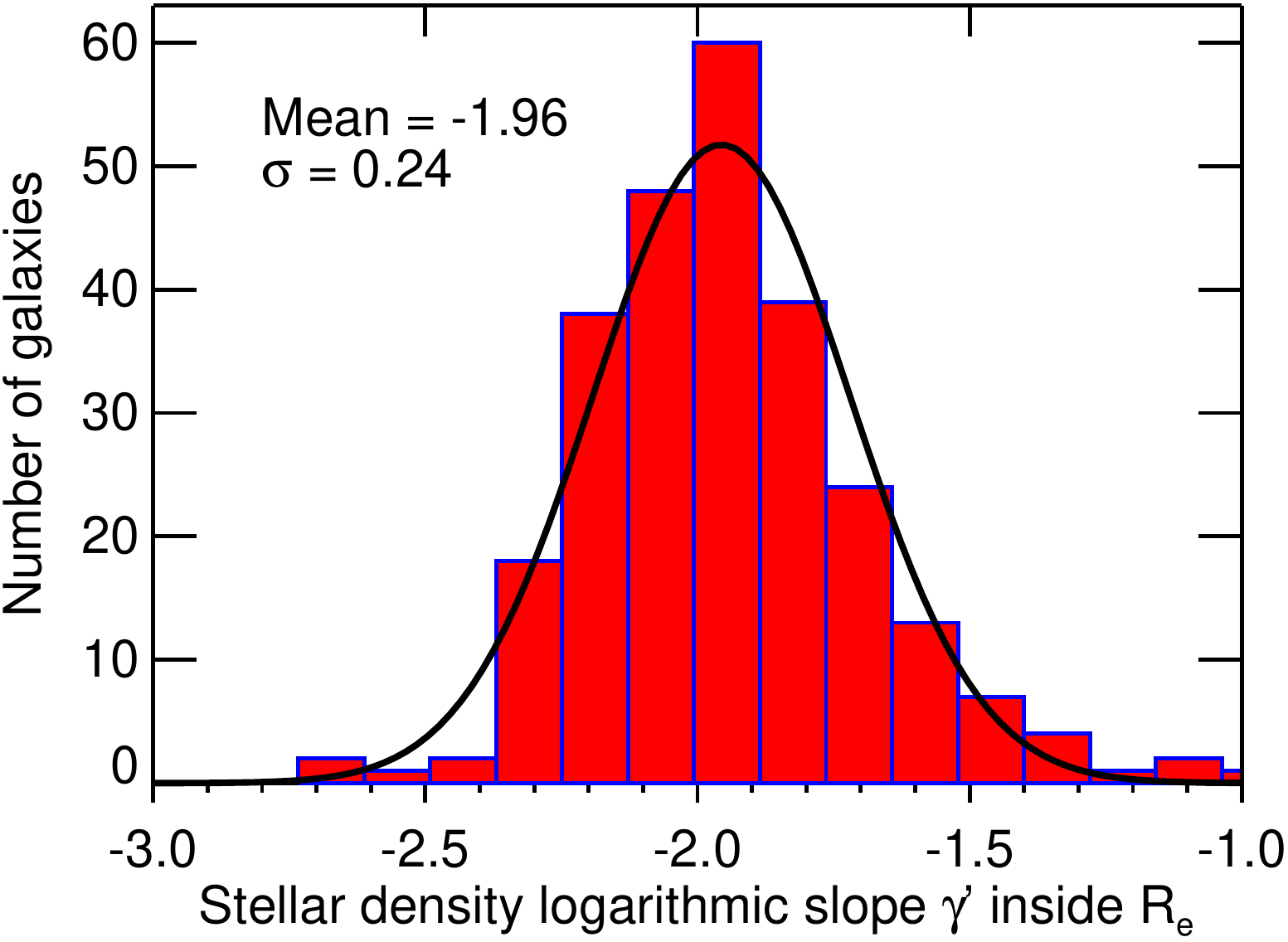}
\caption{Inner slope of contracted dark halos and luminous matter. {\rm Top Panel:} Histogram of the halo slope of contracted halos for all 260 \atl\ galaxies in model (C). The slopes were determined by fitting a power law relation $\rho_{\rm DM}(r)\propto r^{\gamma}$ inside the radius $r<r_s/4$, where we verified the contracted halo profiles are accurately described by a power law. {\rm Bottom Panel:} Histogram of the slope of the deprojected stellar mass density distribution from the MGE models. The slope was fitted inside a spherical radius $r=\re$. Although the stellar density $\rho_{\star}(r)\propto r^{\gamma'}$ inside that radius is not always accurately described by a power-law, on average the {\em stellar} slope peaks with high accuracy at at the `isothermal' value $\gamma'\approx2.0$, with an intrinsic scatter of just $\sigma=0.24$ for our entire sample. \label{fig:contracted_halo_slope}}
\end{figure}

The models we use here were already presented in \citet{Cappellari2012}, where they were used to uncover a systematic variation of the stellar IMF in ETGs. That paper (their table~1) describes six sets of JAM models for all the \atl\ galaxies, making various assumptions on the dark matter halo. Given that the \sauron\ data are typically spatially limited to 1\re\ one cannot expect to be able to robustly characterize the shape of the dark halo out to large radii from them \citep{Mamon2005}. However, as long as the density distribution of the halo is not the same as the one of the stars, we can determine how much room the models allow for a dark matter halo, within the region constrained by the kinematics. The models were summarized in \citet{Cappellari2012}, but we describe them here in some more detail using the same lettering notation as that paper:
\begin{enumerate}[\bf (A)]
\item {\bf Self-consistent JAM model:} Here we assume that the mass distribution follows the light one as inferred from the de-projected MGE. In this case the model has three free parameters. Two parameters are non-linear: (i) the vertical anisotropy $\beta_z=1-\sigma_z^2/\sigma_R^2$ and (ii) the galaxy inclination $i$, which together uniquely specify the shape of the second velocity moment $\langle v^2_{\rm los}\rangle$, which is then linearly scaled by the $(M/L)_{\rm JAM}$ to fit the two-dimensional $V_{\rm rms}$ data. We emphasize that, even though the models do not include a dark halo explicitly, $(M/L)_{\rm JAM}$ does {\em not} represent the stellar $M/L$, as sometimes incorrectly assumed, but the {\em total} one, within a spherical region which has the projected size of our data (see discussion in \refsec{sec:mass_err}). This set of models, like all others, has a central supermassive black hole with mass predicted by the $M_{\rm BH}-\sigma$ correlation \citep{Gebhardt2000bh,Ferrarese2000}, or a black holes mass as published, when available. The supermassive black hole has a minimal effect on $M/L$ in nearly all cases, but we still exclude the central $R<2''$ from the fits, for maximum robustness. Examples of mass-follows-light JAM models are shown in \reffig{fig:jam_vrms}. The inclination and $(M/L)_{\rm JAM}$ of the best fitting models are given in Table~1.

\item {\bf JAM with NFW dark halo:} This set of models adopted the approach introduced by \citet{Rix1997} to reduce the halo to a one-parameter family of models. This approach was already used with axisymmetric JAM models of disk galaxies, as done here, by \citet{Williams2009} and to construct spherical toy models of various stellar systems \citep{Napolitano2005,Tollerud2011}. We assume the halo is spherical and characterized by the two-parameters double power-law NFW profile \citep{navarro96}. We then adopt the halo mass-concentration $M_{200}-c_{200}$ relation \citep{navarro96} as given by \citet{Klypin2011} to make the halo profile a unique function of its mass $M_{200}$. The latter is not a critical assumption: our observations only sample a region well inside the predicted halo break radius, so that all our conclusion are unchanged if we describe the halo with a simple power law density profile $\rho(r)\propto r^{-1}$, as we numerically verified. The resulting JAM models have in this case four parameters: (i) The galaxy inclination $i$ (ii) the anisotropy $\beta_z$, (iii) the {\em stellar} $(M/L)_{\rm stars}$, assumed spatially constant and (iv) the halo virial mass $M_{200}$, defined as the mass within the spherical radius $r_{200}$ at which the average density is equal to 200 times the critical density of the Universe. The $(M/L)_{\rm stars}$ and dark matter fraction $f_{DM}(r=R_e)$ of the best fitting models are given in table~1 of Paper~XX.

\item {\bf JAM with contracted NFW dark halo:}  These models include a halo which is originally assumed to be of NFW form, with concentration specified by its mass via the $M_{200}-c_{200}$ relation as in (B). However, during the fitting process, for every choice of the model parameters, the halo is contracted according to the enclosed stellar mass distribution, which is defined by the (circularized) MGE and the corresponding $(M/L)_{\rm stars}$ parameter. For the contraction we used the prescription of \citet{Gnedin2011}, which is an update of \citet{Gnedin2004}. We verified that our IDL code produces the same output as the C language software \textsc{contra} by \citet{Gnedin2004}, when the same input is given. The resulting JAM model has the same four free parameters $(i,\beta_z,(M/L)_{\rm stars},M_{200})$ as in (B).

\item {\bf JAM with general dark halo (gNFW):} These models include a dark halo that generalizes the NFW profile (see also \citealt{Barnabe2012}), with density:
    \begin{equation}
        \rho_{\rm DM}(r)=\rho_s \left(\frac{r}{r_s}\right)^\gamma
            \left(\frac{1}{2}+\frac{1}{2}\frac{r}{r_s}\right)^{-\gamma-3}.
    \end{equation}
    The density has the same large-radii asymptotic power-law slope $\beta=-3$ as the NFW halo, but it allows for a variable inner slope, which we constrained to the bounds $-1.6<\gamma<0$, by assigning zero probability to the prior $P({\rm model})=0$ (\refsec{sec:bayes}) outside this parameters range. The ranges include a flat inner core $\gamma=0$ and the NFW $\gamma=-1$ as special cases. The upper bound was chosen as the nearly maximum slope we measured for all contracted halos in (C) (top panel of \reffig{fig:contracted_halo_slope}). However, recent simulations suggest that baryonic effects produce flatter halos than these predictions for a broad range of galaxy masses \citep{Duffy2010,Governato2010,Inoue2011,Pontzen2012,Laporte2012,Maccio2012,Martizzi2012}. Note that our adopted maximum halo slope is still generally more shallow than the typical `isothermal' average power slope $\gamma'=2.0$ the we measure for the stellar density alone within 1\re\ (bottom panel of \reffig{fig:contracted_halo_slope}). This fact is important to avoid model degeneracies between the stellar and halo densities.
    This model is the most general of all six and it includes any of the other five models as special cases. It has five free parameters: (i) the galaxy inclination, (ii) the anisotropy $\beta_z$, (iii) the stellar mass $M_{\rm stars}$, (iv) the halo inner slope $\gamma$ and (v) the halo density $\rho_s$ at $r_s$, which we parametrized using the dark matter fraction $f_{DM}(r=R_e)$ to reduce the strong correlation between $\rho_s$ and $\gamma$ during the parameter estimation. The break radius $r_s$ of the halo was not included as a free parameter given that it is (in models E) generally 3--5 times larger than \re\ and it is completely unconstrained by our data. We fixed $r_s=20$ kpc, which is the median value for all models E, but we verified that nearly identical results are obtained if we describe the halo with a simple power-law density profile $\rho(r)\propto r^{-\gamma}$. Examples of model fits are shown in \reffig{fig:jam_mcmc};

\item {\bf JAM with fixed NFW dark halo:} The halo has a NFW profile without any free parameter. During the fitting process the halo mass $M_{200}$ is determined from the enclosed stellar mass $M_{\rm stars}$, which is given by the total luminosity of the MGE model multiplied by its current $(M/L)_{\rm stars}$. This is done using the $M_{200}-M_{\rm stars}$ relation derived by \citet{Moster2010} \citep[see also][]{Moster2012}, which matches the observed galaxy luminosity functions to the simulated halos mass function. However, negligible differences would have been obtained using e.g.\ the similar relations derived by \citet{Behroozi2010} or \citet{Guo2010}. For a given halo mass, the concentration is specified by the $M_{200}-c_{200}$ relation as in (B). The only free model parameters are the three of the stellar component $(i,\beta_z,(M/L)_{\rm stars})$ as in (A). This fixed-halo assumption, in combination however with equally fixed spherical and isotropic \citet{Hernquist1990} galaxy models, was also used by \citet{Auger2010imf} and \citet{Deason2012dm}.

\item {\bf JAM with fixed contracted dark halo:} The halo has a contracted profile without any free parameter. For a given stellar mass, the halo has initially the same NFW form as in (E), but the profile is contracted as in (C) using the prescription of \citet{Gnedin2011}. The only free model parameters are the three of the stellar component $(i,\beta_z,(M/L)_{\rm stars})$ as in (A). This fixed-halo assumption, in combination however with equally fixed spherical and isotropic \citet{Hernquist1990} galaxy models, was also used by \citet{Auger2010imf}.
\end{enumerate}

\subsubsection{Bayesian inference of the JAM model parameters}
\label{sec:bayes}

The determination of the JAM model parameters for the 260 \atl\ galaxies in \citet{Cappellari2012} was done using Bayesian inference \citep{gelman2004bayesian}. The same approach was adopted using JAM models in \citet{Barnabe2012} in combination with gravitational lensing. From Bayes theorem, the posterior probability distribution of a model, with a given set of parameters, given our data is
\begin{equation}
    P({\rm model}\, |\, {\rm data})\propto P({\rm data}\, |\, {\rm model})\times P({\rm model}).
\end{equation}
Here we make the rather common assumption of Gaussian errors, in which case the probability of the data, for a given model is given by
\begin{equation}
    P({\rm data}\, |\, {\rm model})\propto \exp\left(-\frac{\chi^2}{2}\right),
\end{equation}
with
\begin{equation}
    \chi^2 = \sum_j \left(\frac{ \langle v^2_{\rm los}\rangle_j^{1/2} - V_{{\rm rms},j} }{ \Delta V_{{\rm rms},j} }\right)^2.
\end{equation}
We further assume a constant {\em noninformative} prior $P({\rm model})$ for all variables within the given bounds.

\begin{figure*}
\includegraphics[width=\textwidth]{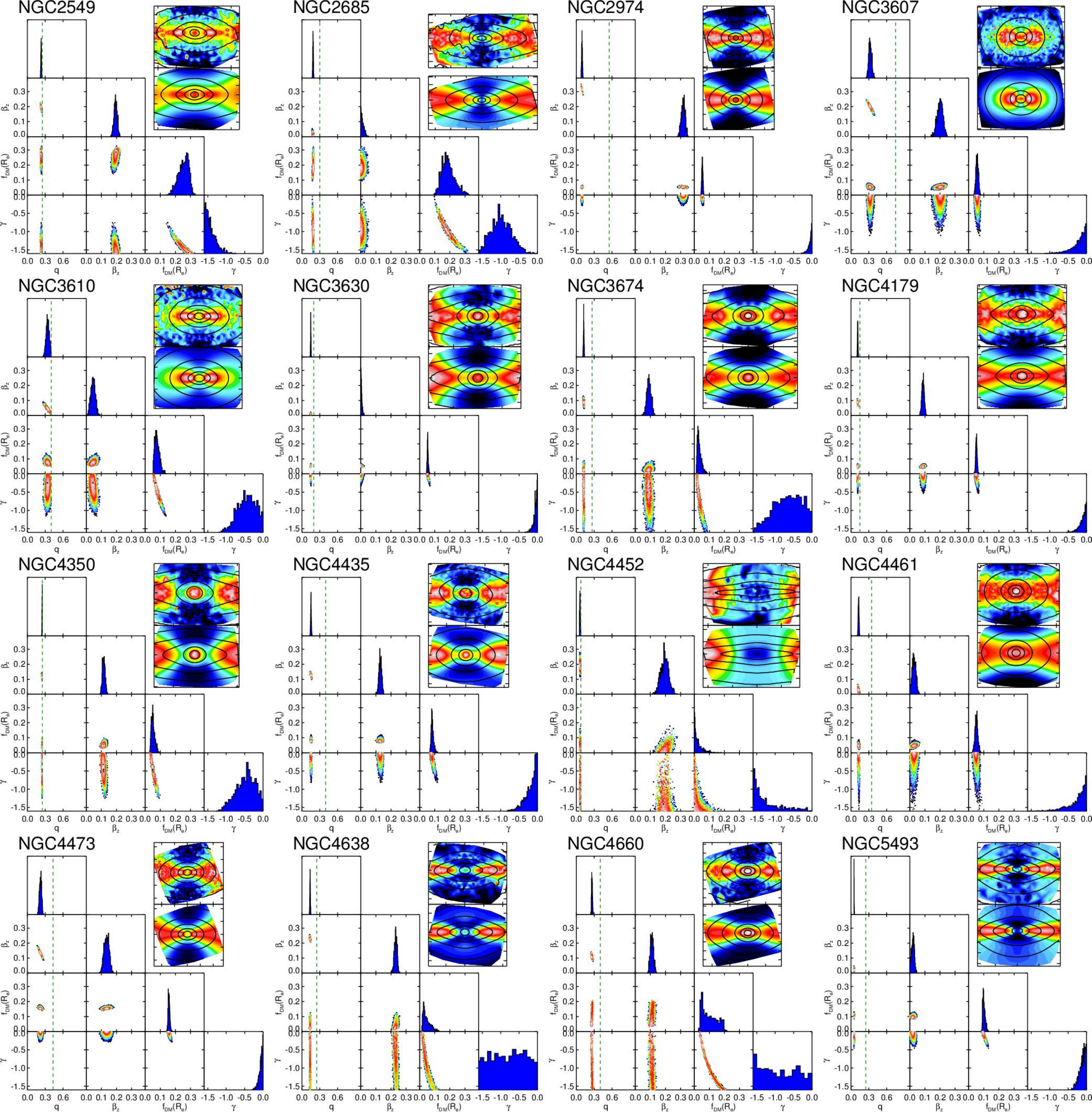}
\caption{Examples of JAM dynamical modelling with general dark halo using an Adaptive Metropolis approach. Each panel shows the corner-plots of the posterior probability distribution for the non-linear model parameters $(q,\beta_z,f_{\rm DM},\gamma)$, of galaxy models (D), marginalized over two dimensions (colour contours) and one dimension (blue histograms). The symbols are coloured according to their likelihood: white corresponds to the maximum value and dark blue to a $3\sigma$ confidence level. The vertical dashed green line indicates the maximum allowed $q$, which corresponds to an edge-on view. For each combination of the non-linear parameters, the linear parameter $(M/L)_{\rm stars}$ is fit to the data. We assumed ignorant (constant) priors on all model parameters. The name of the galaxies is written at the top of each panel. The symmetrized $V_{\rm rms}$ \sauron\ data, and the best-fitting model are shown on the right (as in \reffig{fig:jam_vrms}). This plot illustrates a variety of situations and shapes of the kinematic field: (i) some models (NGC~2685, NGC~3610, NGC~3674, NGC~4350) have best fitting halo parameters within the explored parameters boundaries; (ii) others (NGC~2974, NGC~3607, NGC~3630, NGC~4179, NGC~4435, NGC~4461, NGC~4473, NGC~5493) prefer a flat $\gamma\approx0$ inner halo slope; (iii) others (NGC~4638, NGC~4660) have nearly unconstrained halo slope; (iv) others (NGC~2549, NGC~4452) prefer steep halo slopes at the boundary $\gamma=-1.6$ of our allowed parameter range. In all cases the halo slope is weakly constrained by the \sauron\ data, but the dark matter fraction is tightly constrained by the data to be small ($f_{\rm DM}\la25\%$ in these examples). Only allowing the inner dark halo slope to be as steep as the characteristic stellar density slope $\gamma'\approx-2.0$ (\reffig{fig:contracted_halo_slope}) could significant dark matter be included within 1\re\ in some of the models, while still reproducing the kinematic observations.
\label{fig:jam_mcmc}}
\end{figure*}

The calculation of the posterior distribution is performed using the {\em adaptive} \citet{metropolis1953equation} (AM) algorithm of \citet{haario2001adaptive}. The AM method adapts the multivariate Gaussian proposal distribution during the Markov chain Monte Carlo sampling, in such a way that the Gaussian proposal distribution has the same non-diagonal covariance matrix as the posterior distribution accumulated so far by the algorithm. This natural idea is similar to what is routinely done e.g. in the determination of cosmological parameters, where the covariance matrix of the posterior is calculated after a burn-in phase \citep[e.g.][]{Dunkley2005}. However, the adaptive approach converges much more rapidly as the proposal distribution starts approaching the posterior already after a few points have been sampled. We found the adaptive approach absolutely critical for the speed up of our calculation by orders of magnitudes, given the strong degeneracies between the model parameters producing inclined and narrow posterior distributions. Some examples of the posterior distributions obtained with our approach are shown in \reffig{fig:jam_mcmc}. Although the adaptive nature of the AM algorithm makes the resulting chain non-Markovian, their authors have proven that it has the correct ergodic properties \citep{haario2001adaptive} and for this reason it can be used to estimate the posterior distribution as in standard Markov chain Monte Carlo methods \citep{gilks1996markov}.

Moreover, to basically eliminate the burn-in phase of the AM method, we use the efficient and extremely robust \textsc{DIRECT} deterministic global optimization algorithm of \citet{Jones1993} to find the starting location without the risk for the Metropolis stage to be stuck in a possible secondary minimum in multi-dimensional parameter space.

An important addition to the fitting process is an iterative sigma clipping of the kinematics, to remove spurious features in the data like stars or problematic bins at the edge of the \sauron\ field of view. This is important for a sample of the size of \atl, where the quality of every Voronoi bin cannot be assessed manually for all galaxies. After an initial fit the few bins deviating more than 3$\sigma$ of the local rms noise are excluded from the fit and a new fit is iteratively performed, until convergence.

\subsection{Robust fitting of lines or planes to the data}

\subsubsection{Goodness of fit criteria}

The apparently simple task of fitting linear relations or planes to a set of data with errors does not have a well defined and obvious solution and for this reason has continued to generate significant interest. A number of papers have discussed the solution of the corresponding least-squares problem \citep{Isobe1986,Feigelson1992,Akritas1996,Tremaine2002,Press2007}, while more recent works have addressed the problem using Bayesian methods \citep{Kelly2007,Hogg2010}. A popular method is the least-squares approach by \citet{Tremaine2002}, which is an extension of the \textsc{fitexy} procedure described in \citet[section~15.3]{Press2007}. The method defines the best fit of the linear relation $y=a+b (x-x_0)$ to a set of $N$ pairs of quantities $(x_j,y_j)$, with symmetric errors $\Delta x_j$ and $\Delta y_j$, as the one that minimizes the quantity
\begin{equation}
	\chi^2=\sum_{j=1}^N \frac{[a+b (x_j-x_0) - y_j]^2}
	{(b \Delta x_j)^2 + (\Delta y_j)^2 + \varepsilon_y^2}.
\label{eq:chi2line}
\end{equation}
Here $x_0$ is an adopted reference value, close to the middle of the $x_j$ values, adopted to reduce uncertainty in $a$ and the covariance between the fitted values of $a$ and $b$. While $\varepsilon_y$ is the intrinsic scatter in the $y$ coordinate, which is iteratively adjusted so that the $\chi^2$ per degree of freedom $\nu=N-2$ has the value of unity expected for a good fit. As recognized by \citet{Weiner2006}, minimizing the above $\chi^2$ corresponds to maximizing the likelihood of the data for an assumed intrinsic probability distribution of the observables described by the linear relation $y = a + b (x-x_0) + \varepsilon_y$, where $\varepsilon_y$ is the Gaussian scatter projected along the $y$ coordinate, and one assumes a uniform prior in the $x$ coordinate. \refeq{eq:chi2line} is only rigorously valid when the errors in $x$ and $y$ are Gaussian and uncorrelated (have zero covariances). A term $-2b\, {\rm Cov}(x_j,y_j)$ should be included in the denominator if the covariances are known and non-zero \citep[e.g.][]{FalconBarroso2011}. The 1$\sigma$ confidence interval in $\varepsilon_y$ can be obtained by finding the values for which $\chi^2=\nu\pm \sqrt{2\nu}$ as done by \citet{Novak2006}. The apparent asymmetry of \refeq{eq:chi2line} with respect to the $x$ and $y$ variables does not imply we assume only the $y$ variable has intrinsic scatter. In fact the assumed intrinsic distribution has a Gaussian cross section along any direction non parallel to the ridge line $y=a+b (x-x_0)$. The value $\varepsilon_y$ merely specifies the dispersion along the arbitrary $y$ direction. The formula would give completely equivalent results by interchanging the $x$ and $y$ variables if the distribution of $x$ values was uniform and infinitely extended as assumed. Any difference in the fitting results when interchanging the $x$ and $y$ coordinates are due to the breaking of the uniformity assumptions.

\refeq{eq:chi2line} can be generalized to plane fitting by defining the best-fitting plane $z=a+b (x-x_0)+c (y-y_0)$ to a set of $N$ triplets of quantities $(x_j,y_j,z_j)$, with symmetric errors $\Delta x_j$, $\Delta y_j$ and  $\Delta z_j$, as the one that minimizes the quantity
\begin{equation}
	\chi^2=\sum_{j=1}^N \frac{[a + b (x_j-x_0) + c (y_j-y_0) - z_j]^2}
	{(b \Delta x_j)^2 + (c \Delta y_j)^2 + (\Delta z_j)^2 + \varepsilon_z^2},
\label{eq:chi2plane}
\end{equation}
Here $x_0$ and $y_0$ are adopted reference values, close to the middle of the $x_j$ and $y_j$ values respectively, adopted to reduce uncertainty in $a$ and the covariance between the fitted values of $a$, $b$ and $c$. While $\varepsilon_z$ is the intrinsic scatter in the $z$ coordinate, which is iteratively adjusted so that the $\chi^2$ per degrees of freedom $\nu=N-3$ has the value of unity expected for a good fit. As in the two-dimensional case the minimization of \refeq{eq:chi2plane} is equivalent to the maximization of the likelihood of the data, for an underlying probability distribution of the observables described by the relation $z = a + b (x-x_0) + c (y-y_0) + \varepsilon_z$, where $\varepsilon_z$ is the dispersion of the Gaussian intrinsic scatter in the plane, projected along the $z$ coordinate, for a uniform prior in the $x$ and $y$ coordinates and assuming uncorrelated and Gaussian errors in the $x$, $y$ and $z$ observables (zero covariances).  \refeq{eq:chi2plane} reduces to the so called orthogonal plane fit when the measurements errors are ignored and one simply assumes $\Delta x_j=\Delta y_j=\Delta z_j$. This latter form is the one generally used when fitting the Fundamental Plane \citep[e.g.][]{Jorgensen1996,Pahre1998,Bernardi2003fp}.
Contrary to the popular approach, \refeq{eq:chi2plane} allows for intrinsic scatter in the relation, which is important to deriving unbiased parameters \citep{Tremaine2002}.

Recently \citet{Kelly2007} proposed a Bayesian method to treat the linear regression of astronomical data in a statistically rigorous manner, allowing for intrinsic scatter, covariance between measurements and providing rigorous errors on the parameters in the form of random draws from the posteriori distribution \citep[see also][]{Hogg2010}. He pointed out that the \citet{Tremaine2002} approach to linear fitting can lead to biased results in some circumstances. For this reason in all our fits we used both the results and errors derived from \refeq{eq:chi2line} and (\ref{eq:chi2plane}), and the corresponding results obtained with the Bayesian method and software by \citet{Kelly2007}, which was kindly made available as part of the IDL NASA Astronomy Library \citep{Landsman1993}. In all cases differences between the two method where found to be insignificant, in both the fitted values and the errors, confirming the near conceptual equivalence of the two technically very different approaches.

\subsubsection{Least Trimmed Squares robust fits}

A general issue when fitting linear relations to data using least-squares methods is the presence of outliers, which can dominate the $\chi^2$ and bias the parameter recovery. This is the reason why a number of previous studies have determined the parameters of the Fundamental Plane using the more robust method of minimizing absolute instead of squared deviations \citep[e.g.][]{Jorgensen1996,Pahre1998}, at the expense of decreasing the statistical efficiency, namely larger errors on the fitted parameters. An alternative simple solution, which maintains the efficiency of the least-squares method for Gaussian distributions, consists of removing outliers by iteratively clipping points deviating more than 3$\sigma$ from the currently best-fitting relation. A problem with the $\sigma$-clipping approach is that it is not guaranteed to converge to the desired solution in the presence of significant outliers. Alternative robust methods have been proposed \citep[see][section~15.7]{Press2007}. However, they complicate the error estimation and like the standard $\sigma$-clipping do not always converge.

After some experimentation with different robust approaches the only fully satisfactory solution we found is the Least Trimmed Squares (LTS) regression approach of \citet{Rousseeuw1987}. The reason for its success is that the method, as opposed to other robust approaches, finds a {\em global} solution. The approach consists of finding the global minimum to
\begin{equation}
\chi_h^2=\sum_{j=1}^h (r^2)_{j:N},
\end{equation}
where $(r^2)_{1:N}\le (r^2)_{2:N} \le \ldots \le (r^2)_{N:N}$ are the ordered square residuals from the linear regression of a subset of $N/2<h<N$ data points. In other words the LTS method consists of finding the subset of $h$ data points providing the smallest $\chi_h^2$, among {\em all} possible $h$-subsets. It's easy to realize that this approach is robust to the contamination of up to half of the data points, when $h\approx N/2$. This is a computational very expensive combinatorial problem for which however a fast and nearly optimal solution (FAST-LTS) has recently been proposed by \citet{Rousseeuw2006}.

In our implementations\footnote{Available from http://purl.org/cappellari/idl}, which we called \textsc{lts\_linefit} and \textsc{lts\_planefit} for the line and plane case respectively, we combine the robust approach to outliers with a fitting method which allows and fits for intrinsic scatter. We proceed as follows:
\begin{enumerate}
\item We adopt as initial guess $\varepsilon=0$ for the intrinsic scatter in the $y$ (for \textsc{lts\_linefit}) or $z$ coordinate (for \textsc{lts\_planefit});
\item We start by default with $h=(N+p+1)/2$, where $p$ is the data dimension, and use the FAST-LTS algorithm to produce a least-squares fit\footnote{In all the nonlinear fits the minimization was performed with the IDL program \textsc{mpfit} by \citet{Markwardt2009}, which is in an improved implementation of the MINPACK Levenberg-Marquardt nonlinear least-squares algorithm by \citet{more1980minpack}}.
to the set of points characterized by the smallest $\chi_h^2$ (defined by equation \ref{eq:chi2line} or \ref{eq:chi2plane});
\item We compute the standard deviation $\sigma$ of the residuals for these $h$ values and extend our selection to include all data point deviating no more than 2.6 $\sigma$ from the fitted relation (99\% of the values for a Gaussian distribution);
\item We perform a new linear fit to the newly selected points;
\item We iterate steps (iii)--(iv) until the set of selected points does not change any more;
\item We calculate the $\chi^2$ for the fitted points;
\item The whole process (i)--(vi) is iterated varying $\varepsilon$ using Brent's method \citep[section~9.3]{Press2007} until $\chi^2=\nu$.
\item The errors on the coefficients are computed from the covariance matrix;
\item The error on $\varepsilon$ is computed by increasing $\varepsilon$ until $\chi^2=\nu-\sqrt{2\nu}$ (we do not decrease it to avoid problems when $\varepsilon\approx0$).
\end{enumerate}

This method was used to produce all fits in this paper and automatically exclude outliers. Note that although the approach may appear similar to the standard $\sigma$ clipping one, and produces similar results in simple situations, the key difference is that in \textsc{lts\_linefit} and \textsc{lts\_planefit} the clipping is done from the inside-out instead of the opposite. This was found to be the essential feature for the resulting extreme robustness, which was essential in particular to objectively select Virgo members in \reffig{fig:ml_sigma_virgo}.  Once the outliers are removed, the same set of points was used as input to \citet{Kelly2007} Bayesian algorithm.

\subsection{Measuring scaling relations parameters}

\subsubsection{Determination of $L$, \re\ and $r_{1/2}$ from the MGE}
\label{sec:measuring_re}

Galaxy photometric parameters are generally determined using three main approaches: (i) fitting growth curves, where one constructs profiles of the enclosed light within circular annuli and extrapolates the outermost part of the galaxy profile to infinite radius, typically using the analytic growth curve of the $R^{1/4}$ \citep{deVaucouleurs1948} profile (e.g.\ the Seven Samurai: \citealt{Burstein1987} and \citealt{Faber1989}; the RC3: \citealt{deVaucouleurs1991} and \citealt{Jorgensen1995phot}); (ii) fitting an $R^{1/n}$ \citep{Sersic1968} profile \citep[e.g.][]{Graham1997fp}, possibly including an exponential disk \citep[e.g.][]{Saglia1997}, to the circularized profiles and finding the half-light from the models; (iii) fitting flattened two-dimensional models directly to the galaxy images, where the profile of the models is again parameterized by an $R^{1/4}$ \citep[e.g.][]{Bernardi2003fp}, or by an $R^{1/4}$ bulge plus exponential disk \citep[e.g.][]{Gebhardt2003fp,Saglia2010,Bernardi2010}.

Here we have MGE photometric models for all the galaxies in the sample based on the SDSS+INT photometry (Paper~XXI). Due to the large number of Gaussians used to fit the galaxy images, the MGE models provide a compact and essentially non-parametric description of the galaxies surface brightness, which reproduces the observations much more accurately than the simpler bulge and disk models, but more robustly than using the images directly. Our MGE fitting approach is in fact analogue to the popular \textsc{galfit} \citep{Peng2002} software, when it is used to match  every detail of a galaxy image using multiple components. Here we use the MGE models to measure the photometric parameters ($L$ and \re) in our scaling relations as done in \citet{Cappellari2009}. A key difference between this MGE approach and all the ones previously mentioned is that it does not extrapolate the galaxy light to infinite radii. Outside three times the dispersion  $3\times\max(\sigma_j)$ of the largest MGE Gaussian, the flux of the model essentially drops to zero. No attempt is made to infer the amount of stellar light that we may have observed if we had much deeper photometry. For this reason this \re\ must be necessarily smaller than the ones obtained via extrapolation to infinite radii. 

The extrapolation method depends on the assumed form of the unobservable galaxy profile out to infinite radii. One may argue that an extrapolation of the galaxy profile using a \citet{Sersic1968} function should provide a better estimate of the total luminosity (and \re) than the observed luminosity. This is in general likely correct, however the accuracy of the extrapolation depends on galaxy properties in a unknown systematic manner. Our volume-limited sample of ETGs is dominated by fast rotators (Paper~II; Paper~III), characterized by the presence of disks \citep[hereafter Paper~XVII]{Krajnovic2012} and closely linked to spiral galaxies (\citealt{Cappellari2011b}, hereafter Paper~VII; Paper~XX). Given the variety in the outer profiles of spiral galaxies \citep{vanderKruit1981,Pohlen2006} it is unclear how profiles should be extrapolated. Using \re\ and luminosities derived via extrapolation makes any derived trend necessarily assumption dependent. As we show in \refsec{sec:simple_mass_estimators}, the differences between different assumptions are quite significant. One can obtain different trends in scaling relations and reach different conclusions about their interpretation.

We argue that to make progress one should base conclusions on directly observable quantities. So for this work define \re\ as the radius containing half of the {\em observed} light, not half of the ill-defined amount of total light we think the galaxy may have. Of course even our approach does not solve the problem of determining an absolute normalization of \re, and our sizes appear well reproducible only in a relative sense. The only real solution to the problem is to obtain deeper photometry so that \re\ values converge and become essentially independent on the adopted profiles \citep{Kormendy2009,Ferrarese2012}. However for massive galaxies in clusters the distinction between galaxy light and intra-cluster light may become an issue. Earlier indications using deeper MegaCam photometry, which we have acquired for many of the galaxies in our sample \citep[hereafter Paper~IX]{Duc2011}, confirm that \re\ determinations depend sensitively on the depth of the adopted photometry as expected.

If the $x$-axis is aligned with the galaxy photometric major axis, and the coordinates are centered on the galaxy nucleus, the surface brightness of an MGE model at the position $(x',y')$ on the plane of the sky, already analytically deconvolved for the atmospheric seeing effects, can be written as \citep{Emsellem1994}
\begin{equation}\label{eq:surf}
\Sigma(x',y') = \sum_{j=1}^M \Sigma_j
 \exp
\left[
    -\frac{1}{2\sigma^2_j}
    \left(x'^2 + \frac{y'^2}{q'^2_j} \right)
\right],
\end{equation}
where $M$ is the number of the adopted Gaussian components, having peak surface brightness $\Sigma_j$, observed axial ratio $0\le q'_k\le1$ and dispersion $\sigma_k$ along the major axis. The total luminosity of the MGE model is then:
\begin{equation}
L = \sum_{j=1}^M L_j = \sum_{j=1}^M 2\pi \Sigma_j \sigma^2_j q'_j,
\end{equation}
where $L_j$ are the luminosities of the individual Gaussians.

In \citet{Cappellari2009} the effective radius of the MGE model was obtained by circularizing the individual Gaussians of the MGE, while keeping their peak surface brightness. This was achieved by replacing $(\sigma_j,q'_j)$ with $(\sigma\sqrt{q'_j},1)$ . The luminosity of the circularized MGE enclosed within a cylinder of projected radius $R$ is then
\begin{equation}\label{eq:lumRmge}
L(R) = \sum_{j=1}^M L_j
\left[
    1 - \exp
    \left(
        -\frac{R^2}{2\sigma_j^2  q'_j}
    \right)
\right].
\end{equation}
The circularized effective (half-light) radius \re\ was found by solving $L(R) = L/2$, using a quick interpolation over a grid of $\log R$ values. When the MGE has constant axial ratio $q'_j=q'$ for all Gaussians, this approach finds the circularized radius $\re=\sqrt{a b}=a\sqrt{q'}$ of the elliptical isophote containing half of the analytic MGE light, where $a$ is the major axis of the isophote. This is the quantity almost universally used for studies of scaling relations of ETGs. When the axial ratios of the different Gaussians are not all equal, the approach finds an excellent approximation for the radius $\re=\sqrt{A_e/\pi}$ of a circle having the same area $A_e$ as the isophote containing half of the MGE light. In fact we verified that for all the MGE of the \atl\ sample the two determinations agree with an rms scatter of just 0.17\% and only for four of the flattest galaxies the difference is larger than 3\%.

\begin{figure}
\plotone{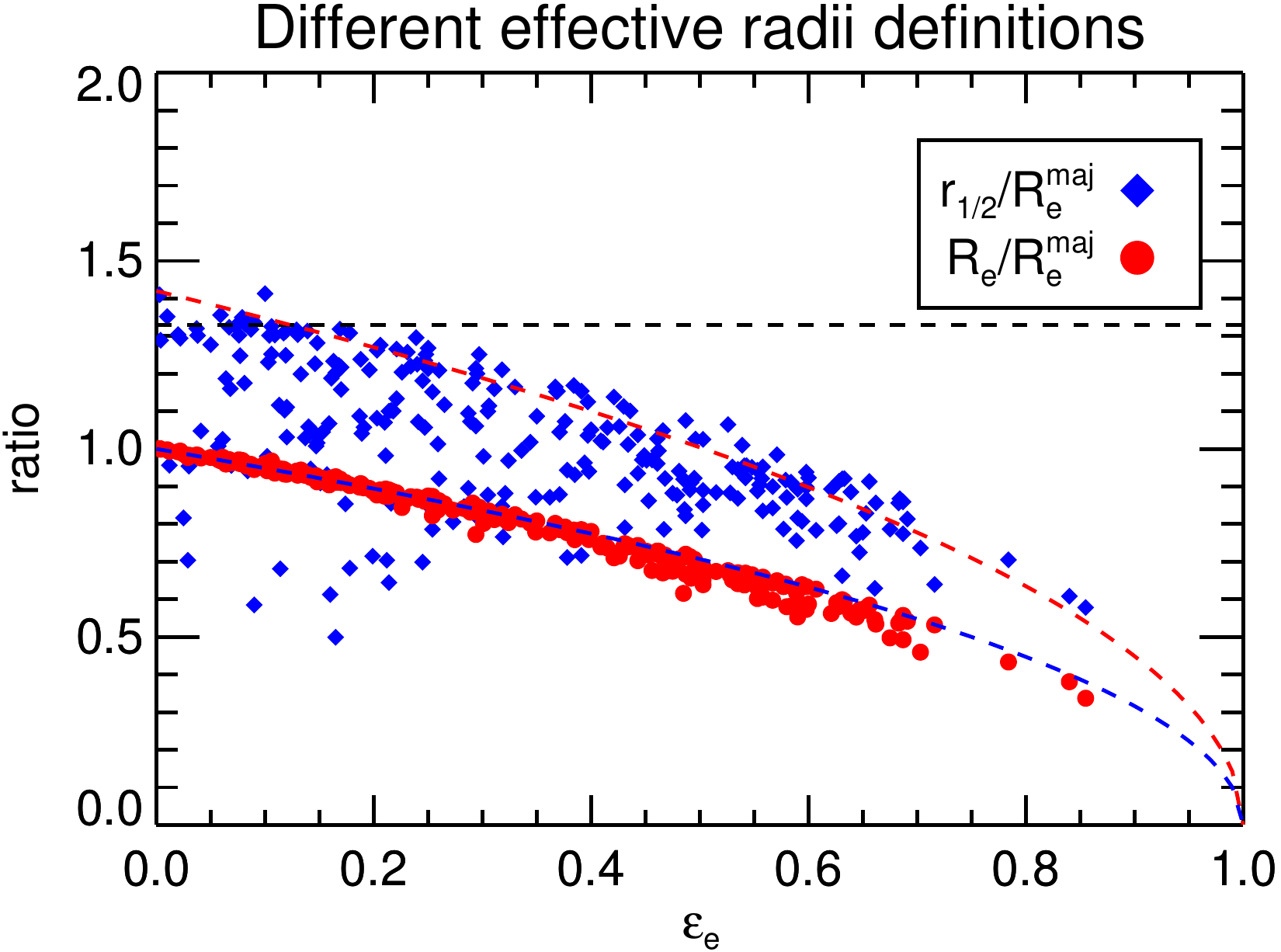}
\caption{Different definitions of \re\ as a function of the galaxy ellipticity. The red filled diamonds are the projected radii \re\ of a cylinder with the same area of the half-light isophote. The blue filled circles are the radii $r_{1/2}$ of a sphere with the same volume as the half-light iso-surface. In both cases the radii are normalized to $R_e^{\rm maj}$, which is the projected semi-major axis of the half-light isophote, having ellipse of inertia of ellipticity $\varepsilon_e$. The red and blue dashed lines are the relations $f(\varepsilon_e)=1.42\sqrt{\varepsilon_e}$ and $f(\varepsilon_e)=\sqrt{\varepsilon_e}$ respectively. The horizontal dashed line marks the theoretical value $4/3$, which approximately applies to a number of simple theoretical profiles.}
\label{fig:different_re_definitions}
\end{figure}

\citet{Hopkins2010} pointed out the usefulness of adopting as size parameter the major axis $R_e^{\rm maj}$ of the half-light isophote instead of the circularized radius \re, when analysing results of simulations. The motivation is that $R_e^{\rm maj}$ is more physically robust and less dependent on inclination. Here we also calculate $R_e^{\rm maj}$ for our observed galaxies as follows.
\begin{enumerate}
\item We construct a synthetic galaxy image from the MGE using \refeq{eq:surf}, with size $\max(\sigma_j)\times\max(\sigma_j)$ (only one quadrant is needed for symmetry);
\item We sample a grid of surface-brightness values $\mu_k=\mu(x_k,0)$  along the MGE major axis, and for each value we calculate the light enclosed within the corresponding isophote;
\item We find the surface brightness $\mu_e$ of the isophote containing half of the analytic MGE total light by solving $L(\mu)=L/2$ using linear interpolation;
\item $R_e^{\rm maj}$ is the maximum radius enclosed inside the isophote $\mu_e$ (the largest $x$ coordinate).
\end{enumerate}
We also calculate the circularized effective radius of the isophote $\re=\sqrt{A/\pi}$ of area $A$ and the effective ellipticity $\varepsilon_e$ of the MGE model inside that isophote as \citep{Cappellari2007}
\begin{equation}
(1-\varepsilon_e)^2=q'^2_e=\frac{\langle y^2 \rangle}{\langle x^2 \rangle} = \frac{\sum_{k=1}^{P} F_k\, y_k^2}{\sum_{k=1}^{P} F_k\, x_k^2},
\label{eq:eps}
\end{equation}
where $F_k$ is the flux inside the $k$-th image pixel, with coordinates $(x_k,y_k)$ and the summation extends to the pixels inside the chose isophote. A similar quantity was calculated from the original galaxy images in Paper~III, but we use here this new determination for maximum consistency between our $\varepsilon_e$ and the ellipticity of the MGE models in the tests of \reffig{fig:different_re_definitions}.

\begin{figure}
\plotone{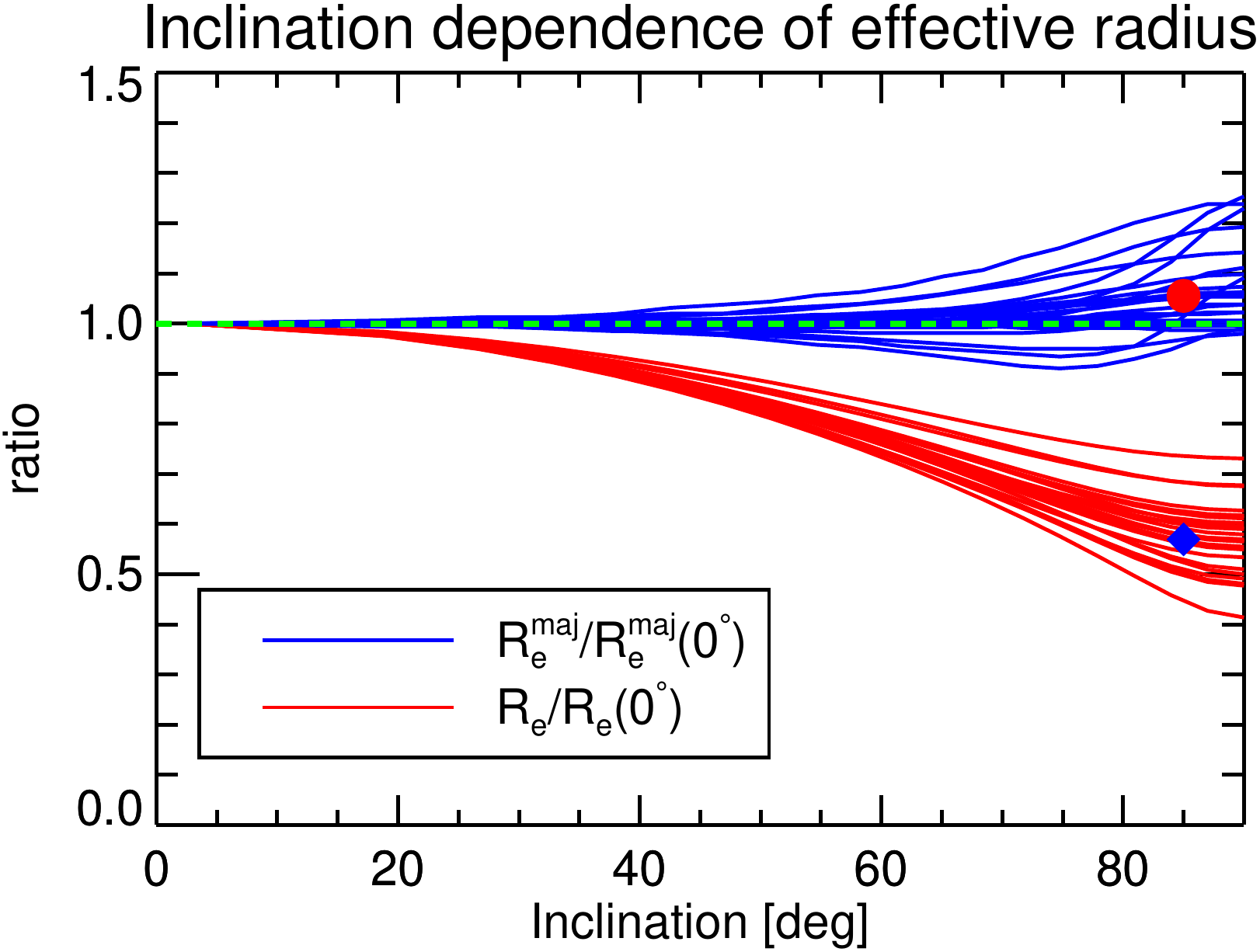}
\caption{Inclination dependence for different definitions of the effective radius. The red lines show the change in the measured circularized \re, normalized to the face-on value, when the inclination is changed from edge-on ($i=90^\circ$) to face-on, for the 26 flattest \atl galaxies. The blue diamond marks the median (43\%) of the maximum variation. The blue lines show the same variation with inclination of the major axis $R_e^{\rm maj}$ of the half-light isophote. The red circle is the median (5\%) of the maximum variation.}
\label{fig:inclination_dependence_re}
\end{figure}

We studied the dependence on inclination of the two definitions of effective radii using the photometry of real galaxies. For this we selected the 26 flattest galaxies in our sample, all having axial ratio $q'<0.4$. These galaxies are likely to be close to edge-on. We assume they are exactly edge-on and we then use the MGE formalism (equations \ref{eq:surf}, \ref{eq:dens} and \ref{eq:qmge}) to deproject the surface brightness and calculate the intrinsic luminosity density. We then project it back on the sky plane at different inclinations, from edge on ($i=90^\circ$) to face on ($i=0^\circ$). At every inclination we calculate the two effective radii \re\ and  $R_e^{\rm maj}$ (\reffig{fig:inclination_dependence_re}). The comparison shows that,  as expected, the \re\ of flattened objects can be much smaller when objects are edge-on than face-on, with a median decrease of 43\% (0.24 dex). The opposite is true for $R_e^{\rm maj}$, but the variations are dramatically smaller, with a median increase of 5\% (0.02 dex). The two effective radii of course are the same for intrinsically spherical objects. The use of $R_e^{\rm maj}$ instead of \re\ is especially useful when one considers that 86\% of the galaxies in \atl\ (and in the nearby Universe) are disk like (Paper~II, III and VII).

In what follows we also need the radius $r_{1/2}$ of a sphere enclosing half of the galaxy light. For this we need to derive the intrinsic galaxy luminosity density from the MGE, assuming the best fitting inclination of the JAM models. A possible deprojection of the observed MGE surface brightness can be derived analytically by deprojecting the individual Gaussians separately \citep{Monnet1992}. The solution is only unique when the galaxy is edge-on \citep{Rybicki1987}. The deprojected luminosity density  $\nu$ is given by
\begin{equation}\label{eq:dens}
\nu(R,z) = \sum_{k=1}^M
\frac{\Sigma_j q'_j}{\sqrt{2\pi}\, \sigma_j q_j} \exp
\left[
    -\frac{1}{2\sigma_j^2}
    \left(R^2+\frac{z^2}{q_j^2} \right)
\right],
\end{equation}
where the individual components have the same dispersion $\sigma_j$ as in the projected case (\ref{eq:surf}), and the intrinsic axial ratio of each Gaussian becomes
\begin{equation}\label{eq:qmge}
    q_j=\frac{\sqrt{q'^2_j-\cos^2 i}}{\sin i},
\end{equation}
where $i$ is the galaxy inclination ($i=90^\circ$ being edge-on). To calculate $r_{1/2}$ from the intrinsic density of \refeq{eq:dens} one can proceed analogously to the approach used to measure the circularized \re. This is done by making the three-dimensional MGE distribution spherical, while keeping the same total luminosity and peak luminosity density of each Gaussian. This is achieved by replacing $(\sigma_j,q_j)$ with $(\sigma\, q_j^{1/3},1)$. The light of this new spherical MGE enclosed within a sphere of radius $r$ is given by
\begin{equation}\label{eq:lumr_mge}
L(r) = \sum_{j=1}^M L_j
    \left[{\rm erf}(h_j) - 2h_j\exp(-h_j^2)/\sqrt{\pi}\right],
\end{equation}
with $h_j=r/(\sqrt{2}\,\sigma_j\, q_j^{1/3})$ and erf the error function. And the half-light spherical radius $r_{1/2}$ is obtained by solving $L(r)=L/2$ by interpolation. As in the projected case, when all Gaussians have the same $q_j=q$, which means the density is stratified on similar oblate spheroids, the method gives the geometric radius $r_{1/2}=(a b c)^{1/3}=a\, q^{1/3}$, where $a$ is the semi-major axis of the spheroid. While when the $q_j$ are different, this radius provides a very good approximation to the radius $r_{1/2}=[3V_e/(4\pi)]^{1/3}$ of a sphere that has the same volume $V_e$ of the iso-surface enclosing half of the total galaxy light.

In \reffig{fig:different_re_definitions} we compare the three definitions of \re\ as a function of the observed effective ellipticity $\varepsilon_e$ of the MGE, for all the galaxies in the \atl\ sample. Even though the galaxy isophotes are in most cases not well approximated by ellipses, and the galaxies are intrinsically not oblate spheroids, the ratio between \re\ and $R_e^{\rm maj}$ follows the relation for elliptical isophotes. When the galaxies are very close to circular on the sky \re\ and $R_e^{\rm maj}$ agree by definition. The situation is very different regarding the relation between $r_{1/2}$ and $R_e^{\rm maj}$. In this case, when the galaxy is edge-on, there is a simple ratio $r_{1/2}/R_e\approx1.42$, but when the galaxies have lower inclinations, large variations in the ratio are possible, so that $r_{1/2}$ cannot be inferred from the observations, without the knowledge of the galaxy inclination, which generally require dynamical models. The situation is of course much simpler for spherical objects, in which case $r_{1/2}/R_e\approx1.42$ as in the edge-on case. For comparison \citet{Hernquist1990} found the theoretical value $r_{1/2}/R_e\approx1.33$ for his spherical models, while \citet{Ciotti1991} has shown that for a $R^{1/m}$ model the ratio is confined between 1.34--1.36, when $m=2-10$, and the same applies to other simple profiles \citep{Wolf2010}. As expected our ratio is slightly larger, given that our models, like real galaxies, do not extend to infinite radii. For flatter models the cylindrical and spherical circularized radii are approximately related as $\re/R_e^{\rm maj}=\sqrt{\varepsilon_e}$, which one would expect for elliptical isophotes while the ratio $r_{1/2}/\re$ remains approximately constant.

\subsubsection{Comparing effective and gravitational radius}
\label{sec:rgrav}

For an isolated spherical system in steady state one obtains from the scalar virial theorem \citep{Binney2008}
\begin{equation}
    M=\frac{r_g \langle v^2 \rangle_\infty}{G},
\end{equation}
where $r_g$ is defined as the gravitational radius, which depends on the total and luminous mass distribution, $M$ is the galaxy total luminous plus dark mass and $\langle v^2 \rangle_\infty$ is the mean-square speed of the galaxy stars, integrated over the full extent of the galaxy. In the spherical case $\langle v^2 \rangle_\infty=3\langle \sigma_{\rm los}^2 \rangle_\infty$ and
\begin{equation}\label{eq:virial}
    M=3\frac{r_g \langle \sigma_{\rm los}^2 \rangle_\infty}{G}.
\end{equation}
This formula is {\em rigorously} independent of anisotropy and only depends on the radial profiles of luminous and dark matter \citep[section~4.8.3]{Binney2008}.

When the spherical system is self-consistent ($L(r)\propto M(r)$) the gravitational radius can be easily calculated as
\begin{equation}\label{eq:rgrav}
r_g = \frac{2L^2}{\int_0^\infty [L(r)/r]^2 {\rm d}r}.
\end{equation}
Here we evaluate this expression using a single numerical quadrature via \refeq{eq:lumr_mge}, from the same spherical deprojected MGE we used in the previous Section to calculate $r_{1/2}$. The MGE is obtained by deprojecting the observed surface brightness at the JAM inclination and subsequently making the MGE spherical while keeping the same peak stellar density and luminosity of every Gaussian. In this way our calculation of $r_g$ is rigorously accurate when the MGE is already spherical, while the formula provides a good approximation for flattened galaxies.

\begin{figure}
\plotone{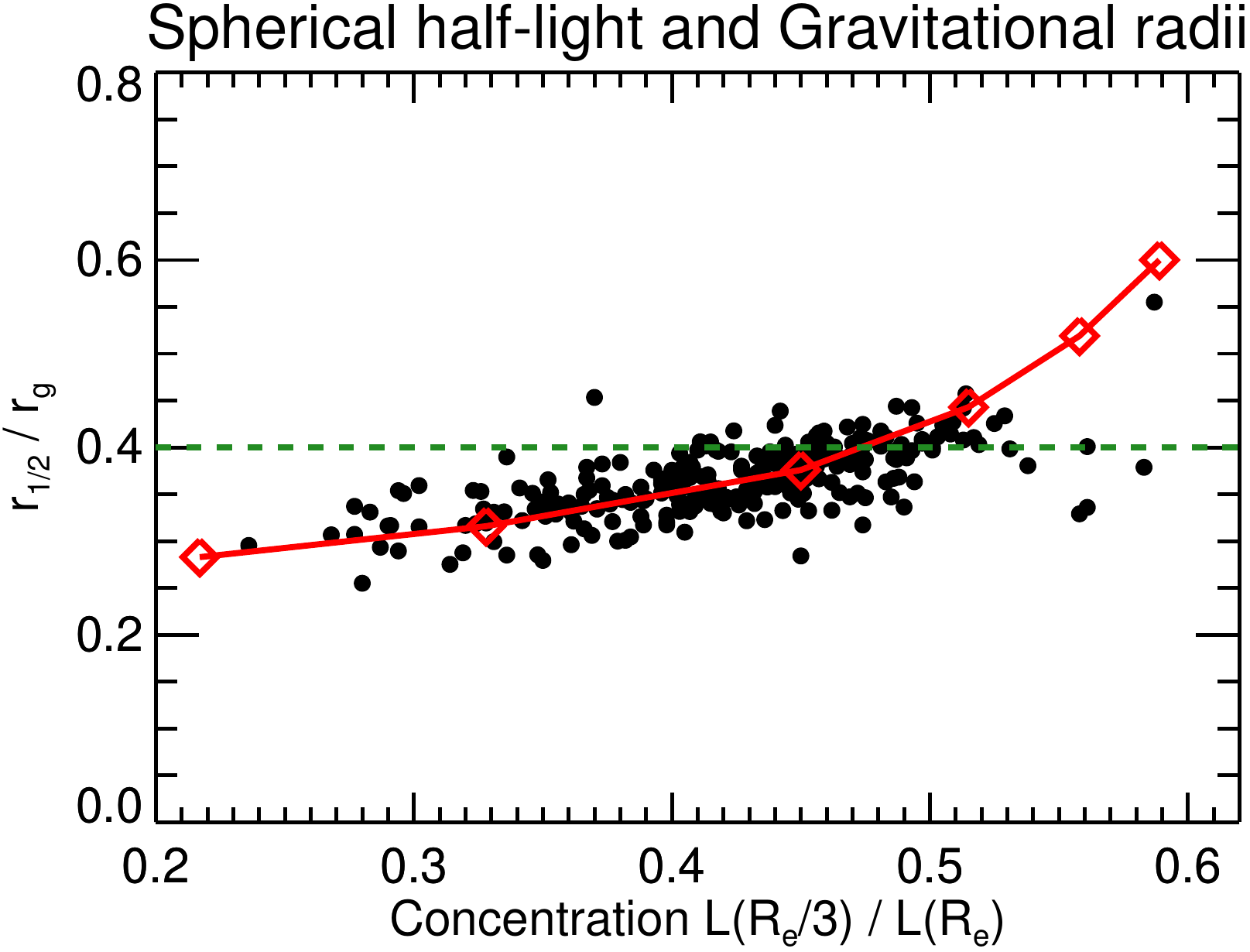}
\caption{The black filled circles mark the ratio $r_{1/2}/r_g$ between the radius of the half-light sphere and the gravitational radius for all the galaxies in the sample. For comparison the solid red line indicates the same ratio for a spherical galaxy with an $R^{1/m}$ surface brightness  profile. From left to right the red diamonds mark the locations $m=1,2,4,6,8,10$ respectively. The green dashed horizontal line indicates the theoretical value for a \citet{Hernquist1990} profile.}
\label{fig:r12_versus_rgrav}
\end{figure}

In \reffig{fig:r12_versus_rgrav} we plot the ratio $r_{1/2}/r_g$, for the full \atl\ sample as a function of the non-parametric Third Galaxy Concentration (TGC) defined in \citet{Trujillo2001} as the ratio between the light $L(\re)=L/2$ enclosed within an isophote of radius \re\ and the one $L(\re/3)$ enclosed within an isophote with radius $\re/3$. \citet{Graham2001conc} have shown that this choice leads to a more robust measure of concentration than popular alternatives \citep[e.g.][]{Doi1993}. We compute the TGC from the circularized MGE using \refeq{eq:lumRmge}, as done for \re. We find a trend in the ratio for the galaxies in our sample that varies between $r_{1/2}/r_g\approx0.3-0.4$ for the range of galaxy concentrations we observed. For comparison we also calculate the TGC and the corresponding $r_g$  for spherical models described by the $R^{1/m}$ profile \citep{Sersic1968}. This was done by constructing analytic profiles, truncating them to $R<4\re$, to mimic the depth of the SDSS photometry, before fitting them with the one-dimensional MGE-fitting procedure of \citet{Cappellari2002mge}. Both TGC and $r_{1/2}/r_g$ span the ranges predicted for profiles with $m=2-6$. Our trend in the ratio is more significant than the generally assumed near constancy around $0.40\pm 0.02$, first reported by \citet{Spitzer1969} for different polytropes, which agrees with the theoretical value $r_{1/2}/r_g=(1+\sqrt{2})/6\approx0.402$ for a \citet{Hernquist1990} profile \citep{Mamon2000,Lokas2001}. However, the variation is indeed rather small, being only at the $\pm15$\% level around a median value of 0.35 in our sample.

The relatively small variations of the ratio between the gravitational and intrinsic $r_{1/2}$ or projected \re\ half-light radii, explain the usefulness of the latter two parameters in measuring dynamical scaling relations of galaxies.
This fact, combined with the rigorous independence on anisotropy, also explains the robustness of a mass estimator like
\begin{equation}\label{eq:m12_virial}
    M_{1/2}=k\frac{r_{1/2} \langle \sigma_{\rm los}^2 \rangle_\infty}{G},
\end{equation}
when the stellar systems can be assumed to be spherical and kinematics is available over the {\em entire extent} of the system, as pointed out by \citet{Wolf2010}. Assuming the measured ratio $r_{1/2}/r_g\approx0.4$ for galaxies with the approximate concentration of an $R^{1/4}$ profile, already in the self-consistent limit the expected coefficient is $k\approx3/0.4/2=3.75$, which is close, but 25\% larger than the corresponding coefficient $k=3$ proposed by \citet{Wolf2010}. However, the ratio $r_{1/2}/r_g$ we empirically measured on real galaxies, does not assume the outermost galaxy profiles are known and can be extrapolated to infinity, so it weakly depends on the depth of the photometry. For example, for a spherical galaxy that follows the $R^{1/4}$ profile to infinity, we obtain $r_{1/2}/r_g=0.456$, which would imply $k=3.29$ in the self-consistent limit. The remaining 10\% difference from \citet{Wolf2010} is easily explained by the small increase of $\langle \sigma_{\rm los}^2 \rangle_\infty$ due to the inclusion of a dark halo.

\subsubsection{Determination of $\sigma_e$}

\label{sec:sigmae}

Unfortunately the quantities $\langle v_{\rm los}^2 \rangle_\infty$, or $\langle \sigma_{\rm los}^2 \rangle_\infty$ are currently only observable via discrete tracers in objects like nearby dwarf spheroidal (dSph) galaxies \citep[e.g.][]{Walker2007}, but it is still not a directly observable quantity in early-type galaxies. Nonetheless \citet{Cappellari2006} showed that in practice $\langle v_{\rm los}^2 \rangle_e$, as approximated by $\sigma_e^2$, which can be empirically measured for large samples of galaxies, can still be used to derive robust central masses when applied to real, non-spherical ETGs, with kinematics extended to about 1\re:
\begin{equation}\label{eq:cap06}
(M/L)(r=\re) \approx 5.0\times\frac{\re \sigma_e^2}{G L},
\end{equation}
where $(M/L)(r=\re)$ is estimated inside an iso-surface of volume $V=4\pi R_{\rm e}^3/3$ (a sphere of radius \re\ if the galaxy is spherical), and $\sigma_e$ is the velocity dispersion calculated within a projected circular aperture of radius \re. In this paper we improve on the previous approach by measuring $\sigma_e$ inside an effective ellipse instead of a circle. The ellipse has area $A=\pi R_e^2$ and ellipticity $\varepsilon_e$. The measurement is done by co-adding the luminosity-weighted spectra inside the elliptical aperture and measuring the $\sigma$ of that effective spectrum using pPXF \citep{Cappellari2004} and assuming a Gaussian line-of-sight velocity distribution (keyword MOMENTS$=$2). Due to the co-addition, the resulting spectrum has extremely high $S/N$ (often above 300) and this makes the measurement robust and accurate. When the \sauron\ data do not fully cover \re\ we correct the $\sigma_e$ to 1\re\ using equation~(1) of \citet{Cappellari2006}. $\sigma_e$ has the big advantage over $\langle v_{\rm los}^2 \rangle_e$ that it can also be much more easily measured at high redshift, as it does not require spatially resolved kinematics. Integrated stellar velocity dispersions have started to become measurable up to redshift $z\approx2$ \citep{Cenarro2009,Cappellari2009,vanDokkum2009nat,Onodera2010,vandeSande2011}.
Moreover the advantage of $\sigma_e$ over the traditional central dispersion $\sigma_c$, is that it is empirically closer to the true second velocity moment $\langle v_{\rm los}^2 \rangle_\infty$ that appears in the virial \refeq{eq:virial} and is directly proportional to mass.
Making the good approximation $(M/L)(r=\re)\approx(M/L)(r=r_{1/2})$, where $r_{1/2}\approx1.33\re$, one can rewrite \refeq{eq:cap06} in a form that is directly comparable to \refeq{eq:m12_virial}
\begin{equation}\label{eq:cap06m12}
M_{1/2}\approx 2.5\times\frac{\re \sigma_e^2}{G} \approx 1.9\times\frac{r_{1/2}\, \sigma_e^2}{G}.
\end{equation}
Note that the empirical coefficient 1.9 is significantly smaller than the value around 3.0 one predicts when using $\langle \sigma_{\rm los}^2 \rangle_\infty$ in \refeq{eq:m12_virial} and we will come back to this point in \refsec{sec:fp_to_vp}.

\section{Results}

\subsection{Uncertainty in the scaling relations parameters}

\subsubsection{Errors in $L$, \re\ and $\sigma$}

In the study of galaxy scaling relations formal errors on $L$, \re\ and $\sigma$ are often adopted, as given in output by the program used for their extraction. These errors assume the uncertainties are of statistical nature. However, in many realistic situations the systematic errors are significant, but difficult to estimate. In this work, the availability of a significant sample of objects, with similar quantities measured via independent data or methods, allow for a direct comparison of quantities. This external comparison permits us to include systematic errors into our adopted errors, instead of just using formal or Monte Carlo errors.

In Paper~XXI we compare the total magnitude $M_r$ of the MGE model, as derived from the SDSS$+$INT $r$-band photometry to various other sources in the literature. We conclude that our total $M_r$ are accurate at the 10\% level, in the relative sense. This is the error we adopted in what follows. This accuracy is comparable to other state-of-the-art photometric surveys. 

\begin{figure}
\centering
\includegraphics[width=0.75\columnwidth]{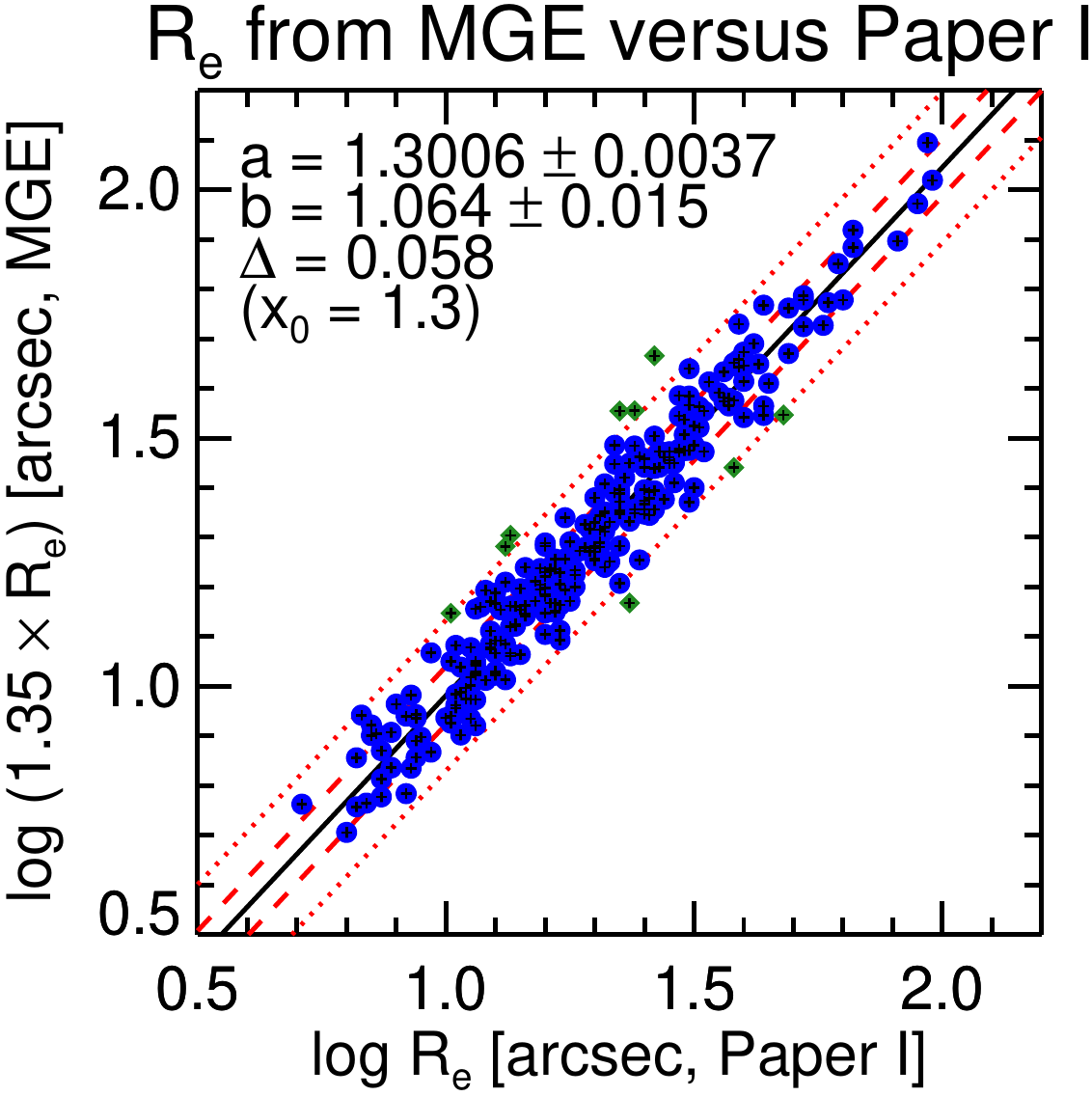}
\caption{Testing the relative accuracy of size measurements. Comparison between the \re\ from 2MASS plus RC3, matched to RC3 as described in Paper~I, and the \re\ from the MGEs. For a good match the MGE values have been {\em increased} by a significant factor 1.35. In what follows the effective radii will always already include this multiplicative factor. The coefficients of the best-fitting relation $y=a+b (x-x_0)$ and the corresponding observed scatter $\Delta$  in $y$ are shown at the top left of the plot. The two red dashed and dotted lines mark the $1\sigma$  bands (enclosing 68\% of the values for a Gaussian distribution) and $2.6\sigma$ (99\%) respectively. The outliers automatically excluded from the fit by the \textsc{lts\_linefit} procedure are shown as green diamonds.}
\label{fig:mge_reff_accuracy}
\end{figure}

A comparison between the circularized half-light radii \re\ of Paper~I and the circularized \re\ from the $r$-band MGE is shown in \reffig{fig:mge_reff_accuracy}. In this case the rms scatter is of $\Delta=0.058$ dex, which would imply errors of $\Delta/\sqrt{2}=10\%$ in the individual \re. This is the error we adopt for our \re\ determination. This must still be a firm upper limit to the errors, given that any relative variations, among galaxies, in the colour gradients in $r$ and $K_s$ will increase the scatter. Remarkably in this case our scatter between SDSS $r$-band and 2MASS $K_s$ bands, for the entire sample, is as small as the best agreement (0.05 dex) reported by \citet[their fig.~8]{Chen2010}, when comparing their determinations versus those of \citet{Janz2008}, using the very same SDSS $g$-band photometry and curve-of-growth technique. We are not aware of other published independent \re\ determinations from different data that agree with such a small scatter, and for such a large sample. The rms scatter we measure is twice smaller that the \citet{Chen2010} comparison in the same band between SDSS and ACSVCS. Our scatter is also twice smaller than a similar comparison we performed in Paper~I between the \re\ of 2MASS and RC3. We interpret the excellent reproducibility of our MGE \re\ values, and the agreement with the values of Paper~I, to the fact that in both 2MASS and our MGE models the total luminosities are {\em not} computed via a extrapolation of the profile to infinity, but simply measured from the data. This result is a reminder of the fact that extrapolation is a dangerous practice, which should be avoided whenever possible. 

A very important feature of \reffig{fig:mge_reff_accuracy} is the significant offset by a factor 1.35 between the MGE \re\ and the values of Paper~I, with the MGE values being smaller. In what follows all our MGE effective radii will always already include this multiplicative factor. The values of Paper~I where determined from a combination of 2MASS \citep{Skrutskie2006} and RC3 \citep{deVaucouleurs1991} \re\ measures. But they were scaled to match on average the values of the RC3 catalogue, which were determined using growth curves extrapolated to infinity. The RC3 normalization agree within 5\% with the \sauron\ determinations in \citep{Cappellari2006,Kuntschner2006,FalconBarroso2011}. Part of the 1.35 offset is simply due to the extrapolated light in an $r^{1/4}$ profile, outside the region where our galaxy extend on the SDSS or INT images. But the source of the remaining offset is unclear and confirms the difficulty of determining \re. For comparison in Paper~I we showed that the 2MASS and RC3 values correlate well, but have an even more significant offset of a factor 1.7!

\begin{figure}
\centering
\includegraphics[width=0.75\columnwidth]{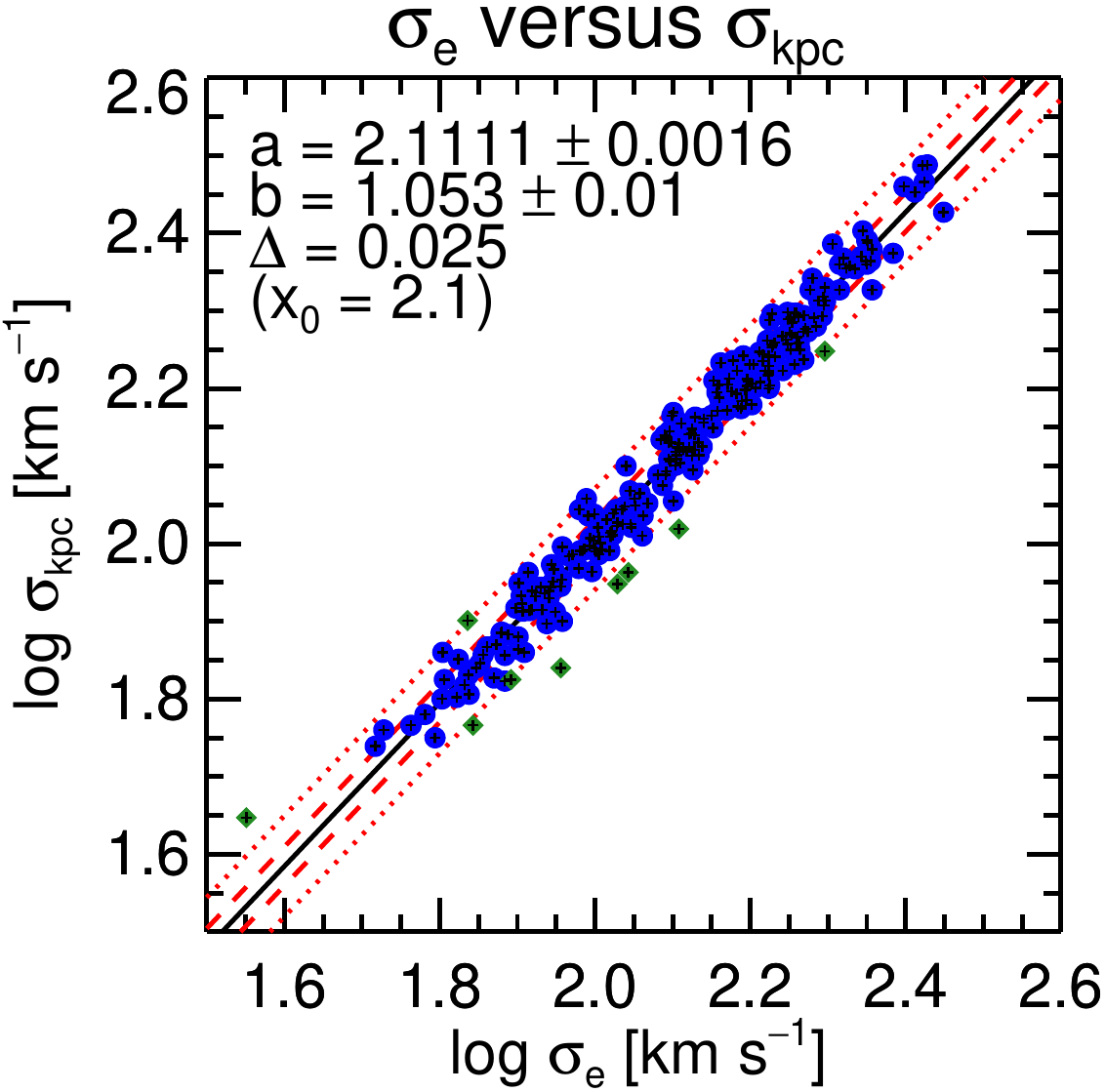}
\includegraphics[width=0.75\columnwidth]{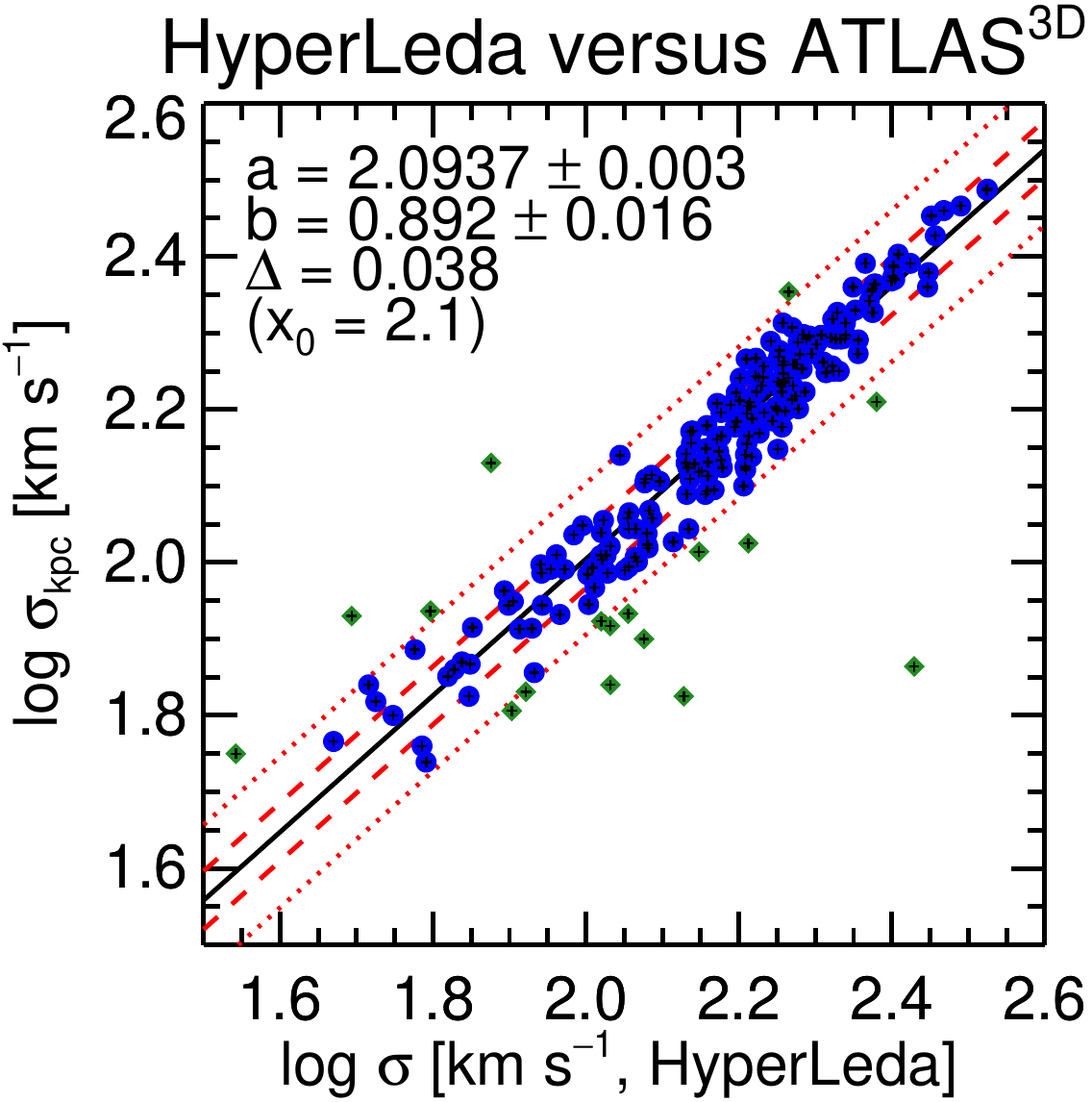}
\caption{Testing the relative accuracy of $\sigma_e$ determinations. {\em Top Panels:} Same as in \reffig{fig:mge_reff_accuracy} for the comparison between the dispersion $\sigma_e$, as measured with pPXF from the spectrum inside an elliptical aperture of area $A=\pi R_e^2$, and the quantity $\sigma_{\rm kpc}$ measured on a spectrum extracted inside a fixed circular aperture of radius $R=1$ kpc. {\em Bottom Panels:} Same as in the top panel, for the comparison between our $\sigma_{\rm kpc}$ and the central $\sigma$ provided by the HyperLeda database.}
\label{fig:sigma_e_accuracy}
\end{figure}

Various comparisons of the accuracy of kinematic quantities have been performed in the literature \citep[e.g.][]{Emsellem2004}. The general finding is that the measurements of the galaxies velocity dispersion can be reproduced at best with an accuracy of $\approx5\%$, mainly due to uncertainties in the stellar templates and various systematic effects that are difficult to control. Here in \reffig{fig:sigma_e_accuracy} we test the internal errors of our kinematic determination by comparing $\sigma_e$ against the velocity dispersion $\sigma_{\rm kpc}$ measured within a circular aperture of radius $R=1$ kpc (close to the radius $R=0.87$ kpc adopted in \citealt{Jorgensen1995spec}). This aperture is always fully contained in the observed \sauron\ field-of-view. We measure an rms scatter of $\Delta=0.025$ dex between the two quantities, which corresponds to a $1\sigma$ error of 4\% in each value. The two values do not measure the same quantity, as the two adopted apertures and fitted spectra are different, and for this reason both the actual velocity dispersion and the stellar population change in the two pPXF fits. For this reason the observed scatter provides a firm upper limit to the true internal uncertainties in $\sigma_e$. However, in what follows we still assume a conservative error of 5\% in $\sigma_e$ and $\sigma_{\rm kpc}$, to account for possible systematics. The same choice was made e.g. in \citet{Tremaine2002} and \citet{Cappellari2006}. We further compared our $\sigma_{\rm kpc}$ values againts the literature $\sigma$ compilation in the HyperLEDA database \citep{Paturel2003}, for 207 galaxies in common with our sample. A robust fit between the logarithm of the two quantities eliminating outliers with \textsc{lts\_linefit} gives an observed rms scatter of 9\% ($\Delta=0.038$ dex), likely dominated by the heterogeneity of the HyperLEDA values, and no significant offset (1\%) in the overall normaliztion. Apart from placing a very firm upper limit to our errors, this provides an external estimate of the typical uncertainties in the HyperLEDA values.

\subsubsection{Errors in mass or $M/L$}
\label{sec:mass_err}

\begin{figure}
\centering
\includegraphics[width=0.75\columnwidth]{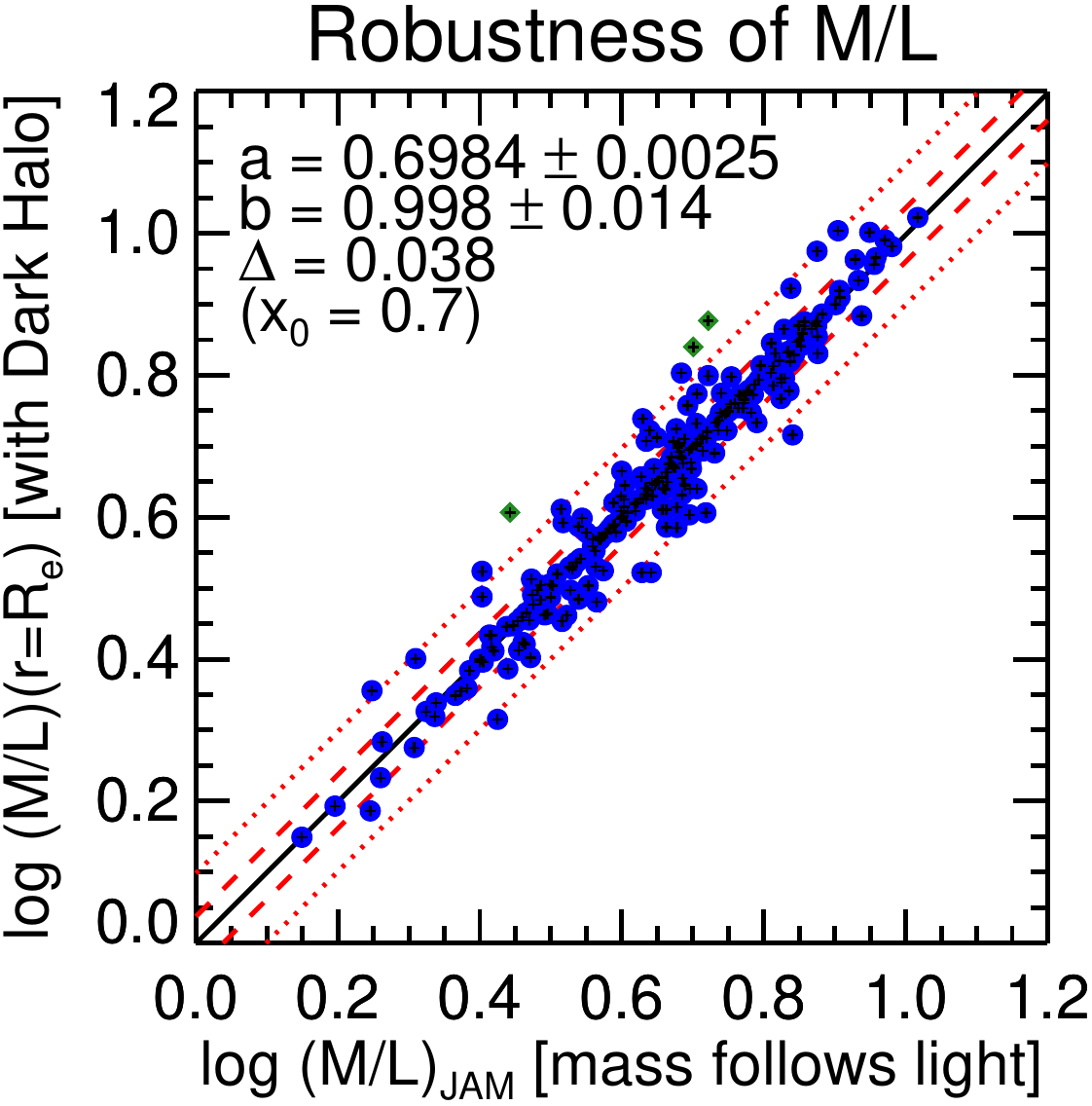}
\includegraphics[width=0.75\columnwidth]{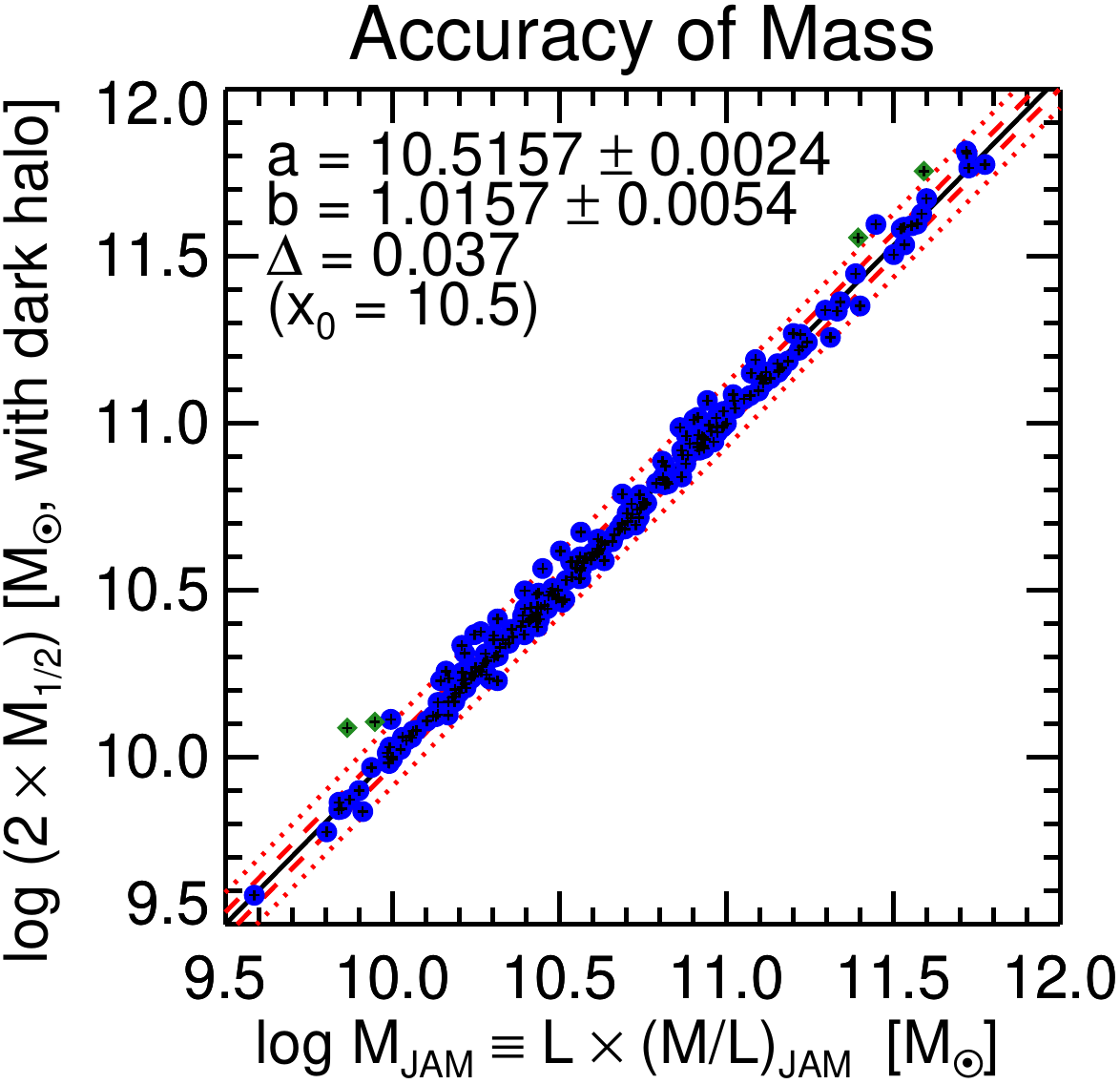}
\caption{Accuracy of $M/L$ and mass. {\em Top Panel:} Same as in \reffig{fig:mge_reff_accuracy} for the comparison between the $(M/L)_{\rm JAM}$ of the best-fitting self-consistent (total mass follows light) models, and the $(M/L)_e$, integrated within an iso-surface of volume $V=4\pi R_{\rm e}^3/3$ (for a spherical galaxy a sphere of radius $r=\re$), including the contribution of both the stellar and the dark matter component. There is no bias or systematic offset between the two determinations, which are consistent with an error of $\Delta/\sqrt{2}=6\%$ in each quantity. {\em Bottom Panel:} Same as the top panel for the comparison between the total mass of the self-consistent JAM model and twice the mass $M_{1/2}$ within the half-light iso-surface, for the model with dark matter halo.}
\label{fig:ml_accuracy}
\end{figure}

To obtain an estimate of our mass and $M/L$ errors for the full sample, we proceed similarly to \citet{Cappellari2006}, namely we compare mass determinations using two significantly different modelling approaches. In \refsec{sec:jam} we described the six modelling approaches that were presented in \citet{Cappellari2012} and we also use in this paper. For this test we compare the self-consistent model (A) and the models (B) which include a NFW halo with mass as free parameter. For the model with NFW halo we then compute the $(M/L)_e\equiv(M/L)(r=\re)$ by numerically integrating the luminous and dark matter distribution of the models. The total $M/L$ enclosed within an iso-surface of volume $V=4\pi R_{\rm e}^3/3$ is defined as follows
\begin{equation}
(M/L)(r=\re) \equiv \frac{L(\re)\times (M/L)_{\rm stars} + M_{\rm DM}(\re)}{L(\re)},
\end{equation}
where $M_{\rm DM}$ is the mass in the dark halo. This quantity is compared with the $(M/L)_{\rm JAM}$ of the self-consistent model in the top panel of \reffig{fig:ml_accuracy}. The agreement is excellent, with no offset or systematic trend, and an rms scatter $\Delta=0.038$ dex, consistent with errors of $\Delta/\sqrt{2}=6\%$ in each quantity. This value is the same we estimated as modelling error in \citet{Cappellari2006} and confirms the original estimate of the random modelling uncertainties. There is no evidence for any significant trend or systematic offset.

Importantly this result clarifies a misconceptions regarding the use of self-consistent models to measure the $M/L$ inside $r\approx\re$ in galaxies. Self-consistent models, like the one used in \citet{Cappellari2006}, do {\em not} underestimate the total $M/L$ as it is sometimes stated \citep[e.g.][section~3.7]{Dutton2011imf}. Even though the model with dark halo has a total galaxy mass typically an order of magnitude larger inside the virial radius, and has a dramatically different mass profile at large radii, the model still measures an unbiased {\em total} $M/L$ within a sphere of radius $r\approx\re$, corresponding to the projected extent of the kinematical data. The robustness in the recovery of the enclosed total mass, in the region constrained by the data, even in the presence of degeneracies in the halo profile, was already pointed out by \citet{Thomas2005} and is demonstrated here with a much larger sample.

Of course the self-consistent $(M/L)_{\rm JAM}$ is larger than the purely stellar one $(M/L)_{\rm stars}$ if dark matter is present, according to the relation
\begin{equation}
 (M/L)_{\rm JAM} \approx (M/L)(r=\re) = \frac{(M/L)_{\rm stars}}{1 - f_{\rm DM}(r=\re)},
\label{eq:mljam_versus_mlstasrs}
\end{equation}
where the fraction of dark matter contained within an iso-density surface of mean radius \re\ is defined as 
\begin{equation}\label{eq:dm_fraction}
f_{\rm DM}(r=\re) \equiv \frac{M_{\rm DM}(\re)}{L(\re)\times (M/L)_{\rm stars} + M_{\rm DM}(\re)}.
\end{equation}
The difference between $(M/L)_{\rm JAM}$ and the stellar $M/L$ inferred from population models can then be used to give quantitative constraints on the dark matter content and the form of the IMF, as done in \citet{Cappellari2006}. Moreover the self-consistent models do not imply or require the dark mass to be negligible inside $r\approx\re$ as sometimes stated \citep[e.g.][]{Thomas2011}. Although a number of galaxies has non-negligible dark matter fraction, the total (luminous plus dark) $M/L$ within 1\re\ is still accurately recovered by the simple self-consistent models, without detectable bias. This makes the self-consistent models well suited to determine unbiased total $M/L$ within 1\re\ at high redshift \citep{vanderMarel2007,vanderWel2008vdM,Cappellari2009}, where high-quality integral-field stellar kinematics still cannot be obtained and dark matter fractions cannot be extracted.

Using integral-field data the error in this measure of enclosed mass is as small as the one that can be obtained from strong lensing studies. The important difference between the two techniques is that the lensing results measure the total mass inside a projected cylinder (or elliptical cylinder), while the stellar kinematics gives the total mass inside a spherical (or spheroidal) region. The lensing mass should be larger than the dynamical one if dark matter is present in the galaxy. The difference between these two quantities provides a measure of the dark matter content along the LOS and can be exploited to get some constraints on the dark matter profiles \citep{Thomas2011,Dutton2011swells}.

In the bottom panel of \reffig{fig:ml_accuracy} we compare the mass $M_{\rm JAM}\equiv L\times (M/L)_{\rm JAM}$, which we use extensively in this and in other papers of this \atl\ series, with the total mass $M_{1/2}$ enclosed within an iso-surface enclosing half of the total light, which is sometimes advocated to compare observations to galaxy simulations \citep[e.g.][]{Wolf2010}. The plot illustrates the equivalence of the two quantities, for all practical purposes. It clarifies the physical meaning of $M_{\rm JAM}$:
\begin{equation}
M_{\rm JAM}\approx2\times M_{1/2}.
\end{equation}

\begin{figure}
\plotone{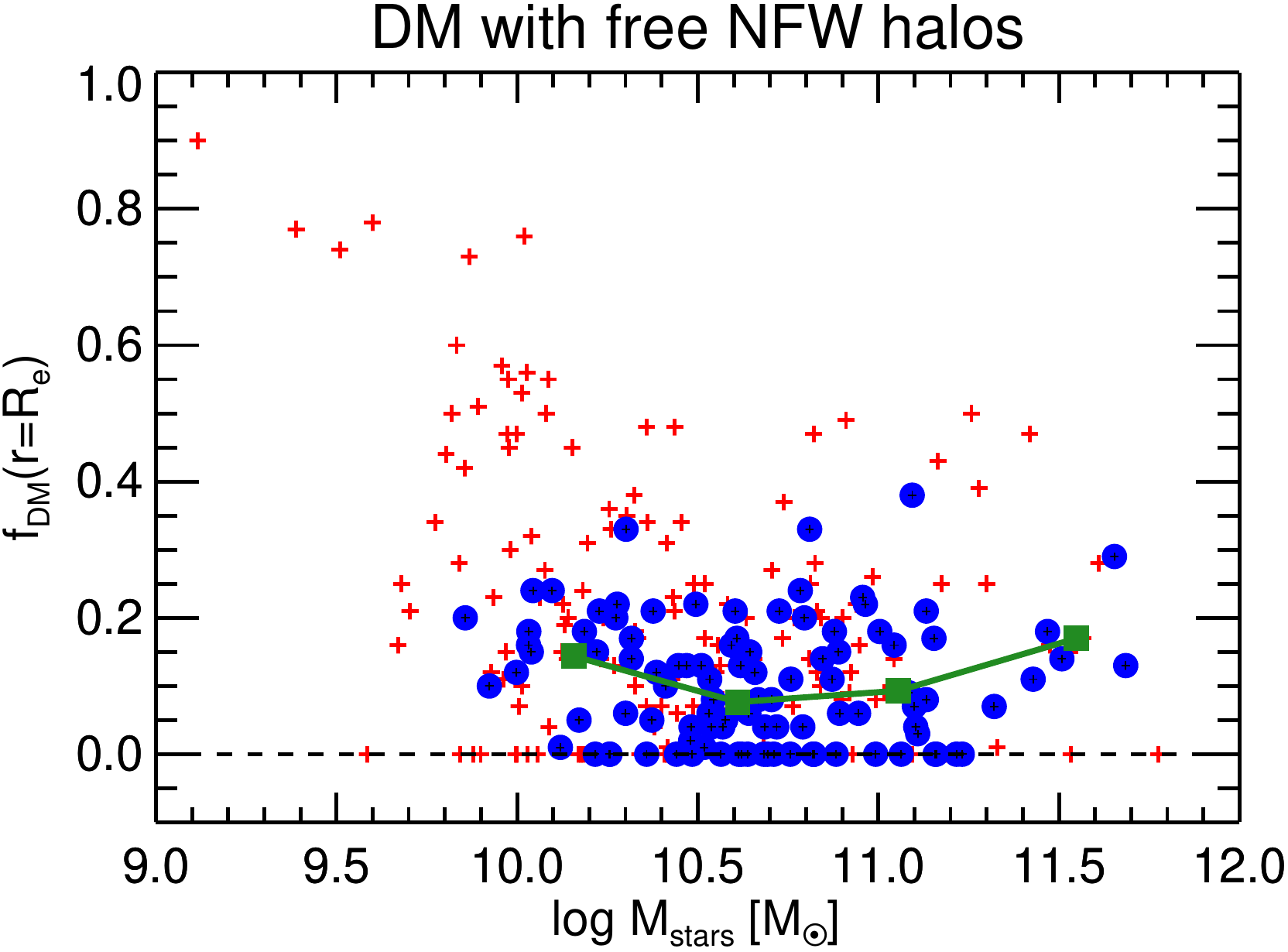}
\plotone{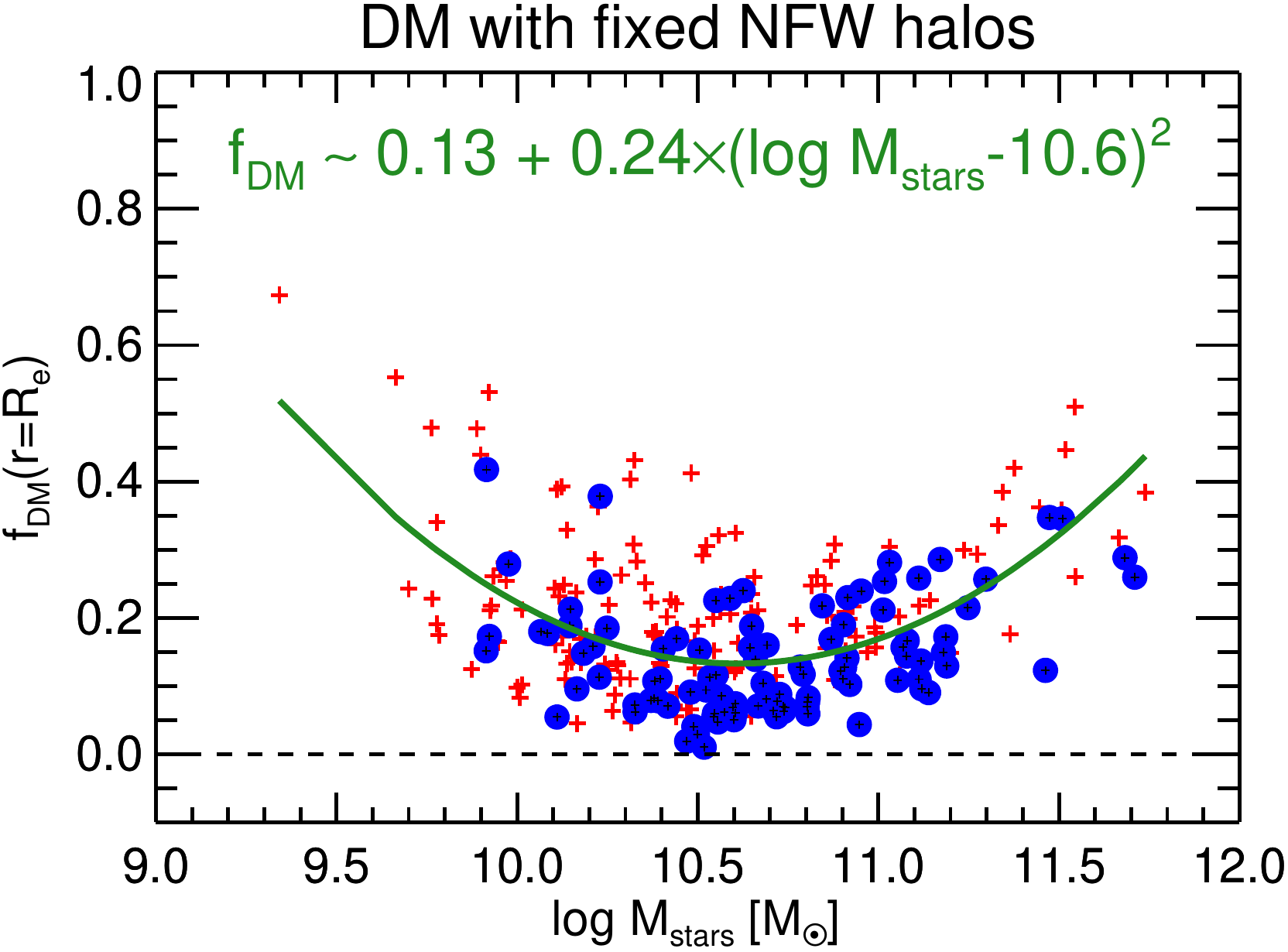}
\caption{Dark matter fraction for \atl\ galaxies. The open circles indicate the fraction $f_{\rm DM}$ of dark matter enclosed within the iso-surface of volume $V=4\pi R_{\rm e}^3/3$ (in the spherical case within a sphere of radius \re), for the best-fitting JAM models, as a function of the galaxy stellar mass $M_{\rm stars}$ inferred by the models. The blue filled circles are for the subset of 103 galaxies with the best models and data (${\rm qual}>1$ in Table~1), while the red crosses indicate less good model fits or inferior data. The {\em Top Panel} corresponds to the results for models (B), with a NFW halo having mass as free parameter. The median is $f_{\rm DM}=13\%$ for the full sample and $f_{\rm DM}=9\%$ for the best models. In a number of cases the model without dark matter is preferred. The solid green line indicates the mean for four mass bins. The {\em Bottom Panel} is the same as the top one, for the set of models (E) which has a cosmologically-motivated NFW halo.  The median $f_{\rm DM}=17\%$ for all models. The green line is a robust parabolic fit (written in the figure) to all the data. The robust result is that dark matter fractions are small, for halo slopes not steeper than NFW. $f_{\rm DM}<21\%$ in 90\% of the best models.}
\label{fig:dark_matter_fraction}
\end{figure}

The JAM models with dark halo additionally provide an estimate of the dark matter fraction $f_{\rm DM}$ (\refeq{eq:dm_fraction}) enclosed within the region constrained by the data $r=\re$. For the galaxies where our kinematics does not cover 1\re, our $f_{\rm DM}$ will be more uncertain. The results is presented, as a function of galaxy stellar mass $M_{\rm stars}$ in \reffig{fig:dark_matter_fraction} for the set of models (B), with a NFW halo, with mass as free parameter, and for the set of models (E), which have a cosmologically-motivated NFW halo, uniquely specified by  $M_{\rm stars}$. We find a median dark matter fraction for the \atl\ sample of  $f_{\rm DM}=15\%$ for the full sample and $f_{\rm DM}=9\%$ for the best (${\rm qual}>1$ in Table~1) models (B) and 17\% with models (E). These value are broadly consistent, but on the lower limit, with numerous previous stellar dynamics determinations inside 1\re\ from much smaller samples and larger uncertainties: \citet{Gerhard2001} found $f_{\rm DM}=10-40\%$ from spherical dynamical modelling of 21 ETGs; \citet{Cappellari2006} inferred a median $f_{\rm DM}\approx30\%$ by comparing dynamics and population masses of 25 ETGs, and assuming a universal IMF; \citet{Thomas2007,Thomas2011} measured $f_{\rm DM}=23\pm17\%$ via axisymmetric dynamical models of 16 ETGs; \citet{Williams2009} measured a median fraction $f_{\rm DM}=15\%$ with JAM models of 15 ETGs, as done here, but with more extended stellar kinematics to $\approx2-3$ \re; The results of \citet{Tortora2009} are not directly comparable, as they used spherical galaxy toy models and inhomogeneous literature data from various sources, however they are interesting because they explored a sample of 335 ETGs, comparable to ours, and report a typical $f_{\rm DM}=30\%$ by comparison with stellar population.

The quite small $f_{\rm DM}$ that we measure seems also consistent with the fact that the strong lensing analysis of the about 70 galaxies of the SLACS sample \citet{Bolton2006} finds a logarithmic slopes for the {\em total} (luminous plus dark matter) density close to isothermal. Subsequent re-analyses of their data all confirmed a trend $\rho_{\rm tot}(r)\propto r^{-2.0}$, with an intrinsic scatter of $\approx0.2$ \citep{Koopmans2006,Koopmans2009,Auger2010,Barnabe2011}. In \reffig{fig:contracted_halo_slope} we derive the same slope and intrinsic scatter for the stellar density alone, inside a sphere of radius $r=\re$. This fact seems to suggest that dark matter does not play a significant role in galaxy centres and that the measured isothermal density slope is essentially due the stellar density distribution. Only a very steep dark matter slope close to isothermal $\rho_{\rm DM}(r)\propto r^{-2.0}$ like the average stellar distribution could allow for significant dark matter fractions, while still being consistent with these observations. We are not aware of any theoretical or empirical evidence for these very steep dark matter cusps in galaxies.

\subsection{The classic Fundamental Plane}

Since the discovery of the Fundamental Plane (FP) relation between luminosity, size, and velocity dispersion, in samples of local elliptical galaxies \citep{Faber1987,Dressler1987,Djorgovski1987}, numerous studies have been devoted to the determination of the FP parameters either including fainter galaxies \citep{Nieto1990}, fast rotating ones \citep{Prugniel1994}, or lenticular galaxies \citep{Jorgensen1996}. The dependency of the FP parameters have been investigated  as a function of the photometric band \citep{Pahre1998,Scodeggio1998} or redshift \citep{vanDokkum1996}. Moreover galaxy samples of more tha $10^4$ galaxies have been studied \citep{Bernardi2003fp,Graves2009b,Hyde2009fp}. In this section, before presenting our result, we study the consistency of our FP parameters with previous studies.

\begin{figure}
\includegraphics[width=\columnwidth]{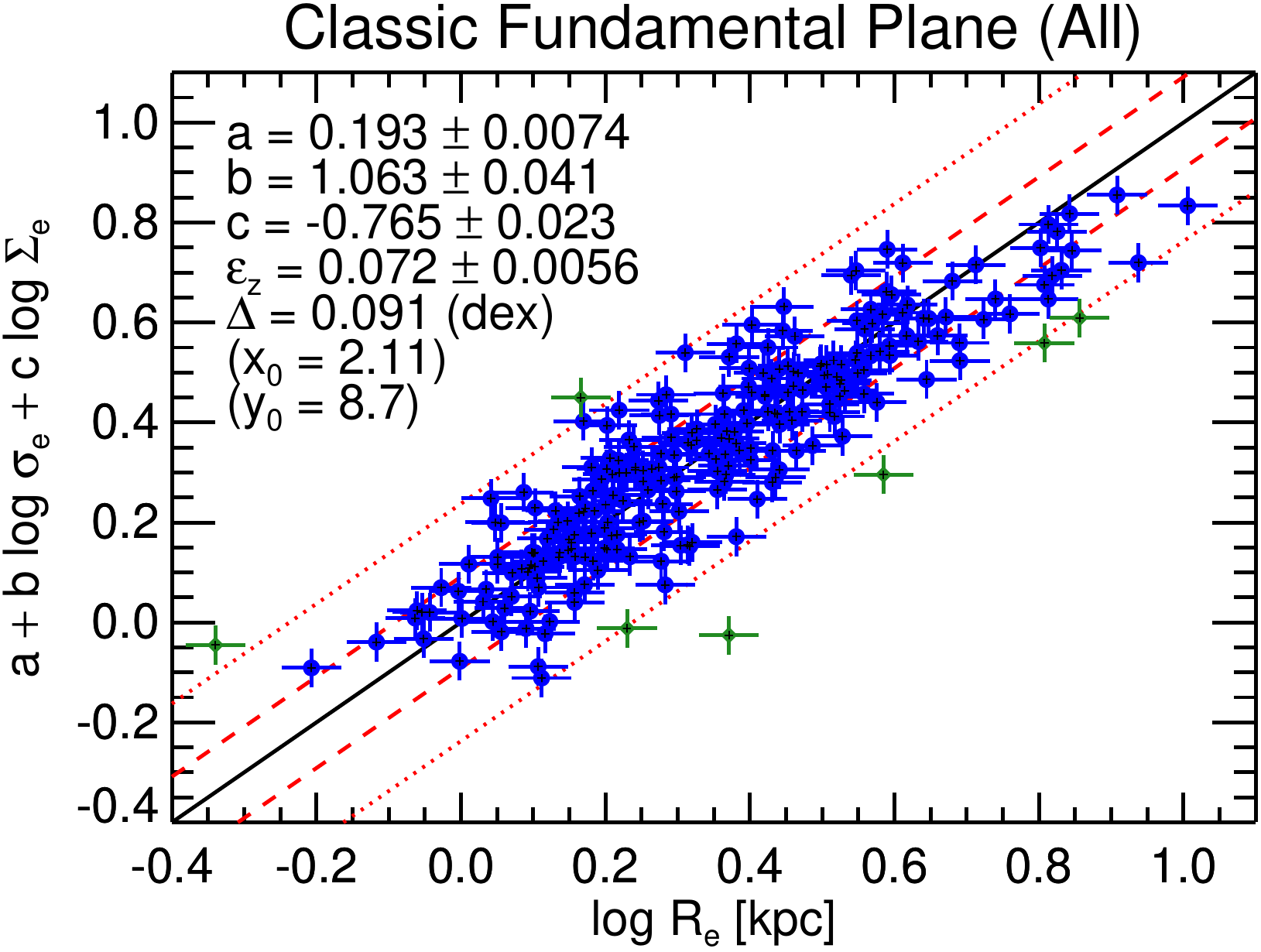}
\includegraphics[width=1.07\columnwidth]{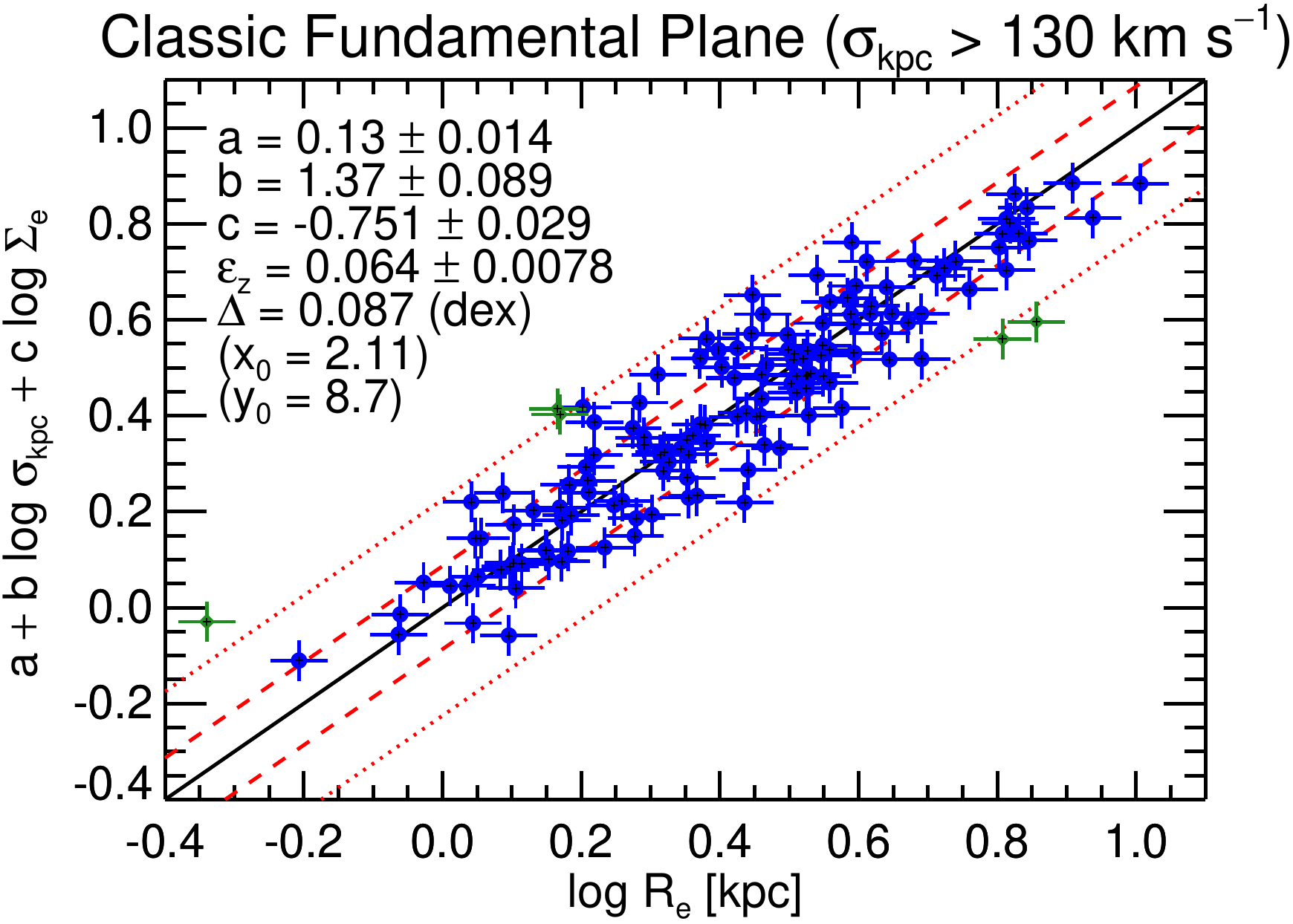}
\includegraphics[width=1.06\columnwidth]{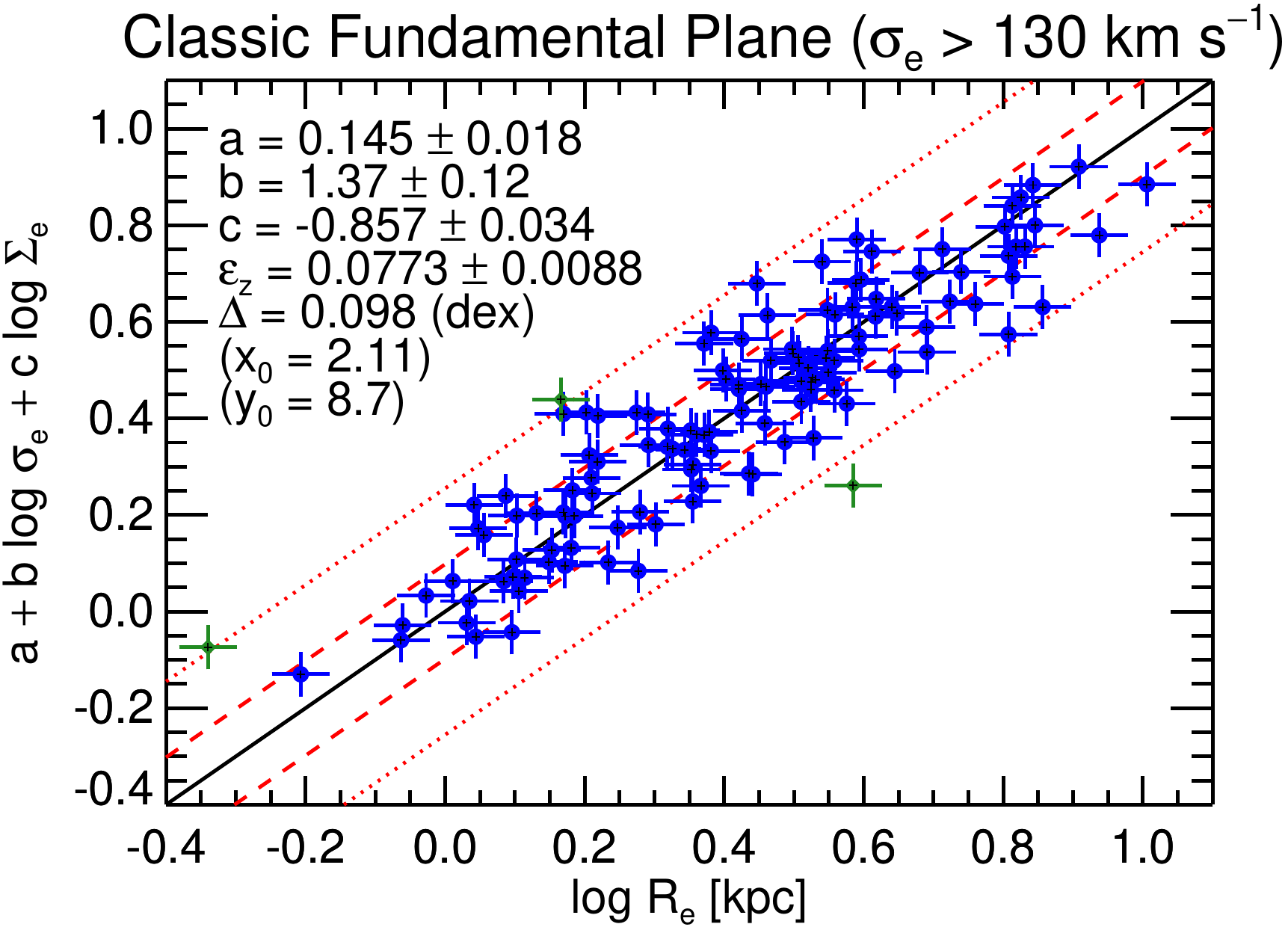}
\caption{Classic fundamental plane. {\em Top Panel:} edge-on view of the FP for all the \atl\ galaxies. The coefficients of the best-fitting plane $z=a+b x+c x$ and the corresponding observed scatter $\Delta$ are shown at the top left of the plot. The two dashed lines mark the $1\sigma$  bands (enclosing 68\% of the values for a Gaussian distribution) and $2.6\sigma$ (99\%). The outliers excluded from the fit by the \textsc{lts\_planefit} procedure are shown with green symbols. The errors are the projection of the observational errors, excluding intrinsic scatter.
{\em Middle Panel:} Same as in the top panel, with $\sigma_{\rm kpc}$ measured within a circle of radius $R=1$ kpc. Only galaxies with $\sigma_{\rm kpc}>130$ \kms\ are included.
{\em Bottom Panel:} Same as in the top panel using $\sigma_e$, but with Only including galaxies with $\sigma_e>130$ \kms.}
\label{fig:classic_fundamental_plane}
\end{figure}

Nearly all previous studies have used as variables the logarithm of the effective radius \re, the effective surface brightness $\Sigma_e$ and the (central) velocity dispersion $\sigma$. One of the reasons for this choice comes from the emphasis of the FP for distance determinations. Both $\Sigma_e$ and $\sigma$ are distance independent, so that all the distance dependence can be collected into the \re\ coordinate by writing the FP as
\begin{equation}
\log\re = a + b \log \sigma + c \log \Sigma_e.
\end{equation}
In the top panel of \reffig{fig:classic_fundamental_plane} we present the edge-on view of our \atl\ FP, obtained with the \textsc{lts\_planefit} routine, where we use as velocity dispersion $\sigma_e$ (\refsec{sec:sigmae}) as done in \citet{Cappellari2006} and \citet{FalconBarroso2011}, but here measured within an elliptical rather than circular isophote. Our best-fitting parameters $b=1.063\pm0.041$ and $c=-0.765\pm0.023$ are formally quite accurate, but significantly different from what is generally found by other studies: the median of the 11 determinations listed in table~4 of \citet{Bernardi2003fp} is $b=1.33$ and $c=-0.82$, with an rms scatter in the values of $\sigma_b=0.12$ and $\sigma_c=0.03$. The observed scatter we measure $\Delta\approx0.091$ dex in $\log\re$ is very close to what has been found by other studies (e.g. \citealt{Jorgensen1996} find 0.084).

To understand the possible reason of this disagreement we test the sensitivity of our estimate to the sample selection and the size of the kinematical aperture used for the $\sigma$ determinations. For this we measure the velocity dispersion $\sigma_{\rm kpc}$ inside a circular aperture with radius $R=1$ kpc (close to the radius $R=0.87$ kpc adopted in the classic study by \citealt{Jorgensen1995spec}). We also select the massive half of our sample by imposing a selection $\sigma_{\rm kpc}>130$ \kms. The resulting FP is shown in the middle panel of \reffig{fig:classic_fundamental_plane}, and now both the fitted values and the observed scatter agree with previous values. For comparison we also show in the bottom panel of \reffig{fig:classic_fundamental_plane} the determination of the FP parameters, when using $\sigma_e$ instead of $\sigma_{\rm kpc}$, but keeping the same selection of the massive half of our \atl\ sample $\sigma_e>130$ \kms. These values are also consistent with the literature. This illustrates the importance of sample-selection and $\sigma$ extraction in the derivation of FP parameters. The increase of $b$ as a function of the lower $\sigma$ cut-off of the selection is fully consistent with the same finding by \citet{Gargiulo2009} and \citet{Hyde2009fp} and we refer the reader to the latter paper for a more complete study of the possible biases in the FP parameters due to sample selection. The reason for the sensitivity of the FP parameters to the selection, is a result of the fact that the FP is not a plane, but a warped surface, as we demonstrate in Paper~XX by studying the variation of the $(M/L)_{\rm JAM}$ on the MP. So that the FP parameters depend on the region of the surface one includes in the fitting. This was also tentatively suggested by \citet{dOnofrio2008}.

Having shown that with our sample and method we can derive results that are consistent and at least as accurate as previous determinations, we now proceed to study the Mass Plane, by replacing the traditionally used stellar luminosity with the total dynamical mass. A similar study was performed by \citet{Bolton2007mp,Bolton2008}, and updated by \citet{Auger2010}, using masses derived from strong lensing analysis. They also call their plane the ``Mass Plane''. Although our studies are closely related, one should keep in mind that, while the lensing masses are measured within a projected cylinder of radius $R=\re/2$, parallel to the LOS, and for this reason they include a possible contribution of dark matter at large radii, our dynamical masses are measured within a sphere of radius $r=\re$.

\subsection{From the Fundamental Plane to the Mass Plane}
\label{sec:fp_to_vp}

\begin{figure}
\plotone{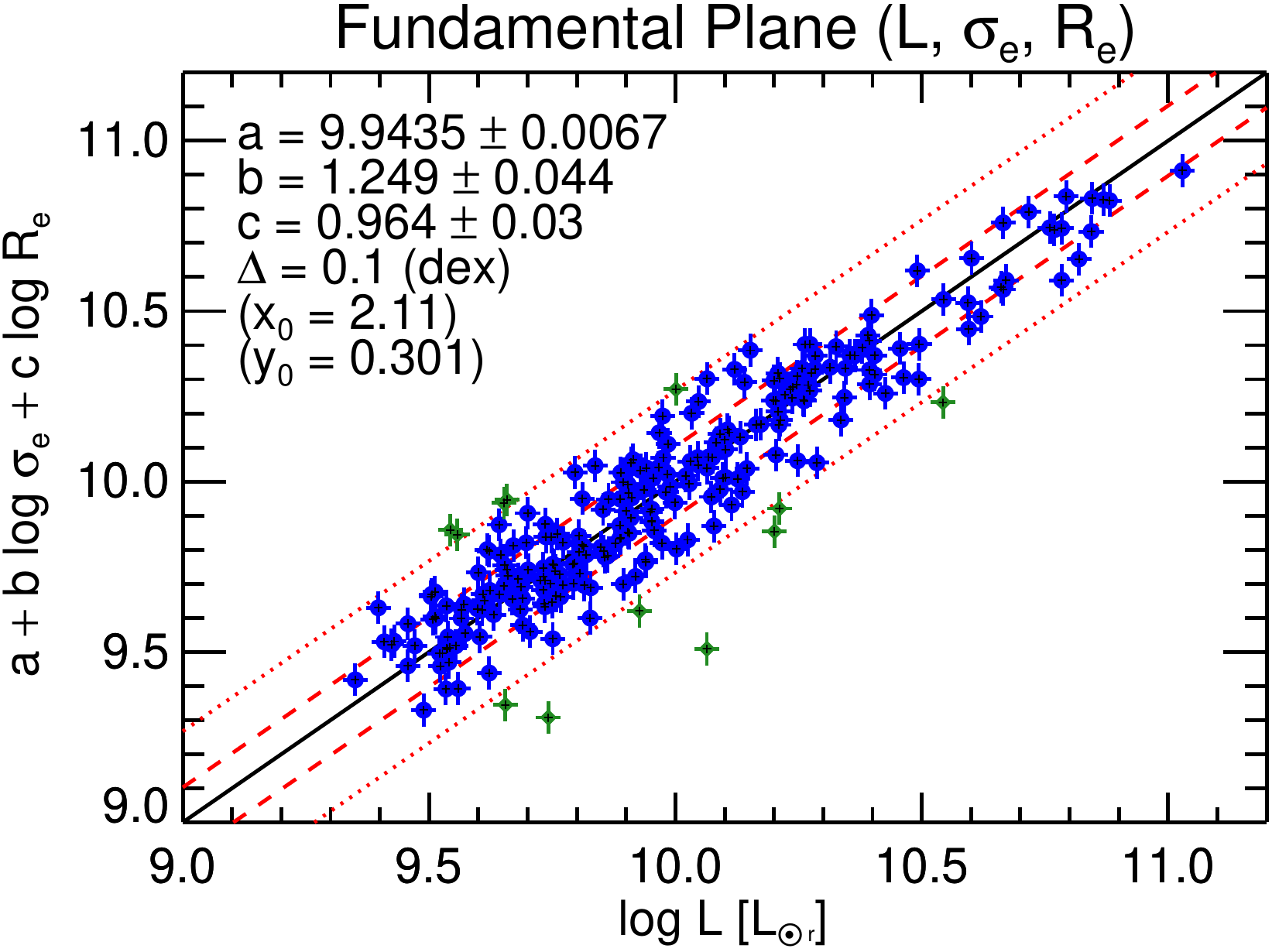}
\plotone{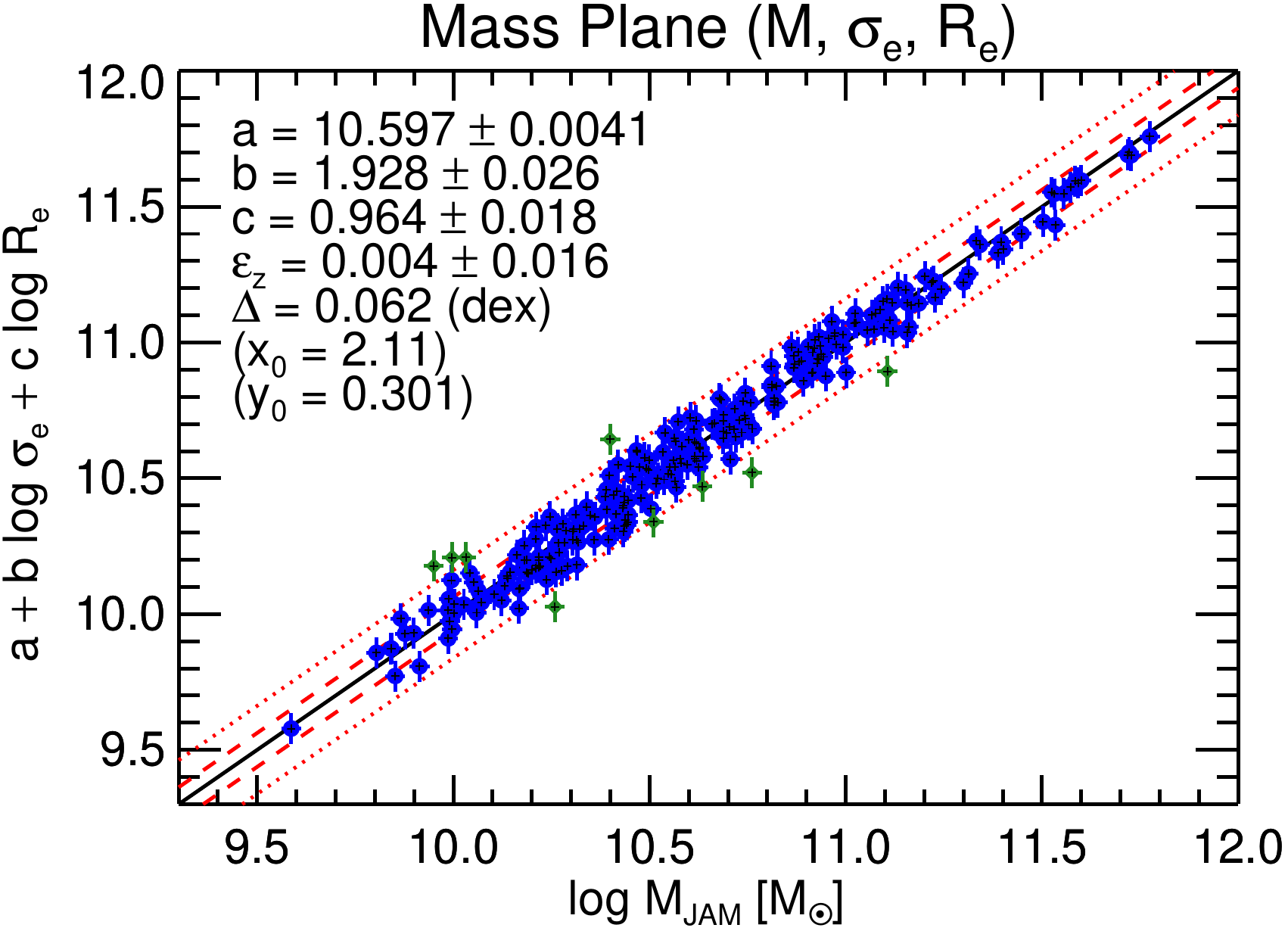}
\plotone{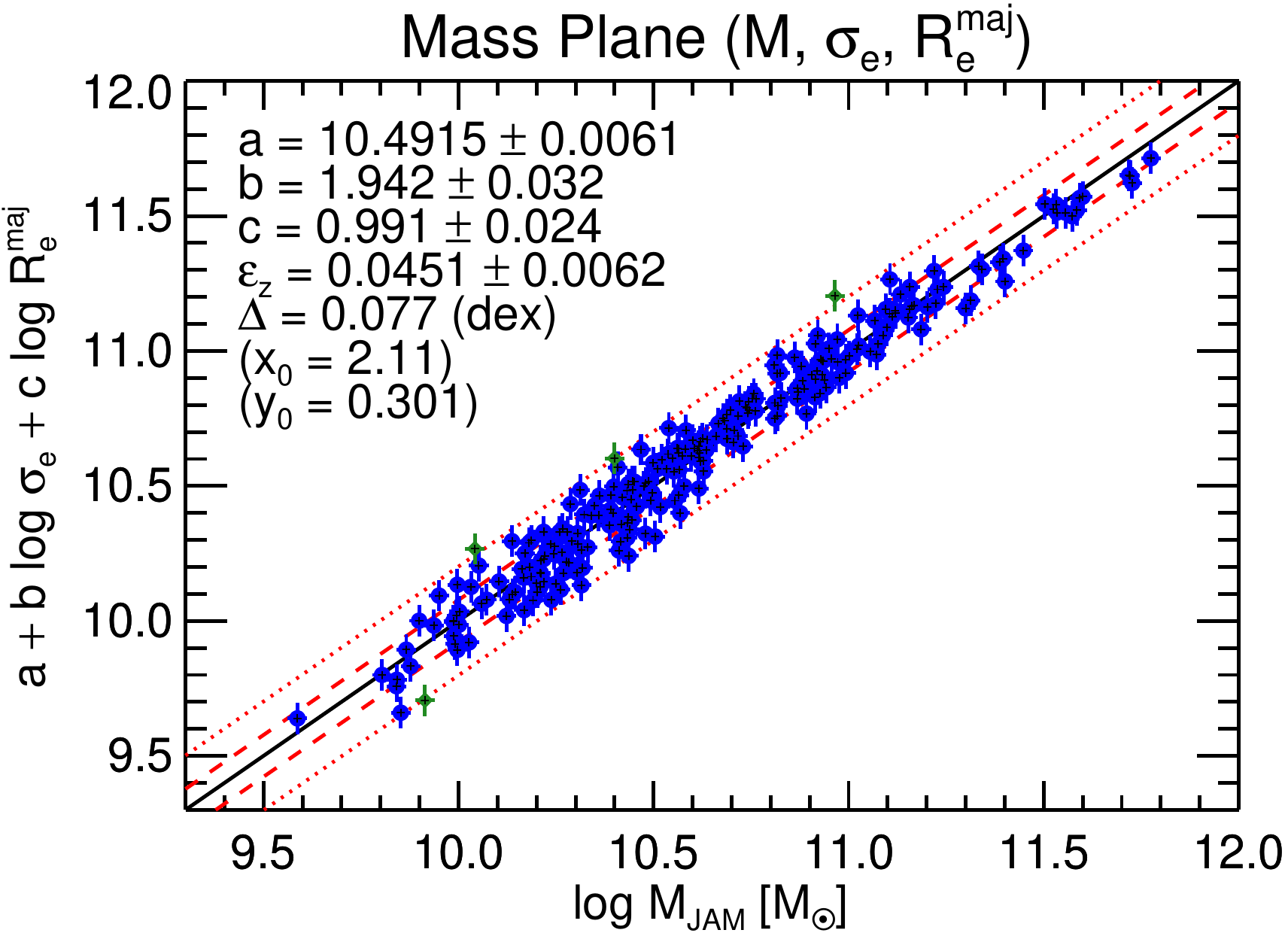}
\caption{From the Fundamental Plane to the Mass Plane. {\em Top Panel:} Edge on view of the FP. Symbols and lines are as in \reffig{fig:classic_fundamental_plane}. {\em Middle Panel:} Edge-on view of the MP. Note the decrease in the scatter, when making the substitution $L\rightarrow M$, and the variation in the coefficients, starting to approach the virial ones $b=2$ and $c=1$. {\em Bottom Panel:} Same as in the middle panel, but using as effective radius the major axis  $R_e^{\rm maj}$ of the effective isophote rather than the circularized radius. The scatter increases slightly, but the tilt is further decreased and now is consistent with the virial prediction.}
\label{fig:from_fp_to_vp}
\end{figure}

The classic form for the FP is ideal when the FP is used to determine distances. However, a different form seems more suited to studies where the FP is mainly used as a mass or $M/L$ estimator. For this we  rewrite the FP as
\begin{equation}
\log \left(\frac{L}{L_{\odot,r}}\right) = a + b\, \log \left(\frac{\sigma_e}{130\,\kms}\right) + c\, \log \left(\frac{\re}{2\, {\rm kpc}}\right).
\end{equation}
Here we normalized the $\sigma_e$ and \re\ values by the approximate median of the values for our sample, to reduce the covariance in the fitted parameters and the error in $a$. Using $L$ instead of $\Sigma_e$ has the advantage that it reduced the covariances between the pairs of observables $(\Sigma_e,\re)$. Here in fact, as opposed to when $\Sigma_e\equiv L/(2\pi R_{\rm e}^2)$ is used, there is no explicit dependence between the three axes, which become independently measured quantities. The new fit to the FP is shown in the top panel of \reffig{fig:from_fp_to_vp}. In agreement with all previous authors the fitted parameters are very different from the values $b=2$ and $c=1$ expected in the case of the virial \refeq{eq:cap06}. The relation shows a negligible increase in the observed rms scatter, from $\Delta=0.091$ dex (23\%) to $\Delta=0.10$ (26\%). This may be due to the reduced covariances between the adopted quantities: the new scatter is now a better representation of the true scatter in the FP relation.

In the bottom panel of \reffig{fig:from_fp_to_vp} we show for comparison the relation obtained by replacing the total galaxy luminosity with the dynamical mass \begin{equation}
	M_{\rm JAM}\equiv L\times (M/L)_{\rm JAM} \approx 2\times M_{1/2} \approx M_{\rm stars},
\end{equation}
where $(M/L)_{\rm JAM}$ is the total (luminous plus dark) dynamical $M/L$ obtained using self-consistent JAM models (A), $L$ is the total galaxy luminosity and $M_{1/2}$ is the total mass within a sphere of radius $r_{1/2}$ enclosing half of the total galaxy light, where $r_{1/2}\approx1.33\re$ (\citealt{Hernquist1990,Ciotti1991,Wolf2010}; \reffig{fig:different_re_definitions}). The correctness of the $M_{\rm JAM}\approx 2\times M_{1/2}$ approximation is illustrated in the bottom panel of \reffig{fig:ml_accuracy}. While the $2\times M_{1/2} \approx M_{\rm stars}$ approximation is due to the relatively small amount of dark matter enclosed within $r=r_{1/2}$ (\reffig{fig:dark_matter_fraction}). This is only approximately true, generally within 20\%, but much larger errors are generally made when determining stellar masses from stellar population models, due the assumption of a universal IMF, which was recently shown not to represent real galaxies \citep{vanDokkum2010,Cappellari2012}. None of our conclusions is affected by the last approximation, which only serves to allow for comparisons of our results to previous similar studies that use stellar mass as parameter.

Two features are obvious from the plot: (i) There is a dramatic reduction of the observed scatter from $\Delta=0.10$ (26\%) to $\Delta=0.062$ (15\%). This shows without doubt that a major part of the scatter in the FP is due to variations in the $M/L$, in agreement with independent results from strong lensing \citep{Auger2010}; (ii) The $b$ coefficient substantially increase and is now much closer to the virial value $b=2$, while the $c$ coefficient remains nearly unchanged. The coefficients become consistent with the virial ones when using the effective radius $R_e^{\rm maj}$, which is insensitive to projection effects, instead of \re. This confirms that much of the deviation of the FP from the virial predictions is due to a systematic variation in $M/L$ along the FP, not to non-homology in the luminosity profiles or kinematics, also in agreement with previous dynamical \citep{Cappellari2006} and strong lensing results \citep{Bolton2008,Auger2010}.

Galaxies are seen at random orientations so that projection effects should affect the measured $\sigma_e$. Given that the velocity ellipsoid in ETGs is generally not too far from a sphere \citep{Gerhard2001,Cappellari2007,Thomas2009}, the velocity dispersion changes weakly with inclination, while the LOS velocity varies as $V=v\sin i$, where $i$ is the galaxy inclination and $v$ is the edge-on ($i=90^\circ$) velocity. In this work we have an estimate of the galaxy inclination for every galaxy in our sample, as measured via the JAM models. Although the inclination may not be always accurate, \citet{Cappellari2008} showed that it agrees with the inclination inferred from dust disks, for a sample of four galaxies. Here we extend the comparison to an additional sample of 22 galaxies with regular dust disks. The JAM inclination is found to always agree within the relative errors, with the inclination inferred from the dust disks, assumed to be circular and in equilibrium in the galaxies equatorial plane. Moreover our tests using JAM to recover the inclination of N-body simulated galaxies also shows excellent agreement between the recovered values and the known ones (Paper~XII). Our estimator of the `deprojected' second velocity moment is then defined as
\begin{equation}
\langle v_{\rm rms}^2 \rangle_e = \langle v^2 + \sigma^2 \rangle_e \equiv
\frac{\sum_{k=1}^P F_k (V_k^2/\sin^2 i + \sigma_k^2)}{\sum_{k=1}^P F_k},
\end{equation}
where $i$ is the inclination of the best-fitting JAM models (A), $V_k$ and $\sigma_k$ are the stellar velocity and dispersion, extracted via pPXF adopting a Gaussian line-of-sight velocity distribution, and $F_k$ is the flux contained within that bin, for the $P$ Voronoi bins \citep{Cappellari2003} falling within the `effective' ellipse of major axis $R_e^{\rm maj}$ and ellipticity $\varepsilon_e$ (Table~1). We found that $\langle v_{\rm rms}^2 \rangle_e$ agrees with $\sigma_e$ with an rms scatter of $\Delta=0.025$ dex, consistent with our random errors. $\langle v_{\rm rms}^2 \rangle_e$ did not improve any of our correlations with respect to the much simpler and robust $\sigma_e$, which also has the key advantage of not requiring spatially-resolved IFU kinematics. For this reason we will not present any relation using $\langle v_{\rm rms}^2 \rangle_e$.

The result of this exercise clearly shows that the existence of the FP is due to the fact that galaxies can be remarkably well approximated by virialized stellar systems with an $M/L$ that varies systematically with their properties. These facts have been clearly realized since the discovery of the FP \citep{Faber1987} and have been generally assumed in most recent studies \citep[see][for a full discussion]{Ciotti2009}. The new findings on the tilt of the FP agree with a similar study of scaling relations in ETGs using accurate dynamical models and integral-field kinematics of a sample of just 25 galaxies \citep{Cappellari2006} and with independent confirmations using strong gravitational lensing \citep{Bolton2007mp,Bolton2008,Auger2010}. Galaxy structural non-homology has a minor effect at best, when the determination of galaxy scaling parameters is pushed to the maximum accuracy and an attempt is made to remove the most important biases. 

The level of accuracy at which the simple structural and dynamical homology approximation holds is not entirely expected however, given the apparent complexity of galaxy photometry and kinematics. Of course the dynamical models assume equilibrium and rigorously satisfy the virial equations. One may think that a tight relation is a necessary feature of the approach. This is however not correct. It is true in fact that the  models satisfy the scalar virial equation $2T + W=0$ by construction, where $T$ is the total kinetic energy and $W$ is the total potential energy. However, given the complex multi-component nature of galaxies, the presence of bars, the importance of projection and the fact that the potential energy should include dark matter, it is far from obvious that one should be able to define any simple empirical measure of projected radius on the galaxy, and a measure of velocity dispersion within a limited region, so that the virial equation can be written in the simple form $M_{1/2}=k\, \sigma^2 R/G$ (designed for spherical homologous systems), with fixed exponents and nearly constant coefficient for the entire population!

\subsection{Simple mass estimators}
\label{sec:simple_mass_estimators}

\begin{figure}
\centering
\includegraphics[width=0.75\columnwidth]{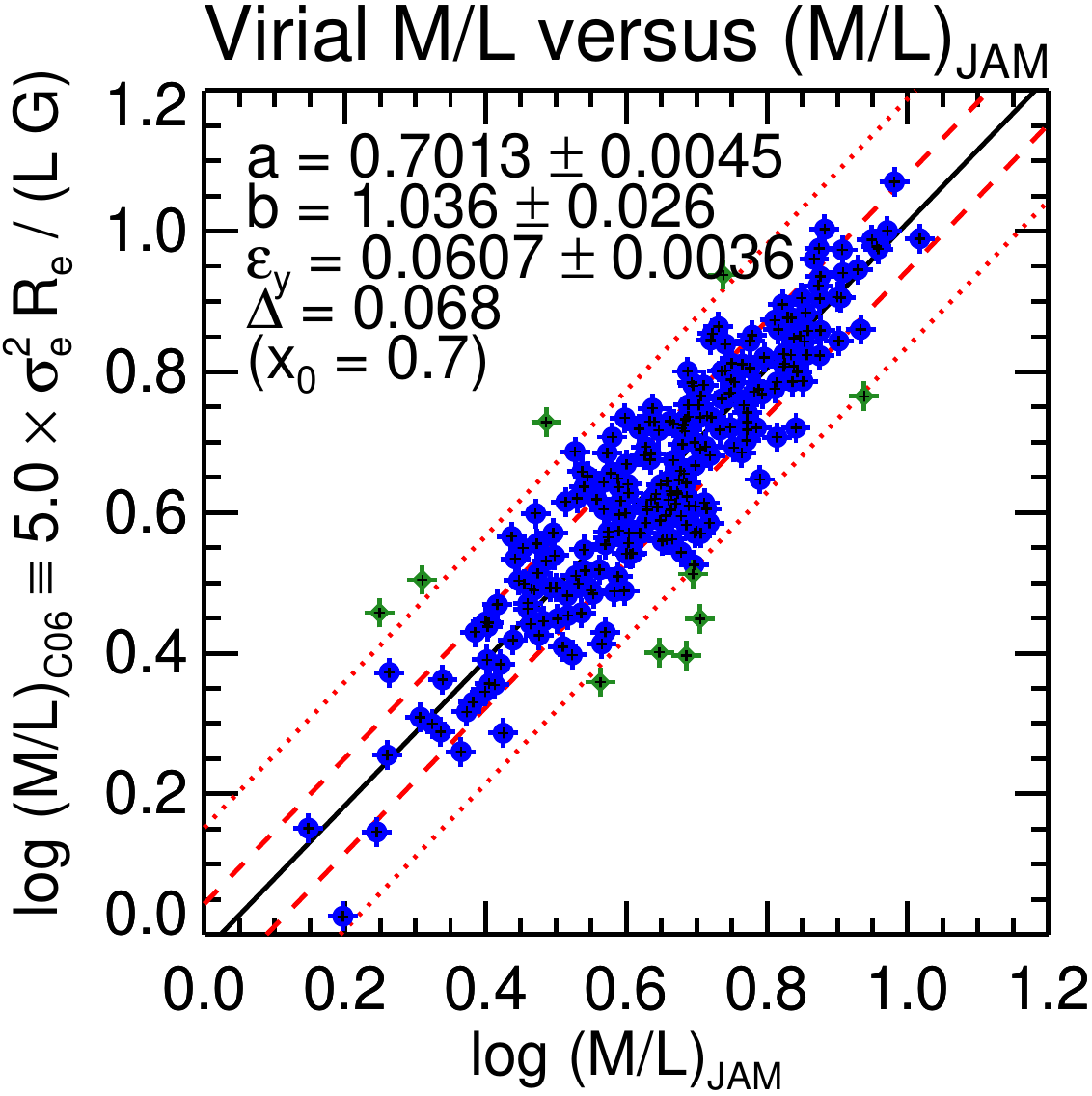}
\caption{Accuracy of the simple virial estimate. Comparison between the virial estimator of \citet{Cappellari2006} and the more accurate JAM values. The inferred  rms errors in the estimation of $M/L$ are 15\%. Symbols and lines are as in \reffig{fig:classic_fundamental_plane}.}
\label{fig:comparing_c06_jam_ml}
\end{figure}

In \reffig{fig:comparing_c06_jam_ml} we present a direct comparison between the new $(M/L)_{\rm JAM}$ estimates, which approximate the total $M/L$ within an iso-surface with volume $V=4\pi R_{\rm e}^3/3$, and the simple virial estimate of \refeq{eq:cap06} from \citet{Cappellari2006}. Considering the modelling errors of 5\% in $M/L$ estimated in this paper, we infer an error of 15\% in the virial estimates. This shows that, although the virial estimates do not suffer from strong biases, they provide errors about a factor 3 larger, even when using our good data.

Our finding does not seem to agree with the small systematic offsets recently reported by \citet{Thomas2011}. The disagreement may be an effect of small sample statistics and larger errors, given that they studied only 16 objects and did not use integral-field data. However, even more likely is that the difference they find may be due to a systematic difference in their \re\ determination, with respect to the \sauron\ ones. Our new empirical confirmation of the scaling of the coefficient in \citet{Cappellari2006}, even in the presence of dark matter, also emphasizes the importance of using virial coefficients that are calibrated to the extent of the available kinematic data. The coefficient $k=3.75$ given by \citet{Spitzer1969} or $k=3$ proposed by \citet{Wolf2010} for \refeq{eq:m12_virial} should not be used to estimate central masses in early-type galaxies, where stellar kinematics out to at most a couple of \re\ is available and the corresponding value $k\approx1.9$ of \refeq{eq:cap06m12} applies. The difference of the two coefficients is due to the fact that, while the estimator of \citet{Wolf2010} is a theoretical one, designed for spherical geometry, very extended kinematics, and assumes galaxy profiles are known to infinite radii, the one by \citet{Cappellari2006} is an empirical one, designed for quantitative measures of masses in the central regions of ETGs. Both estimators are useful in their own range of applicability, but they should not be used interchangeably, unless one can tolerate systematic biases of $\approx60\%$ in the absolute mass normalization.

\begin{figure}
\centering
\includegraphics[width=\columnwidth]{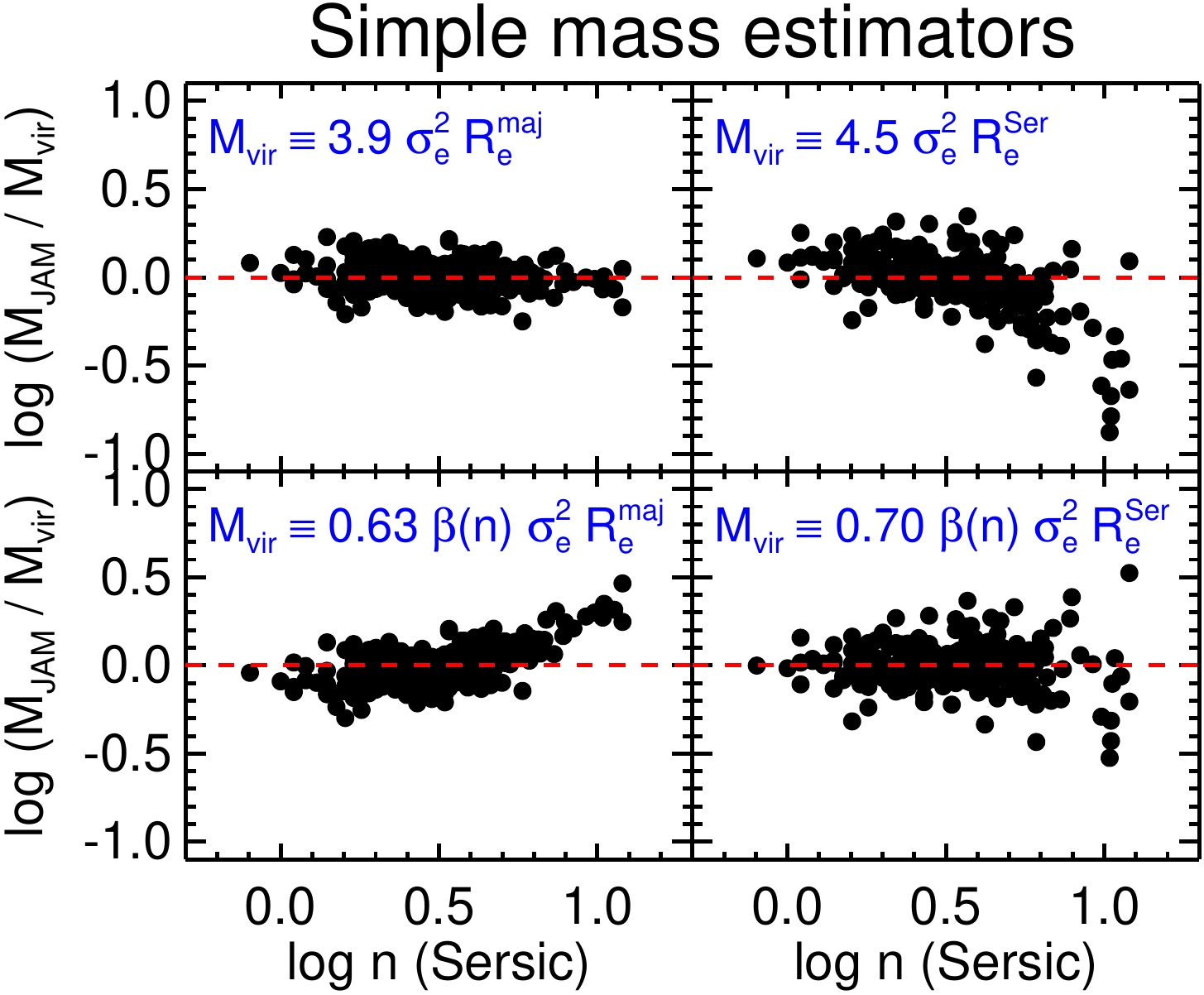}
\caption{Comparing simple mass estimators. The ratio between the mass predicted by different estimators (written in blue in the panels), all based on the scalar virial equation, is compared to the rigorous determination using JAM models, and plotted as a function of the Sersic index $n$ from Paper~XVII. The best estimator is the one in the top left panel, which measures \re\ from the data without extrapolation, and uses a fixed virial coefficient. When \re\ is measured from a Sersic fit to the profile extrapolated to infinite radii, the virial coefficient $\beta(n)$ needs to vary as a function of $n$, but the scatter in the recovered mass is large, especially for large $n$ (bottom right panel). Here we used the theoretical prediction for $\beta(n)$ of equation~(20) in \citet{Cappellari2006}}
\label{fig:simple_mass_estimators}
\end{figure}

In \reffig{fig:simple_mass_estimators} we compare the ability of different simple mass estimators $M_{\rm vir}$, all based on the scalar virial equation, to properly reproduce $M_{\rm JAM}$. We show trends as a function of the \citet{Sersic1968} index $n$ obtained for our galaxies by fitting a single Sersic profile to the entire galaxy (for both E and S0 galaxoes) and given in table C1 of Paper~XVII.
Our preferred estimator, which uses a fixed virial coefficient and the semi-major axis $R_e^{\rm maj}$ of the effective isophote, reproduces $M_{\rm JAM}$ better than any alternative one. It has no detectable trend with the Sersic index and presents the smallest scatted (0.08 dex rms; top left panel). The best-fitting coefficient of this estimator is smaller that the value 5.0 determined in \citet{Cappellari2006}. This accounts for the fact that $R_e^{\rm maj}$ is systematically larger than the circularized radius $R_e$. In the bottom left panel we change the virial coefficient $\beta(n)$ according the the predictions of spherical isotropic models with a Sersic profile \citep[e.g.][]{Prugniel1997fp,Bertin2002fp}. For this we adopt the expression in equation~(20) of \citet{Cappellari2006}, which was calculated for $\sigma_e$ measured in an aperture of radius $R_e$ as adopted here. The plot shows a clear trend as a function of $n$, with a systematic bias of up to a factor three for the largest $n$. This confirms that when $R_e$ is measured without extrapolation of the data as done here, or in the `classic' way \citep{Burstein1987,Jorgensen1995phot,Cappellari2006}, using growth curves with fixed $n=4$, a constant virial coefficient performs better than one that changes with $n$.

The opposite is true when $R_e^{\rm Ser}$ is the value obtained from Sersic fits of the galaxy profile, assuming the galaxy is described by that functional form out to infinite radii (from table C1 of Paper~XVII). These \re\ can be significantly different from the non-extrapolated values. Given that for both the $R_e^{\rm Ser}$ and $R_e^{\rm maj}$ determinations we used the very same images, the differences are entirely due to the {\em assumed} functional form of the surface brightness beyond the region where we have data.
In this situation, adopting a fixed virial coefficients causes severe biases. The systematic bias in the virial estimator can be essentially removed adopting for $\beta(n)$ the analytic prediction of \citet{Cappellari2006}. However the scatter in this estimator is significantly larger (0.13 dex rms) than the non-extrapolated one. Moreover deviations are particularly large (up to a factor three) at the largest $n$, where a larger fraction of the total galaxy light is not actually observed on the images, but just extrapolated. Also note that an extra factor 0.70 is needed in addition to the theoretically predicted coefficient $\beta(n)$. This factor must be calibrated empirically and makes the absolute normalization of the masses determined with this simple estimator rather uncertain.

These tests illustrate the extreme sensitivity of the reliability of masses estimated using the scalar virial equation, on the technique adopted to measure \re.
They also show the difficulty of obtaining masses that are properly normalized. Ultimately the general unreliability and poor reproducibility of effective radii determined from photometry of different quality is the main limiting factor to a quantitative use of the scalar virial relations to measure accurate masses or $M/L$, when a proper absolute normalization is essential, like in IMF studies of distant galaxies \citep{Cappellari2009}. If different methods or extrapolations, applied to different, but high-quality photometric data of local galaxies, can produce revisions in \re\ by as much a s a factor of two (\citealt{Kormendy2009}; see also \citealt{Chen2010}), more significant biases should be expected when comparing local and high-redshift observations, as already pointed out by \citet{Mancini2010}. When biases in \re\ are present, only dynamical models can still provide robust central masses and $M/L$, due to the near insensitivity of the models to the shape of the outer mass and light profiles \citep{vanderMarel2007,vanderWel2008vdM,Cappellari2009}.


\subsection{The $(M/L)-\sigma_e$ relation}
\label{sec:ml_sigma}

In the previous sections we showed that the existence of the Fundamental Plane can be accurately explained by the virial relation combined with a smooth variation of the $M/L$. Here we study the previously reported correlation $(M/L)\propto\sigma_e^{0.8}$ (in the $I$-band) between the effective velocity dispersion and the dynamical $M/L$ within a sphere of radius \re\ \citep{Cappellari2006,vanderMarel2007}. This relation was previously found to provide the tightest relation among other parameters of scaling relations (dynamical mass, luminosity or size), with an observed scatter of 18\% and an inferred intrinsic one of just $\sim$13\%, when using integral-field kinematics.

\begin{figure*}
\includegraphics[width=0.49\textwidth]{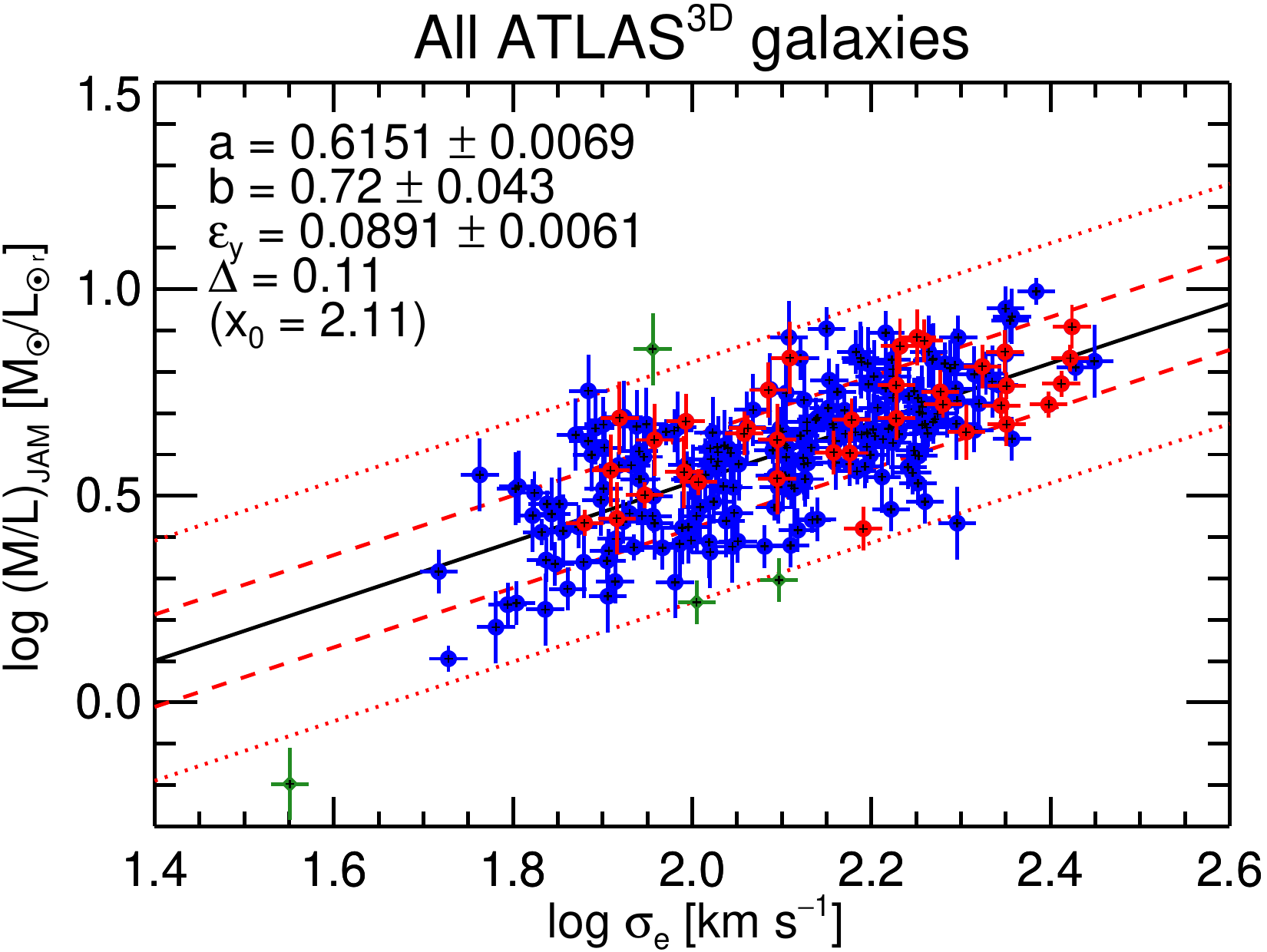}
\includegraphics[width=0.49\textwidth]{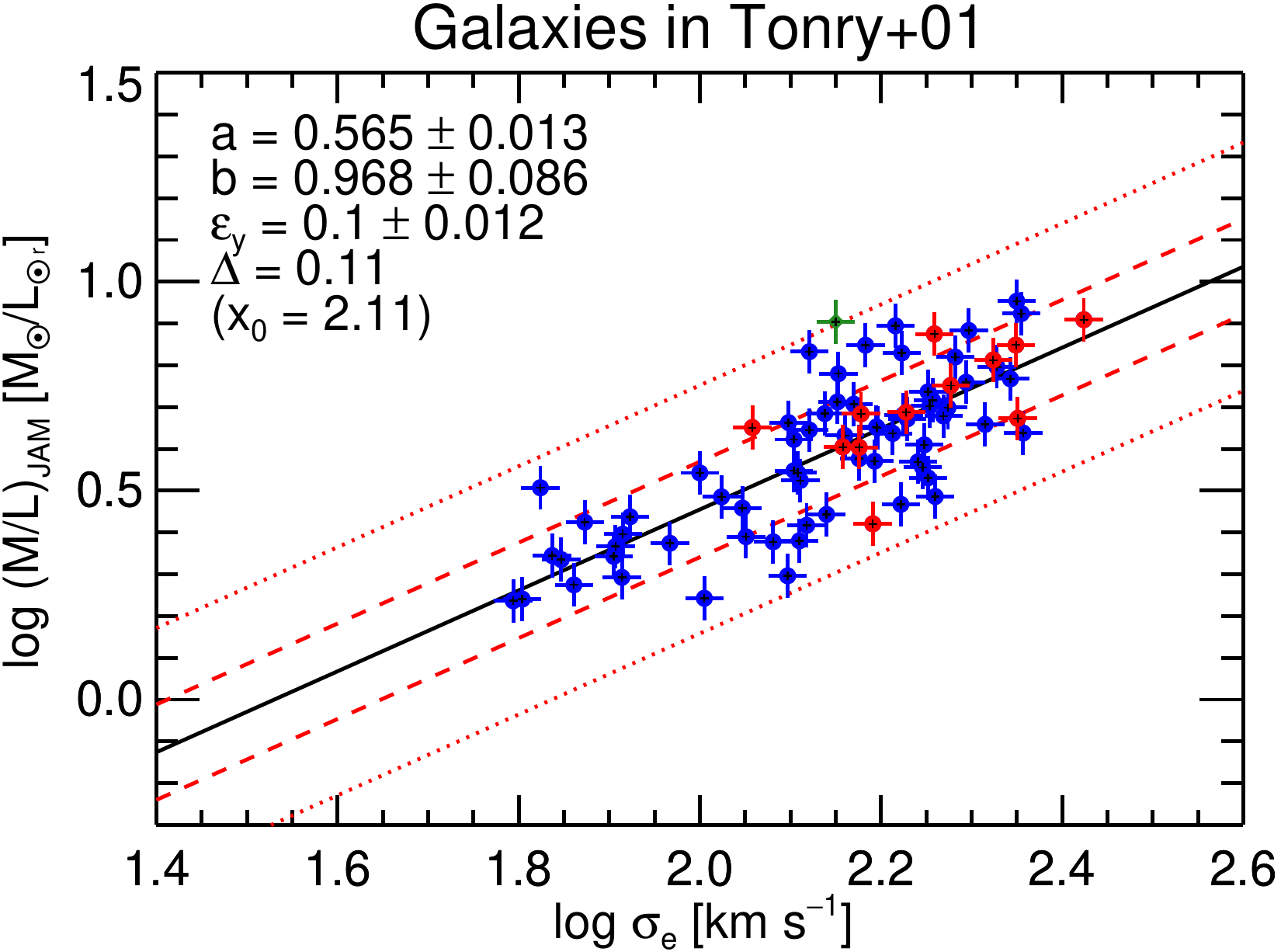}
\includegraphics[width=0.49\textwidth]{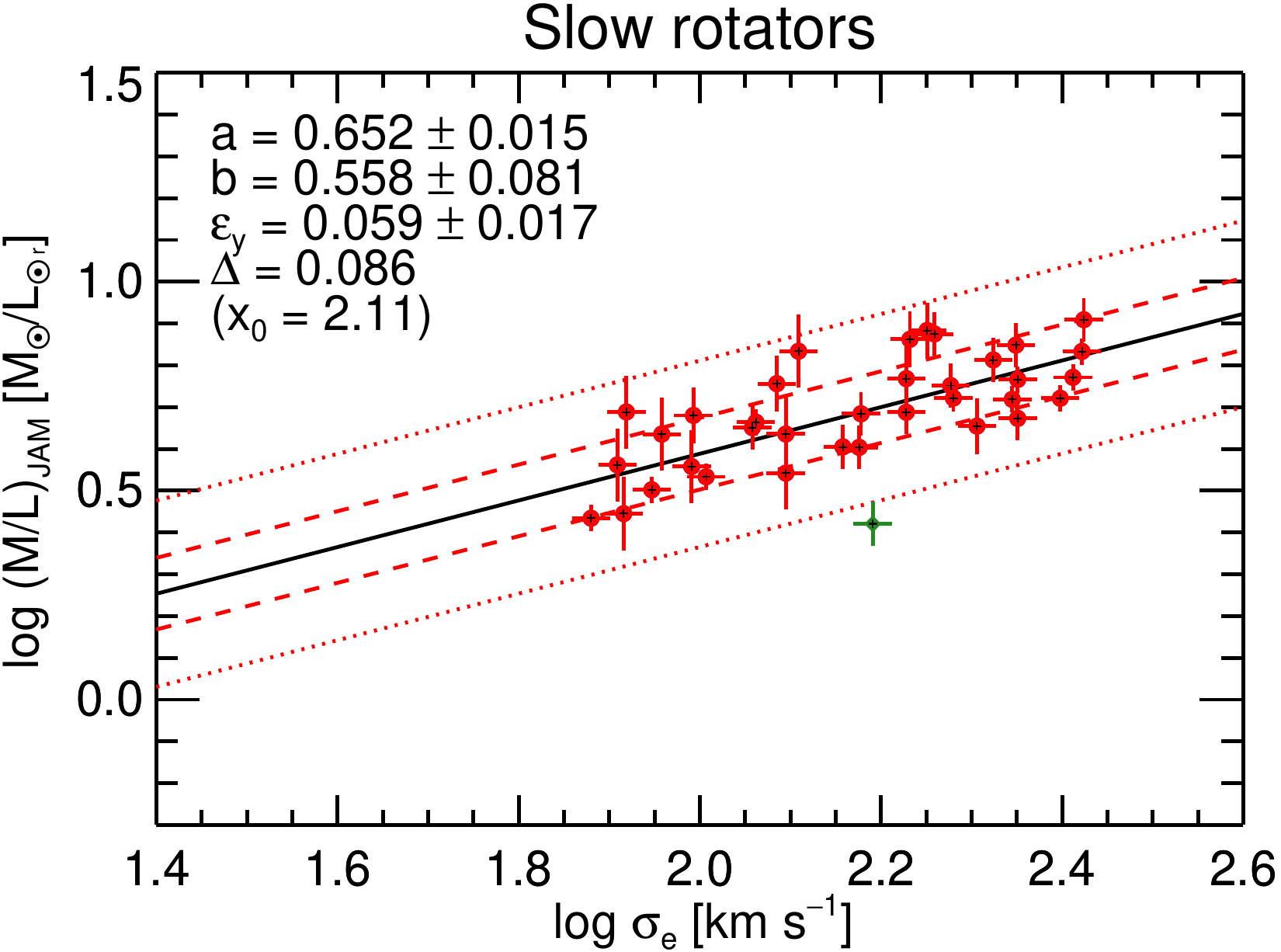}
\includegraphics[width=0.49\textwidth]{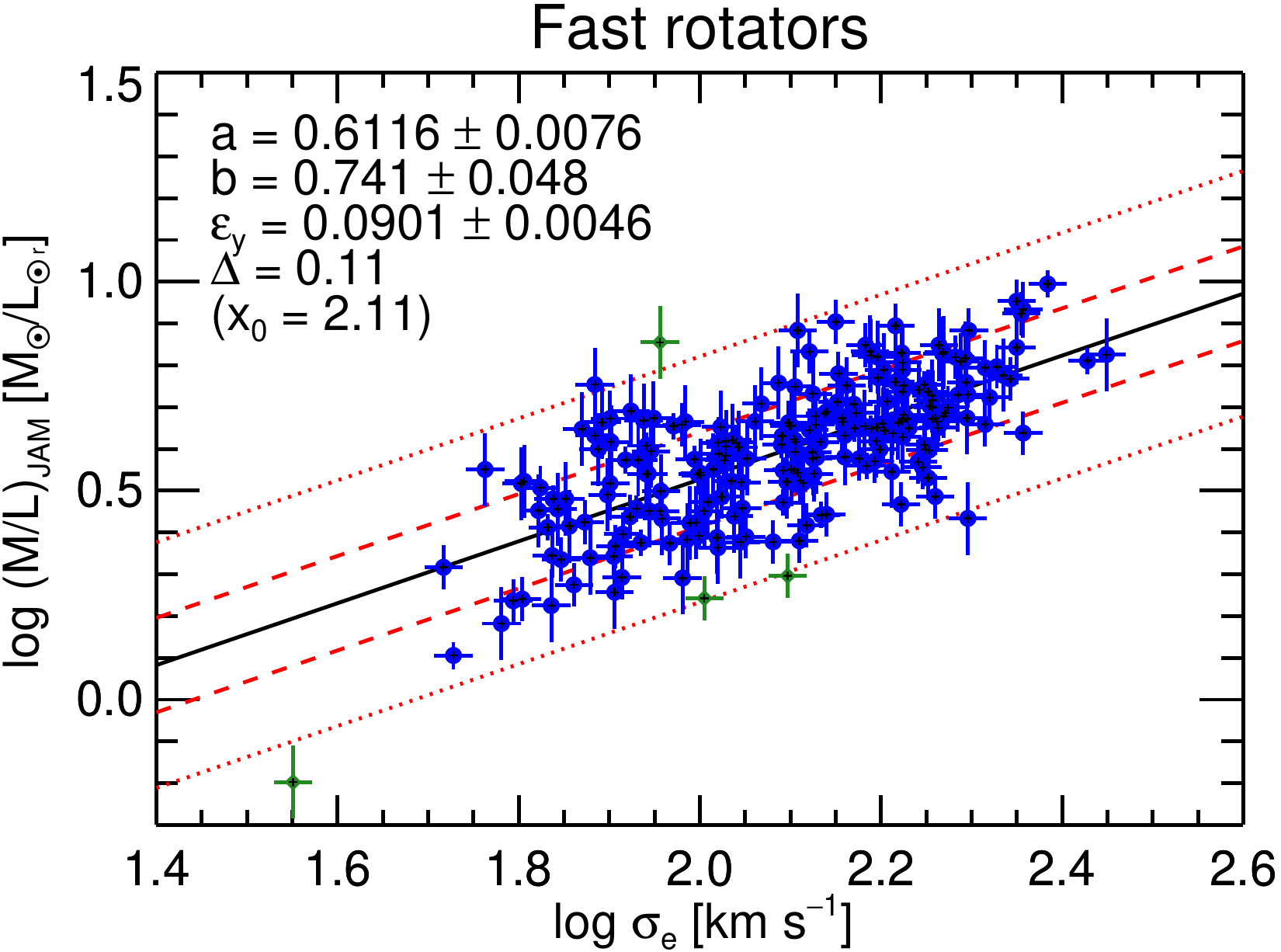}
\includegraphics[width=0.49\textwidth]{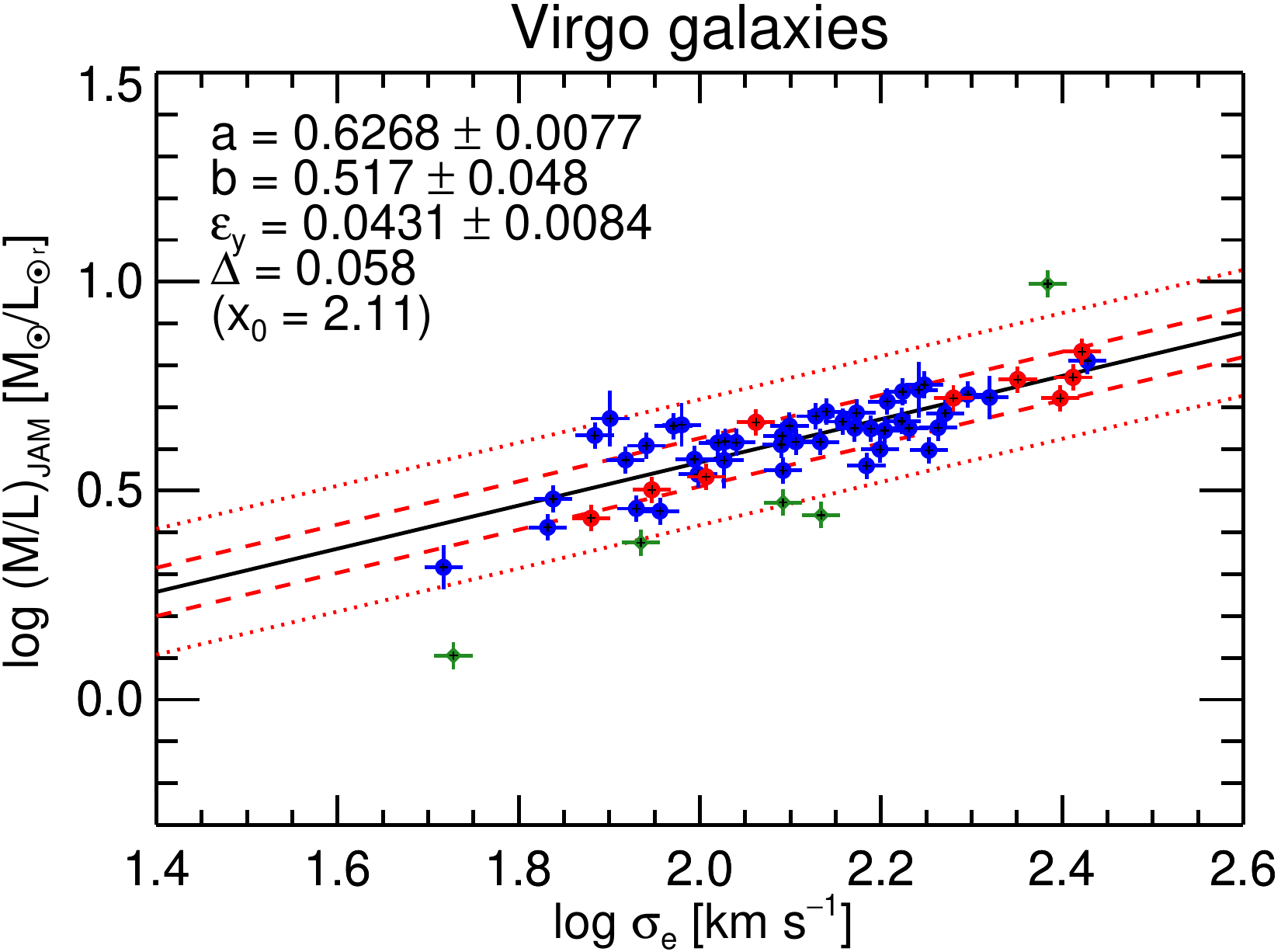}
\includegraphics[width=0.49\textwidth]{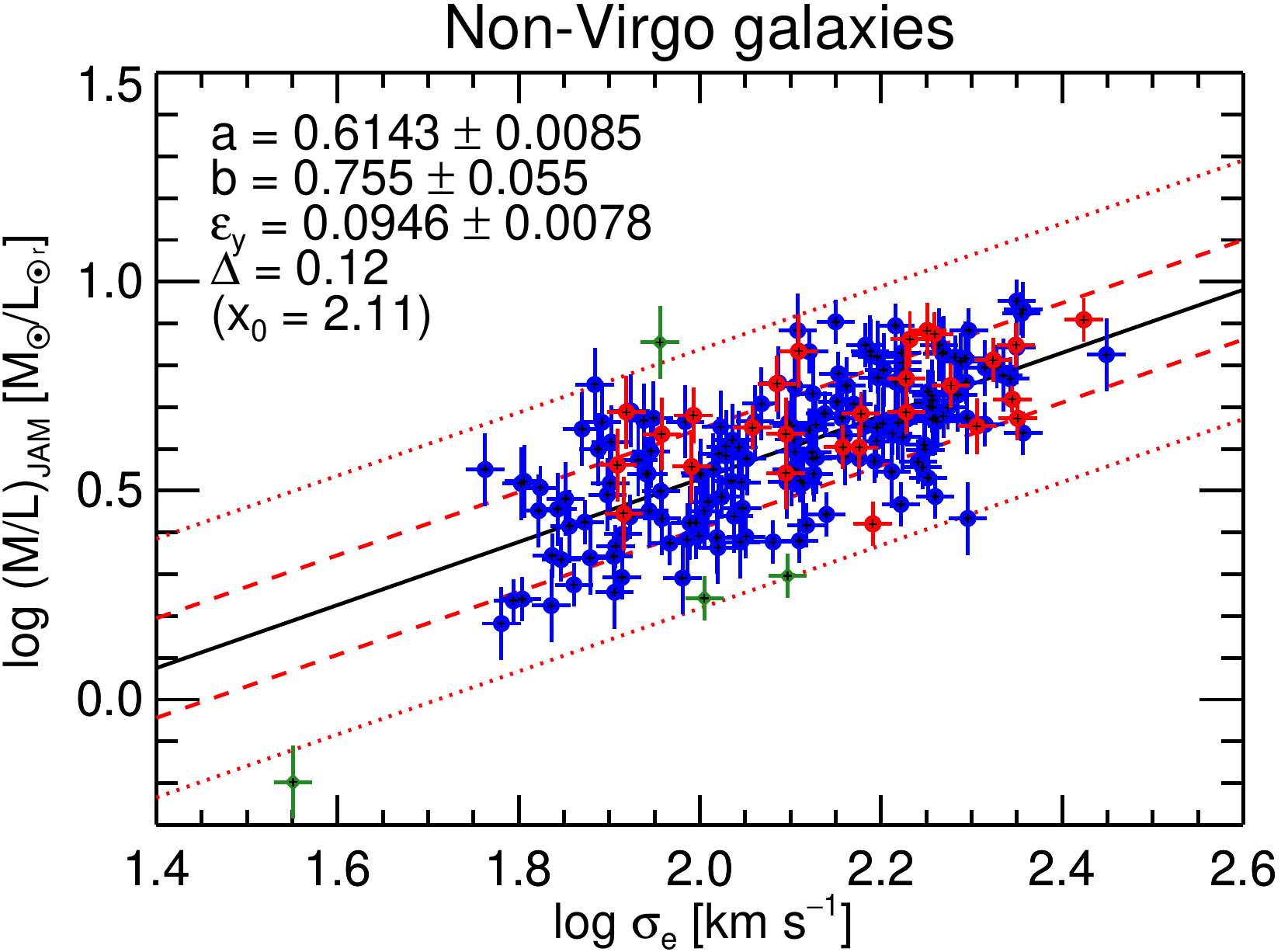}
\caption{The $(M/L)_e-\sigma_e$ relation. From left to right and from top to bottom the relation is shown (i) for all \atl\ galaxies; (ii) for the subset in \citet{Tonry2001}; (iii) for the subset of slow rotators (from Paper~III); (iv) for the subset of fast rotators (from Paper~III); (v) for the subset of galaxies in the Virgo cluster; (vi) for subset not in the Virgo cluster. In all plots the blue symbols are fast rotators , while red symbols are slow rotators. Green symbols are outliers excluded from the fit by \textsc{lts\_linefit}.}
\label{fig:ml-sigma}
\end{figure*}

The $(M/L)-\sigma_e$ relation for the full \atl\ sample is shown in the top-left panel of \reffig{fig:ml-sigma}. Our new relation has an observed scatter of 29\%, from which we infer an intrinsic scatter of 23\%, when combining our 5\% errors in the models with the distance errors for the various subsamples as described in section~2.2 of Paper~I. We adopted as distance errors the median one for each given class of determinations reported in Paper~I, instead of the individual errors, which are not easy to trust in every case, and that are likely dominated by systematics. The scatter is significantly larger than the previously reported one. The new relation has a formally accurate power slope of $b=0.720\pm0.043$, which is a bit shallower than the previous one, based on a sample ten times smaller than the current one. 

To understand the reason for the differences between our $(M/L)-\sigma_e$ slope and previous determinations, in the top-right panel \reffig{fig:ml-sigma} we plot the $(M/L)-\sigma_e$ relation for the subset of 78 galaxies with SBF distances from \citet{Tonry2001}, as done in both \citet{Cappellari2006} and \citet{vanderMarel2007}.
The relation for this subset now steepens and becomes even steeper than the previous determinations. The reason for this is likely related to the fact that the \citet{Tonry2001} subsample is biased towards elliptical galaxies, which tend to be the brightest in our sample. A change in slope is then expected from the curvature of the $(M/L)-\sigma_e$ relation, which is not clearly visible in our range of $\sigma$ values, but is implied by the deviations from our relation when other classes of objects with smaller of larger $\sigma$ are considered \citep{Zaritsky2006,Zaritsky2008,Tollerud2011}. A small but systematic increase in the slope is indeed visible when we select subsamples within different $\sigma$ ranges from our \atl\ sample.
We conclude that the minor difference between our newly fitted value and the previous works is due to the difference in the sample selection. The present sample is not only much large than the one used in previous studies, but also volume-limited so it provides a statistically representative view of the scaling relations in the nearby Universe.

In the middle-left panel of \reffig{fig:ml-sigma} we show the $(M/L)-\sigma_e$ of the 36 slow rotator ETGs defined in Paper~III and for the fast rotators. We confirm a detectable offset in the relation, with the slow rotators having slightly larger $M/L$ than fast rotators, as previously reported in \citet{Cappellari2006}. However, the difference is just at the 9\% level. There is also a change in the slope, with the slow rotators defining a more shallow relation that the full population. We also confirm the smaller scatter in the relation, as reported by \citet{FalconBarroso2011} for the colour-$\sigma$ and FP relations. The slow rotators have an observed scatter of 22\%, and an inferred intrinsic one of 15\% in the $(M/L)-\sigma_e$ relation. This is likely due to the fact that significant amounts of cold gas and star formation, which affect the $M/L$ but not $\sigma$, are in fast rotators (Paper~IV, McDermid et al. in preparation). The relation for the fast rotators (middle-right panel) agrees with the global one, as expected from the fact that they dominate the \atl\ sample.

The dependence of the slope and zero point of the $(M/L)-\sigma_e$ relation on environment effects is shown in the bottom panels of  \reffig{fig:ml-sigma}. As discussed in Paper~VII, most of the environmental differences in the \atl\ sample can be characterized by whether a galaxy belongs to the Virgo cluster or not. The left panel shows the 58 \atl\ galaxies in Virgo. They follow the same shallow relation as the slow rotators, but with the zero point of the global relation. The observed scatter decreases to just 14\%, in part due to the accurate distances from ACSVCS \citep{Mei2007}. However, the  intrinsic scatter $\Delta (M/L)$ also further decreases to just 10\%. This is consistent with the intrinsic scatter measured by \citet{Cappellari2006}, using a radically different set of models and different distance estimates (no ACSVCS), but on a sample that, contrary to the \atl\ sample, was dominated by Virgo galaxies. The decrease in the scatter must be related to the decrease in the fraction of young objects in Virgo (\citealt{Kuntschner2010}; McDermid et al. in preparation). It again confirms that the scatter of the $(M/L)-\sigma_e$ relation is dominated by stellar population (including IMF) effects, as previously demonstrated for the FP. The two results are two ways of looking at the same thing, given that the $(M/L)-\sigma_e$ relation is the projection of the differences between the FP and MP along the $\sigma_e$ axis. For completeness we also show in the bottom-right the relation for non-Virgo galaxies, which dominate the sample and again are consistent, albeit a bit steeper, than the global relation.

\begin{figure}
\plotone{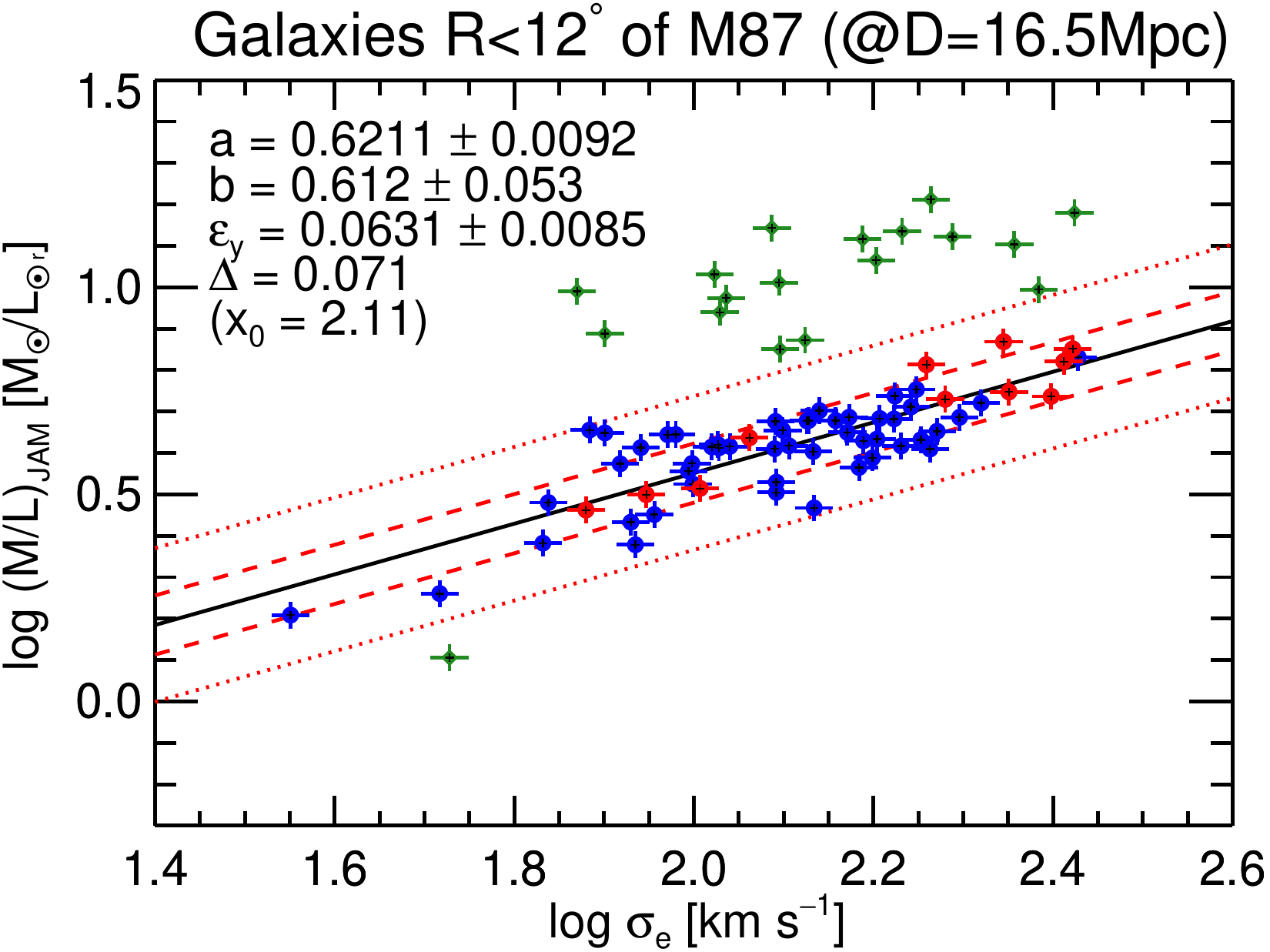}
\plotone{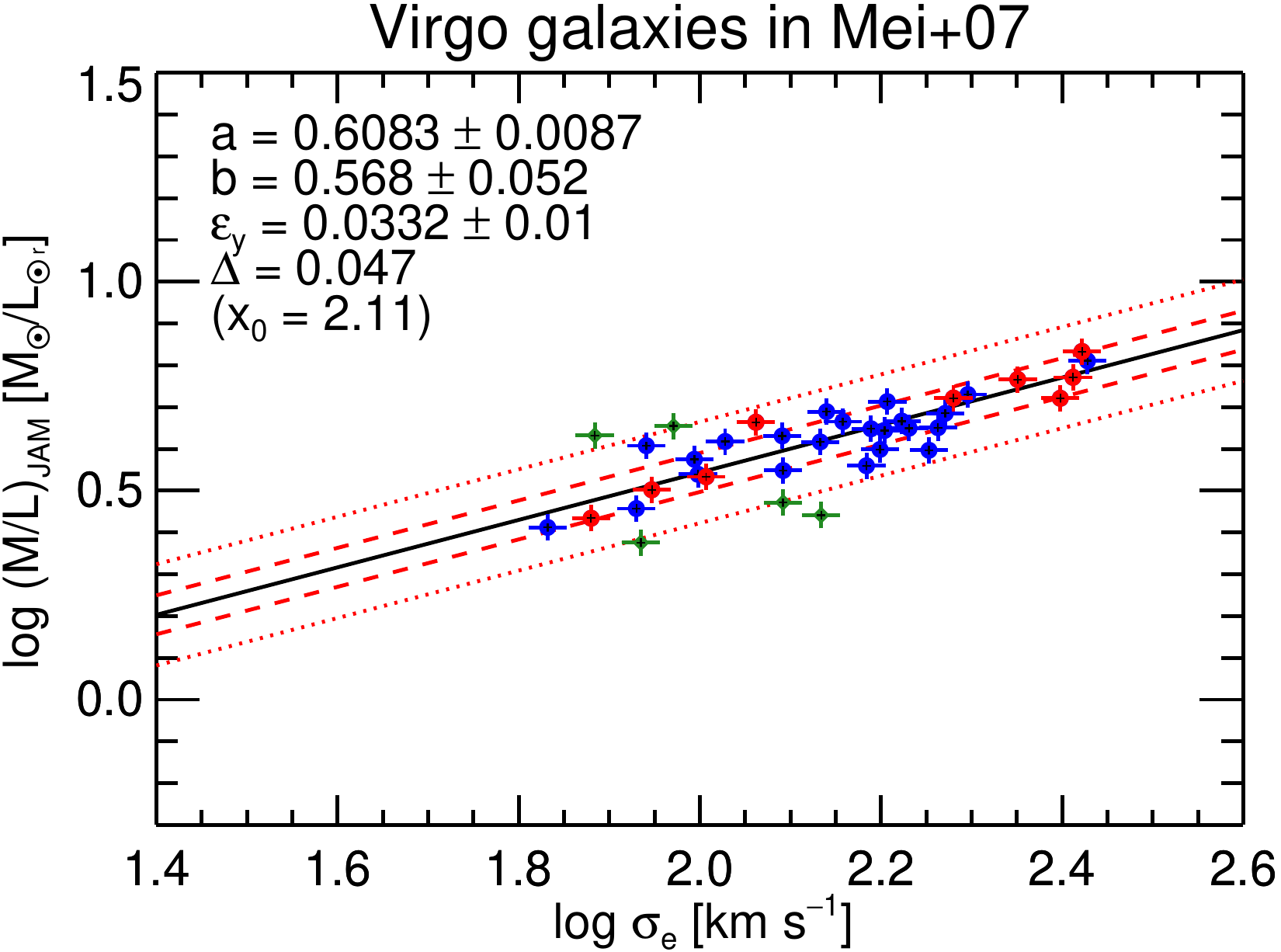}
\caption{Scatter in the $(M/L)_e-\sigma_e$ relation in the Virgo galaxy cluster. {\em Top Panel:} All \atl\ galaxies within 12$^\circ$ of the center of the Virgo cluster have been assigned a fixed distance of $D=16.5$ Mpc. The measured $M/L$ naturally defines a clean $(M/L)_e-\sigma_e$ relation for galaxies belonging to the cluster. The scatter in this relation is due to a combination of the cluster depth and the intrinsic scatter in $M/L$. {\em Bottom Panel:} $(M/L)_e-\sigma_e$ relation for the galaxies in \citet{Mei2007}. The accurate distances produce a quite significant decrease in the observed scatter, down to just 11\%, indicating that both the $(M/L)_{\rm JAM}$ and the SBF distances are significantly more accurate than this value and confirming that the SBF distances are able to resolve the spatial structure of Virgo, along the LOS, as claimed.}
\label{fig:ml_sigma_virgo}
\end{figure}

In the top panel of \reffig{fig:ml_sigma_virgo} we show how the tightness of the $(M/L)-\sigma_e$ relation can be used to cleanly select galaxies belonging to the Virgo cluster. Here we selected all \atl\ galaxies contained within a cylinder of radius of $R=12^\circ$ centred on the Virgo cluster (approximately at the location of the galaxy M87) and assigned to all of them the cluster distance of $D=16.5$ Mpc from \citet{Mei2007}. We then used the \textsc{lts\_linefit} routine to fit a line. Even in the presence of 20 dramatic outliers out of 79 objects, the method is able to robustly converge to a clean relation.\footnote{Other robust method like (i) minimizing the absolute deviation, (ii) using iterated biweight estimates or (iii) M-estimates \citep[section~15.7]{Press2007}, failed to provide a sensible solution to this problem.} The method selects 59 galaxies within the 99\% (2.6$\sigma$) confidence bands from the best-fitting relation. The plot reveals a tight sequence in the $(M/L)-\sigma_e$, which corresponds to galaxies in the Virgo cluster, with an observed scatter of $\Delta (M/L)=0.071$ (18\%). It is reassuring to see that this relation, which uses no individual distance information for the galaxies, agrees both in the slope and zeropoint with the ones for all \atl\ galaxies, even though it has smaller scatter. Galaxies above the relation lie in the background of Virgo, and their difference in distance modulus from Virgo is 2.5$\times$ the difference in $\log (M/L)$ from the best-fitting relations. In this fit we assume that the distance error are due to the $1\sigma$ depth of the Virgo cluster. Adopting the value of $\sigma_D=0.6\pm0.1$ from \citet{Mei2007} we derive an intrinsic scatter in $M/L$ of $\varepsilon_{M/L}=0.063$ dex (16\%).

When we select only the galaxies with SBF distances from the ACSVCS \citep{Mei2007} (bottom panel of \reffig{fig:ml_sigma_virgo}), we find a relation with the same slope, but a decreased observed scatter of $\Delta (M/L)=0.047$ (11\%). For this relatively small, but still statistically significant sample of 32 galaxies, the inferred intrinsic scatter in $M/L$ would be a mere 8\%! Considering that ETGs appear to have very small fractions of dark matter in their central region (\reffig{fig:dark_matter_fraction}), a small scatter in dynamical $M/L$ should be expected from the extreme tightness of the colour-magnitude relation in clusters \citep{Bower1992} and specifically for the ACSVCS galaxies \citep{Chen2010}, given that colour is a direct tracer of the $M/L$ of the stellar population \citep{Bell2001}. Our small scatter finding confirms the remarkable accuracy of the ACSVCS SBF distances and their ability to resolve the cluster structure as claimed. It shows that the intrinsic $(M/L)-\sigma_e$ relation is extremely tight, but its study is limited in our sample by the distance errors. It would be valuable to perform a similar analysis as in the top panel of \reffig{fig:ml_sigma_virgo}, with integral-field data and accurate models, in a cluster like Coma, sufficiently close that good stellar kinematics can be obtained, but sufficiently far that errors in the distance can be virtually ignored. The smaller intrinsic scatter inferred for this sample, with respect to the one in the top panel, suggests that, either they are not drawn from the same population, or the ACSVCS sample in \citet{Mei2007} spans a slightly smaller set of distances within the Virgo cluster, than the \atl\ Virgo sample. The tightness of this correlation also places stringent constraints on the possible intrinsic scatter on the ${\rm IMF}-\sigma$ trend that we discuss in Paper~XX. Any IMF trend must satisfy the small scatter that we observe in this relation.

\subsection{Relation between $\sigma_e$ and the maximum circular velocity}

Previous studies \citep{Zaritsky2006,Zaritsky2008,McGaugh2010,Dutton2011imf} have tried to unify dynamical scaling relations of spiral galaxies and early-type galaxies. For spirals one can measure the rotation velocity of the gas, which appears in the \citet{Tully1977} relation between galaxy luminosity (or mass) and its maximum (asymptotic) circular velocity $\max(V_{\rm circ})$, typically measured from the kinematics of the neutral gas at large radii. For early-type galaxies one can measure the velocity dispersion, which enters the \citet{Faber1976} and Fundamental Plane relations. Unification of the scaling relations is done by converting velocity dispersion into the circular velocity $V_{\rm circ}(R_e^{\rm maj})$ at the half-light radius or into the maximum one $\max(V_{\rm circ})$ adopting constant factors.

Typical conversion factors for $V_{\rm circ}(R_e^{\rm maj})$ used in the literature range from $\sqrt{2}$ to $\sqrt{3}$ \citep{Courteau2007vcirc}. For example \cite{Padmanabhan2004} estimates $k\approx1.65$. While \citet{Schulz2010} adopts $k\approx1.7$ and \citet{Dutton2011imf} uses $k\approx1.54$.

\begin{figure}
\plotone{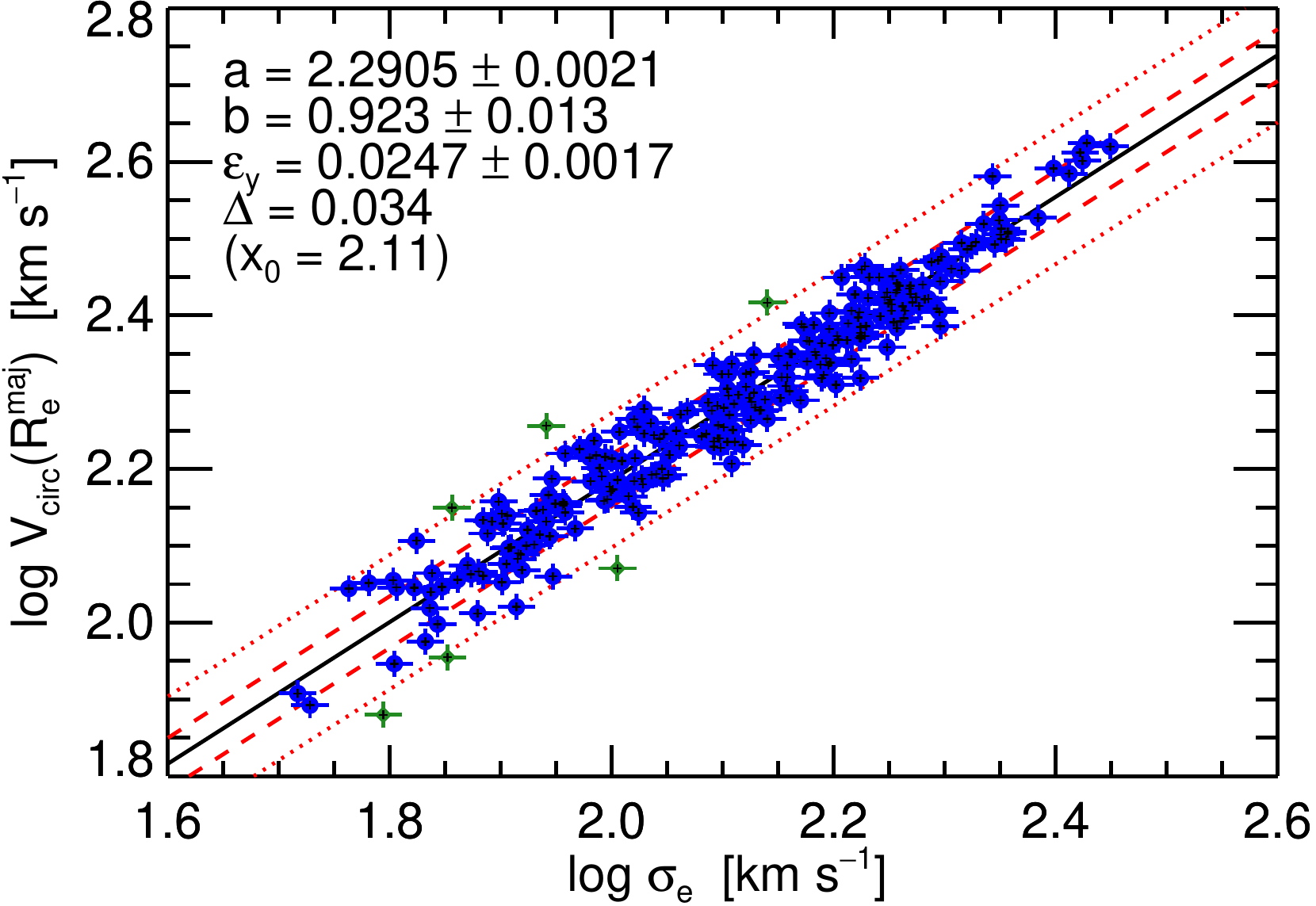}
\plotone{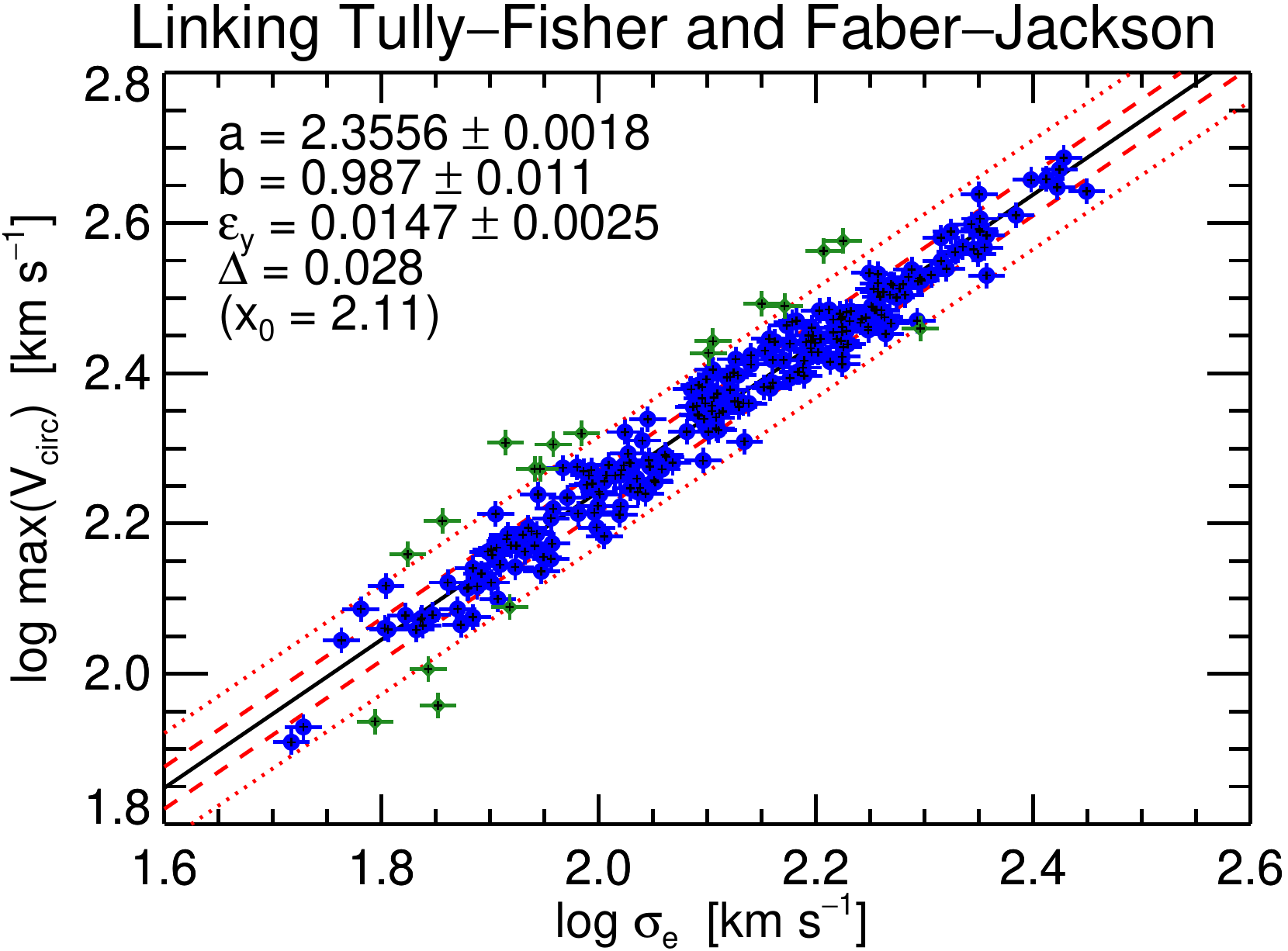}
\caption{Circular velocity $V_{\rm circ}$ versus $\sigma_e$. {\em Top Panel:} correlation between the circular velocity $V_{\rm circ}(1\re)$ inferred from our models at 1\re, and $\sigma_e$. {\em Bottom Panel:} correlation between the peak circular velocity $\max(V_{\rm circ})$ (within 1\re) and $\sigma_e$. 
\label{fig:vcirce_versus_sige}}
\end{figure}

Our dataset provides accurate $\sigma_e$ for all galaxies, together with circular velocities from our dynamical models. This allows for a robust empirical calibration of the relation. The correlation between $\sigma_e$ and $V_{\rm circ}(R_e^{\rm maj})$ is shown in \reffig{fig:vcirce_versus_sige} and the best-fitting relation has the form
\begin{equation}
    V_{\rm circ}(R_e^{\rm maj}) \approx 1.51\times \sigma_e.
\end{equation}
Considering the variety of photometric profile and galaxy flattening in our complete sample of ETGs, it is remarkable that the relation has a scatter of just 8\%, with a weak dependency on $\sigma_e$.

\begin{figure}
\plotone{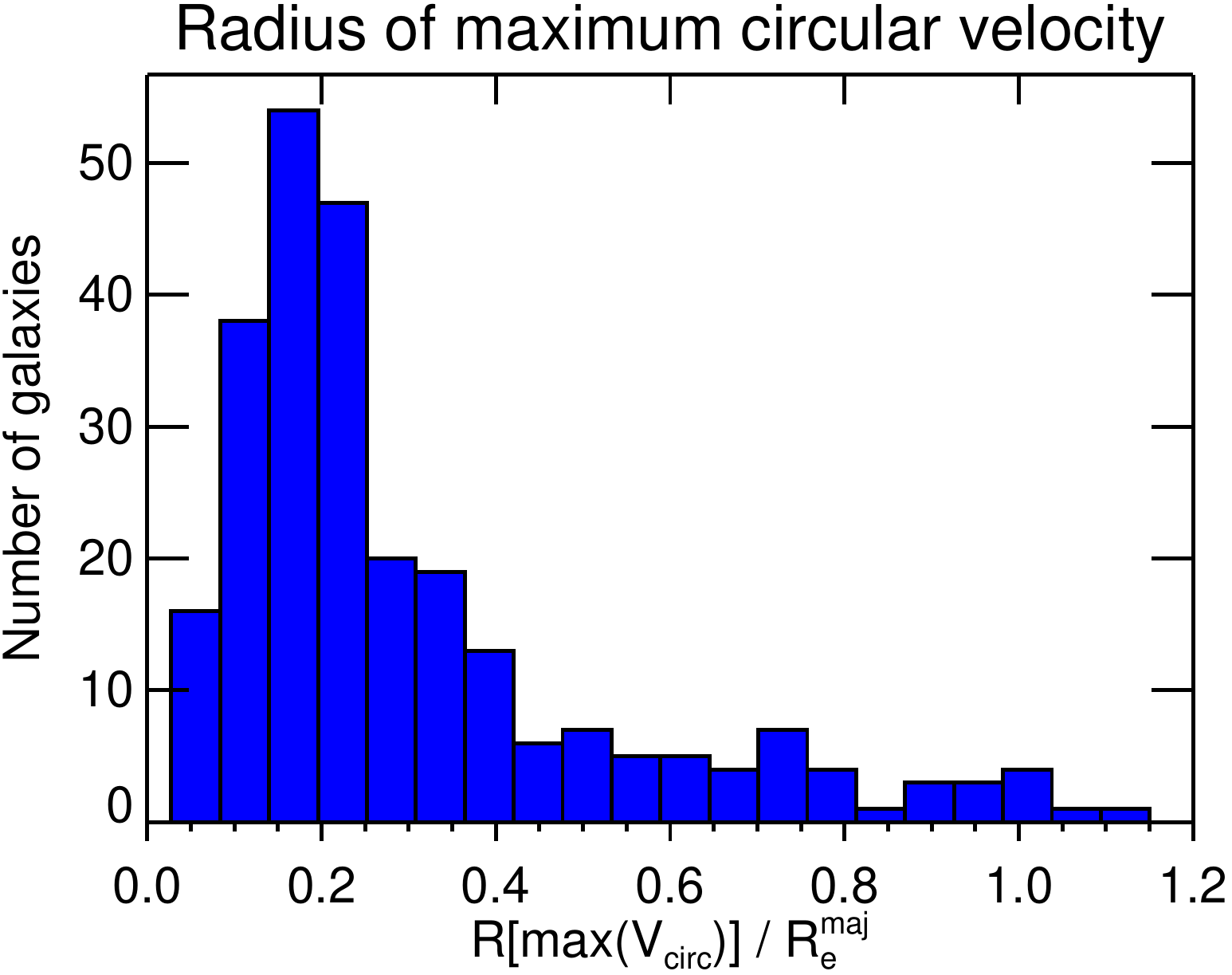}
\caption{Histogram for the distribution of the radius $R/\re$ at which the maximum circular  $\max(V_{\rm circ})$ is reached, as a fraction of the galaxy effective radius \re.
\label{fig:radius_of_vmax}}
\end{figure}

Even slightly tighter is the correlation between $\sigma_e$ and  $\max(V_{\rm circ})$, which has the form
\begin{equation}
    \max(V_{\rm circ}) \approx 1.76\times \sigma_e,
\end{equation}
and an observed scatter of 7\%. Importantly this coefficient show essentially no variation with $\sigma_e$ (the exponent is one within the small errors). The $\max(V_{\rm circ})$ defined here is the peak in the rotation curve within the region where we have stellar kinematics, which is generally within 1\re. 
As shown in \reffig{fig:radius_of_vmax}, the inner maximum in the circular velocity $\max(V_{\rm circ})$ is almost always reached well inside 1\re, with 85\% of the peaks happening at a radius smaller than $\re/2$ and a median radius of just $\re/5$.
At these radii the contribution of the stellar mass totally dominates the total mass. For this reason $\max(V_{\rm circ})$ should not be confused with the asymptotic value of the circular velocity at large radii, where dark matter dominates. The latter is generally used in the \citet{Tully1977} relation (but see \citealt{Davis2011a}). Although the so-called bulge-halo conspiracy \citep{vanAlbada1986} seems to generally make the two peak velocity values similar \citep[e.g.\ see][]{Williams2009}, this fact has never robustly established for a significant sample of ETGs.

\section{Summary}

We construct detailed dynamical models (JAM), based on the Jeans equations and allowing for orbital anisotropy, for the volume-limited and essentially mass-selected \atl\ sample of early-type galaxies. The models fit in detail the two-dimensional galaxy images and reproduce in detail the integral-field stellar kinematics obtained with \sauron\ out to about 1\re, the projected half-light radius. We derive accurate total mass-to-light ratios $(M/L)_e$ and dark matter fractions $f_{\rm DM}$, within a sphere of radius $r=\re$ centred on the galaxies. We infer masses $M_{\rm JAM}\equiv L\times (M/L)_e \approx 2\times M_{1/2}$, where $M_{1/2}$ is the mass within a sphere enclosing half of the galaxy light. We also measure stellar $(M/L)_{\rm stars}$. 

We test the accuracy of our mass determinations by running models with and without dark matter and we find that the enclosed total $(M/L)_e$ is a robust quantity, independent of the inclusion of a dark-matter halo, with an rms accuracy of 5\% and negligible bias. In other words, even using simple mass-follow-light models, one recovers the total enclosed $(M/L)_e$ with good accuracy and small bias. We illustrate the tecniques we use to measure radii and global kinematical quantities from our data, and to robustly fit linear relations or planes to the data, even in the presence of outliers and significant intrinsic scatter. We stress the difficulty of measuring  absolutely calibrated effective radii \re, and we argue againt extrapolation in the profiles, for more reproducible results. Systematic offsets in \re\ determinations are the main limitation for the use of the scalar virial relation for mass estimates, and may affect size comparisons as a function of redshift.

We find that the thin two-dimensional subset spanned by galaxies in the $(M_{\rm JAM},\sigma_e,R_e^{\rm maj})$ coordinates system, which we call the Mass Plane (MP) has an observed rms scatter of 19\%, which would imply an intrinsic one of just 11\%. The MP satisfies the scalar virial relation $M_{\rm JAM}\propto\sigma_e^2 R_e^{\rm maj}$ within our tight errors. However, this is only true if one pays special attention to the methodology employed to determine the galaxy global parameters and in particular, (i) one uses as scale radius the major axis $ R_{\rm e}^{\rm maj}$ of the `effective' isophote enclosing half of the total projected galaxy light (without extrapolating the profile beyond the data), and (ii) one measures the velocity dispersion $\sigma_e$ (which includes rotation and random motions) from a spectrum derived inside that effective isophote. This confirms with unprecedented accuracy previous claims \citep{Cappellari2006,Bolton2008} that galaxies accurately satisfy the virial relations and that the existence of the FP is entirely explained by virial equilibrium plus a systematic variation in the total $(M/L)_e$.

We revisit the $(M/L)_e-\sigma$ relation and measure a marginally shallower observed slope than previously reported. The minor difference can be explained by selection of the sample of galaxies previously used to fit the relations. We find that the correlation depends both on galaxy rotation and environment, in the sense that both for the subsamples of the galaxies in Virgo, or for the subsample of slow rotators,  the relation is more shallow and has a reduced scatter. In the best case, when the most accurate distances are used, the observed scatter drops to 11\% and the intrinsic one is estimated to be a mere 8\%.

We study the correlation between $\sigma_e$ and the circular velocity from the dynamical models. We find that $V_{\rm circ}(R_e^{\rm maj}) \approx 1.51\times \sigma_e$ and $\max(V_{\rm circ}) \approx 1.76\times \sigma_e$. The relations have an observed scatter of 7--8\% and the coefficient is independent on $\sigma_e$.

The accurate global dynamical scaling parameters for the ETGs in the \atl\ sample are used in the companion Paper~XX to explore different projection of the Mass Plane and the variation of galaxy physical parameters.

\section*{acknowledgements}

MC acknowledges support from a Royal Society University Research Fellowship.
This work was supported by the rolling grants `Astrophysics at Oxford' PP/E001114/1 and ST/H002456/1 and visitors grants PPA/V/S/2002/00553, PP/E001564/1 and ST/H504862/1 from the UK Research Councils. RLD acknowledges travel and computer grants from Christ Church, Oxford and support from the Royal Society in the form of a Wolfson Merit Award 502011.K502/jd. RLD also acknowledges the support of the ESO Visitor Programme which funded a 3 month stay in 2010.
SK acknowledges support from the Royal Society Joint Projects Grant JP0869822.
RMcD is supported by the Gemini Observatory, which is operated by the Association of Universities for Research in Astronomy, Inc., on behalf of the international Gemini partnership of Argentina, Australia, Brazil, Canada, Chile, the United Kingdom, and the United States of America.
TN and MBois acknowledge support from the DFG Cluster of Excellence `Origin and Structure of the Universe'.
MS acknowledges support from a STFC Advanced Fellowship ST/F009186/1.
PS is a NWO/Veni fellow.
(TAD) The research leading to these results has received funding from the European
Community's Seventh Framework Programme (/FP7/2007-2013/) under grant agreement
No 229517.
MBois has received, during this research, funding from the European Research Council under the Advanced Grant Program Num 267399-Momentum.
The authors acknowledge financial support from ESO. 
The SAURON observations were obtained at the WHT, operated by the
Isaac Newton Group in the Spanish Observatorio del Roque de
los Muchachos of the Instituto de Astrofisica de Canarias.
We acknowledge the usage 
of the HyperLeda data base (http://leda.univ-lyon1.fr). Funding for
the SDSS and SDSS-II was provided by the Alfred P. Sloan Foundation, the Participating Institutions, the National Science Foundation,
the US Department of Energy, the National Aeronautics and Space
Administration, the Japanese Monbukagakusho, the Max Planck
Society and the Higher Education Funding Council for England.
The SDSS was managed by the Astrophysical Research Consortium 
for the Participating Institutions.

\clearpage
\begin{deluxetable}{lcccccccccccccc}
\tablewidth{0pt}
\tabletypesize{\small}
\setlength{\tabcolsep}{2pt}
\tablecaption{Scaling relations and mass-follows-light JAM models parameters for the \atl\ sample of 260 early-type galaxies.\label{tab:atlas3d_parameters}}
\tablehead{
 \colhead{Galaxy} &
 \colhead{$\log\sigma_e$} &
 \colhead{$R(\sigma)/\re$} &
 \colhead{$\log\sigma_{\rm kpc}$} &
 \colhead{inc} &
 \colhead{$\log(M/L)_{\rm JAM}$} &
  \colhead{$\log V_{\rm circ}^{\rm max}$} &
 \colhead{qual} &
 \colhead{$\log R_e^{\rm maj}$} &
 \colhead{$\log R_e$} &
 \colhead{$\log r_{1/2}$} &
 \colhead{$\log r_g$} &
 \colhead{conc} &
 \colhead{$\varepsilon_e$} &
 \colhead{$\log L$} \\
 \colhead{} &
 \colhead{(\kms)} &
 \colhead{} &
 \colhead{(\kms)} &
 \colhead{(\degr)} &
 \colhead{($M_\odot/L_{\odot r}$)} &
 \colhead{(\kms)} &
 \colhead{} &
 \colhead{(\arcsec)} &
 \colhead{(\arcsec)} &
 \colhead{(\arcsec)} &
 \colhead{(\arcsec)} &
 \colhead{} &
 \colhead{} &
 \colhead{($L_{\odot r}$)}  \\
 \colhead{(1)} &
 \colhead{(2)} &
 \colhead{(3)} &
 \colhead{(4)} &
 \colhead{(5)} &
 \colhead{(6)} &
 \colhead{(7)} &
 \colhead{(8)} &
 \colhead{(9)} &
 \colhead{(10)} &
 \colhead{(11)} &
 \colhead{(12)} &
 \colhead{(13)} &
 \colhead{(14)} &
 \colhead{(15)}
}
\startdata
     IC0560 &       1.879 &        0.97 &       1.885 &   73 &       0.486 &       2.114 &    2 &       1.352 &       1.154 &       1.248 &       1.659 &       0.458 &       0.577 &       9.566 \\
     IC0598 &       1.996 &        1.00 &       1.964 &   75 &       0.439 &       2.215 &    2 &       1.148 &       0.956 &       1.109 &       1.551 &       0.397 &       0.567 &       9.828 \\
     IC0676 &       1.806 &        0.81 &       1.825 &   49 &       0.550 &       2.059 &    0 &       1.460 &       1.396 &       1.477 &       1.989 &       0.366 &       0.259 &       9.660 \\
     IC0719 &       2.108 &        0.98 &       2.019 &   77 &       0.938 &       2.341 &    3 &       1.301 &       1.027 &       1.155 &       1.663 &       0.268 &       0.716 &       9.697 \\
     IC0782 &       1.870 &        1.00 &       1.827 &   67 &       0.670 &       2.086 &    1 &       1.303 &       1.207 &       1.339 &       1.853 &       0.361 &       0.301 &       9.645 \\
     IC1024 &       1.892 &        0.99 &       1.825 &   88 &       0.696 &       2.133 &    2 &       1.332 &       1.078 &       1.276 &       1.807 &       0.236 &       0.687 &       9.471 \\
     IC3631 &       1.551 &        1.00 &       1.647 &   70 &      -0.155 &       1.833 &    0 &       1.235 &       1.062 &       1.188 &       1.635 &       0.388 &       0.535 &       9.742 \\
    NGC0448 &       2.047 &        1.00 &       2.021 &   90 &       0.523 &       2.276 &    2 &       1.235 &       1.002 &       1.203 &       1.664 &       0.375 &       0.656 &       9.919 \\
\enddata
\end{deluxetable}

\label{lastpage}

\end{document}